\documentclass[sn-basic,iicol,pdflatex]{sn-jnl} 
\usepackage{graphicx}
\usepackage{txfonts}
\usepackage{xcolor}
\usepackage{natbib}
\usepackage{hyperref}
\usepackage[normalem]{ulem}
\usepackage{multirow}%

\usepackage{amsmath,amssymb,amsfonts}%
\usepackage{amsthm}%
\usepackage{mathrsfs}%
\usepackage[title]{appendix}%
\usepackage{textcomp}%
\usepackage{manyfoot}%
\usepackage{booktabs}%
\usepackage{algorithm}%
\usepackage{algorithmicx}%
\usepackage{algpseudocode}%
\usepackage{listings}%
\usepackage{orcidlink}

\begin{document} 

\title[The first cut is the \textit{cheapest}: optimizing Athena/X-IFU-like TES detectors resolution by filter truncation]{The first cut is the \textit{cheapest}: optimizing Athena/X-IFU-like TES detectors resolution by filter truncation}


\author*[1]{\fnm{M. Teresa} \sur{Ceballos\href{http://orcid.org/0000-0001-6074-3621}{\includegraphics[width=10pt]{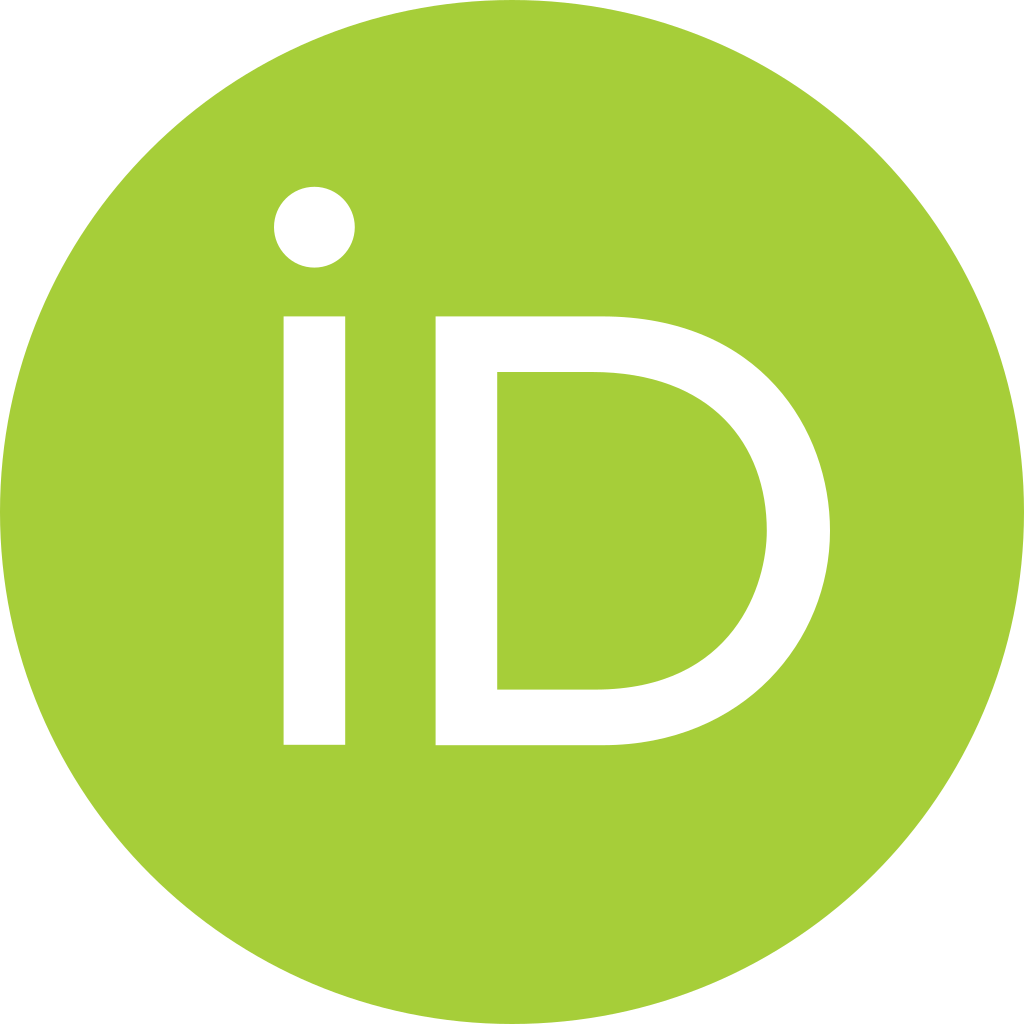}}} }\email{ceballost@unican.es}
\author[2,3]{\fnm{Nicolás} \sur{Cardiel{\href{http://orcid.org/0000-0002-9334-2979}{\includegraphics[width=10pt]{Images/1024px-ORCID_iD.svg.png}}}}}
\author[1]{\fnm{Beatriz} \sur{Cobo{\href{http://orcid.org/0000-0002-7480-9190}{\includegraphics[width=10pt]{Images/1024px-ORCID_iD.svg.png}}}}}
\author[4]{\fnm{Stephen J.} \sur{Smith{\href{http://orcid.org/0000-0003-4096-4675}{\includegraphics[width=10pt]{Images/1024px-ORCID_iD.svg.png}}}}}
\author[4,5]{\fnm{Michael C.} \sur{Witthoeft{\href{http://orcid.org/0000-0001-6237-7776}{\includegraphics[width=10pt]{Images/1024px-ORCID_iD.svg.png}}}}}
\author[6]{\fnm{Philippe} \sur{Peille{\href{http://orcid.org/0000-0002-0695-1662}{\includegraphics[width=10pt]{Images/1024px-ORCID_iD.svg.png}}}}}
\author[7,8]{\fnm{Malcolm S.} \sur{Durkin{\href{http://orcid.org/0000-0001-5640-2525}{\includegraphics[width=10pt]{Images/1024px-ORCID_iD.svg.png}}}}}

\affil*[1]{\orgdiv{Instituto de F\'{\i}sica de Cantabria}, \orgname{CSIC-Universidad de Cantabria}, \orgaddress{\street{Avda Los Castros s/n}, \city{Santander}, \postcode{E-39005}, \country{Spain}}}
\affil[2]{\orgdiv{Departamento de F\'{\i}sica de la Tierra y Astrof\'{\i}sica, Facultad de CC. F\'{\i}sicas}, \orgname{Universidad Complutense de Madrid}, \orgaddress{\street{Plaza Ciencias~1}, \city{Madrid}, \postcode{E-28040}, \country{Spain}}}
\affil[3]{\orgdiv{Instituto de F\'{\i}sica de Part\'{\i}culas y del Cosmos, IPARCOS, Facultad de CC. F\'{\i}sicas}, \orgname{Universidad Complutense de Madrid}, \orgaddress{\street{Plaza Ciencias~1}, \city{Madrid}, \postcode{E-28040}, \country{Spain}}}
\affil[4]{\orgdiv{NASA Goddard Space Flight Centre}, \orgaddress{\street{Greenbelt Road}, \city{Greenbelt}, \postcode{20771}, \state{MD}, \country{USA}}}
\affil[5]{\orgdiv{ADNET Systems, Inc}, \orgaddress{\city{Bethesda}, \state{MD}, \country{USA}}}
\affil[6]{\orgname{CNES}, \orgaddress{\street{18 Av. Edouard Belin}, \city{Toulouse Cedex 9}, \postcode{31401}, \country{France}}}
\affil[7]{\orgname{National Institute of Standards and Technology}, \orgaddress{\city{Boulder}, \state{CO}, \country{USA}}}
\affil[8]{\orgdiv{Department of Physics}, \orgname{University of Colorado}, \orgaddress{\city{Boulder}, \state{CO}, \country{USA}}}

\abstract{
    The X-ray Integral Field Unit (X-IFU) instrument on the future ESA mission Athena X-ray Observatory is a cryogenic micro-calorimeter array of Transition Edge Sensor (TES) detectors designed to provide spatially-resolved high-resolution spectroscopy. The onboard reconstruction software provides energy, spatial location and arrival time of incoming X-ray photons hitting the detector. A new processing algorithm based on a truncation of the classical optimal filter and  called \textit{0-padding}, has been recently proposed aiming to reduce the computational cost without compromising energy resolution. Initial tests with simple synthetic data displayed promising results. 
    
    This study  explores the slightly better performance of the \textit{0-padding} filter and assess its final application to real data. The goal is to examine the larger sensitivity to instrumental conditions that was previously observed during the analysis of the simulations. This \textit{0-padding} technique is thoroughly tested using more realistic simulations and real data acquired from NASA and NIST laboratories employing X-IFU-like TES detectors. Different fitting methods are applied to the data, and a comparative analysis is performed to assess the energy resolution values obtained from these fittings. The \textit{0-padding} filter achieves energy resolutions as good as those obtained with standard filters, even with those of larger lengths, across different line complexes and  instrumental conditions. This method proves to be useful for energy reconstruction of X-ray photons detected by the TES detectors provided  proper corrections for baseline drift and jitter effects are applied. The finding is highly promising especially for onboard processing, offering efficiency in computational resources and facilitating the analysis of sources with higher count rates at high resolution.}


\maketitle

\section{Introduction}
\label{sec:intro} 
The X-ray Integral Field Unit \citep[X-IFU; ][]{barret2023} is a high-resolution cryogenic imaging spectrometer that will be one of the two instruments on-board the ESA's \textit{Athena} mission \citep{Nandra2013}. It  will operate in the 0.2--12~keV band and  provide unprecedented spectral resolution with a Full Width at Half Maximum (FWHM) of 2.5~eV at 7~keV. The X-IFU Focal Plane will contain a large array of Transition Edge Sensors \citep[TES;][]{Smith2021} with several tens of TES per readout channel using a Time Division Multiplexing (TDM) scheme \citep{Durkin2019}. 
The on-board Event Processor \citep{Ravera2014, Ravera2018} hardware will reconstruct the detected events caused by the impact of X-ray photons in the detector to estimate their energy, arrival time and spatial location (based on impact pixel).

Event processing poses a significant challenge, demanding a delicate balance between achieving high energy resolution from photons and minimizing computational costs. The effectiveness of selected algorithms for working in event processing must be optimized to prevent degradation of the energy resolution caused by detector non-linearity.

Numerous studies have investigated different algorithms to characterize the energy of detected photons by X-IFU \citep{Ceballos2019, Cobo2020, CardielCDF, Ceballos2022}. Building upon our previous findings in \citet{Ceballos2022}, this paper offers a more comprehensive understanding of the ongoing efforts to identify the most suitable strategy for maximizing energy resolution with this instrument.

At the core of reconstructing X-IFU events lies the classical optimal filtering technique \citep{Szymkowiak1993}. This method involves digitizing time stream data into fixed-length records, which are then utilized to construct the signal and noise components of the filter.

In order to construct the filter, two steps are followed \citep[see e.g.][]{Boyce1999, Doriese2009}. Firstly, the Discrete Fourier Transform (DFT) of the average of multiple pulse records is calculated to create the signal portion. Meanwhile, the noise portion is generated by averaging the spectra of several pulse-free records. The $f\!=\!0$~Hz bin of the DFT, typically containing a slowly varying and arbitrary offset, is usually discarded to achieve a final filter that is zero-summed. This zero-summed filter is crucial as it effectively rejects the signal baseline during processing.

When working in the time domain, the most accurate estimate of photon energy is obtained by computing the scalar product of the data pulse and an optimal filter. This straightforward approach provides a proportional estimate of the photon energy
\begin{equation}
\label{eq:energy_estimation_integral}
\hat{E} = k \int{d(t)\,o\!f(t) \;\mathrm{d}t}, 
\end{equation}
where $d(t)$ is the pulse data, $o\!f(t)$ is the time domain expression of the optimal filter and $k$ is the normalization factor to give $\hat{E}$ in units of energy
\begin{equation}
\label{eq:normalization_optimal_filter}
k = \int\frac{\langle|N(f)|^2\rangle}{S(f) \cdot S^*(f)} \;\mathrm{d}f.
\end{equation}

The matched filter (a normalized model pulse shape, $S(f)$) and the noise spectrum ($N(f)$) are used to initially build the optimal filter in frequency domain as
\begin{equation}
O\!F(f) = \frac{S^*(f)}{\langle|N(f)|^2\rangle}.
\end{equation}

The optimal filtering technique relies on the assumption that all pulses are scaled versions of a single template, which is not valid for non-linear detectors like Athena/X-IFU. Therefore, $\hat{E}$ serves strictly as an energy estimator that requires correction to obtain the final energy. This correction involves  applying a gain scale obtained from filtering pulse templates measured at different calibrated energies (see Sect.~\ref{sec:datarecon}). To underscore the distinction between real energies and estimated (or reconstructed) "pseudo-energies," we will employ (k)eV units for the former and (k)$\widehat{\rm eV}$ for the later.

The energy resolution of the instrument, determined after event reconstruction, is measured by the FWHM of the Gaussian broadening resulting from the instrumental setup and reconstruction algorithm, in addition to the Lorentzian natural profiles of the lines in a typical X-ray complex.

As the average value of the filtered pulse is set to~0 (specifically the $f\!=\!0$~Hz bin), the number of samples used in the discrete expression of the data pulse and filter can influence the final energy resolution achieved through the optimal filter \citep{Doriese2009}. Increasing the record length can improve resolution, but it comes with the trade-off of higher computational demands as well as more sensitivity to low frequency fluctuations. Furthermore long filters cannot be built at high count rates due to the temporal proximity of the photon arrival.

In a previous study \citep{Cobo2020}, we aimed to reduce the on board computing operations by exploring optimal filters of varying lengths and comparing their performance in terms of energy resolution.

The filters under investigation were:
\begin{itemize}
\item FULL: This filter uses a pulse template obtained by maximizing the length of the data records ($N_{\rm FULL}$ samples).
\item SHORT: The pulse template in this filter is  constructed using shorter pulses ($N_{\rm SHORT}$ samples), specifically half the length of the record, to save computational resources.
\item \textit{0-padding}: A modified version of the FULL filter truncated to half its length in the time domain. This approach is equivalent to \textit{0-padding} the data pulses in the scalar product of filter and pulse (Sect.~\ref{sec:ener-recon-detail}).
\end{itemize}

The analysis was conducted using synthetic monochromatic data at 6~keV simulated with the X-IFU official simulator \mbox{\texttt{xifusim} \citep{xifusim2022}}.

The primary finding indicated that the \textit{0-padding} technique outperformed both the SHORT and FULL filters in terms of energy resolution. Remarkably, it outperformed the FULL filter (which is currently the baseline filter for high resolution events) despite being only half its length. This result suggests that \textit{0-padding} offers a viable alternative for reducing the computational burden associated with optimal filtering.

However, to apply these findings in real-life scenarios it was crucial to extend the analysis to more representative simulations, including photons from a typical X-ray line complex with controlled simulated energy resolution.

Moreover, the initial analysis revealed that the \textit{0-padding} filter is more sensitive to variations in instrumental conditions, especially changes in bath temperature leading to baseline drifts. Therefore, it was essential to test this approach using real laboratory data before considering \textit{0-padding} as an optimization or even a feasible alternative to the current baselined reconstruction algorithm.

This paper presents the energy resolution results obtained by applying the \textit{0-padding} filter to a realistic X-IFU simulation of the Mn~K$\alpha$  line complex and to TES (X-IFU-like) real data from the Goddard Space Flight Center (GSFC) and the National Institute of Standards and Technology (NIST) laboratories. We compare these results with those obtained using the FULL optimal filter, which serves as the baseline method at these laboratories, as well as with those obtained with the SHORT filter, equal in length to the \textit{0-padding} filter. Additionally, we assess the performance and potential systematic effects of different analysis algorithms, along with some external factors that could influence the results.

It is important to highlight that although the initial motivation of this work stems from the effort to find the optimal algorithm for reconstructing energy for the X-IFU instrument, the results presented are applicable not only to  data on X-IFU-type detectors but can also be useful for other present or future TES detectors.

This work benefited significantly from the use of the software SIRENA \citep{Sirena2019, sirenaASCL} (Software IFCA for Reconstruction of EveNts for Athena X-IFU)\footnote{Available at \url{https://sirena.readthedocs.io/}}, a package developed to reconstruct the energy of the incoming X-ray photons after their detection in the X-IFU TES detector.

Section~\ref{sec:zero-padding-explanation} explores mathematically the possible reasons behind the better performance of the \mbox{\textit{0-padding}} filter. Section~\ref{sec:mnka_sims} describes the simulations of the Mn~K${\alpha}$ line complex with \mbox{\texttt{xifusim}} and the performance of the filters on these simulated data. Section~\ref{sec:data} provides a description of the laboratory data utilized in the analysis.  In section~\ref{sec:datarecon}, the real-data reconstruction process is presented, and section~\ref{sec:fitting} describes and compares the two techniques utilized to fit the energy distribution and retrieve the energy resolution. Section~\ref{sec:methods} presents the results of the filter comparison in terms of the measured energy resolution. The analysis to other line complexes at energies different from the standard Mn~K$\alpha$ complex from which the optimal filters are built is described in section~\ref{sec:other_complexes}. Finally, section~\ref{sec:conclusions} summarizes the main conclusions of this work.

\section{Insights into the effectiveness of 0-padding }
\label{sec:zero-padding-explanation}

\subsection{Energy reconstruction in detail}
\label{sec:ener-recon-detail}

In practical terms, the application of Eq.~(\ref{eq:energy_estimation_integral}) to compute the reconstructed energy of a pulse is evaluated through a discrete sum expressed as follows
\begin{equation}
\label{eq:energy_estimation_dot_product}
    \hat{E} = \sum_{i=1}^{N_{\rm final}} d(t_i)\cdot \widetilde{o\!f}(t_i).
\end{equation}
In this equation $d(t_i)$ represents the discretized pulse, sampled at $N_{\rm final}$ time values denoted as $t_i$. Likewise $\widetilde{o\!f}(t_i)$ corresponds to the optimal filter in the time domain. For the sake of simplicity in the notation, we incorporate the normalization factor from Eq.~(\ref{eq:normalization_optimal_filter}) into $\widetilde{o\!f}(t_i)$. This makes it  clear that Eq.~(\ref{eq:energy_estimation_dot_product}) is essentially a dot product calculation between the discretized pulse and the optimal filter, considering them as vectors.

\begin{figure*}[tb]
    \centering
    \includegraphics[width=0.58\linewidth]{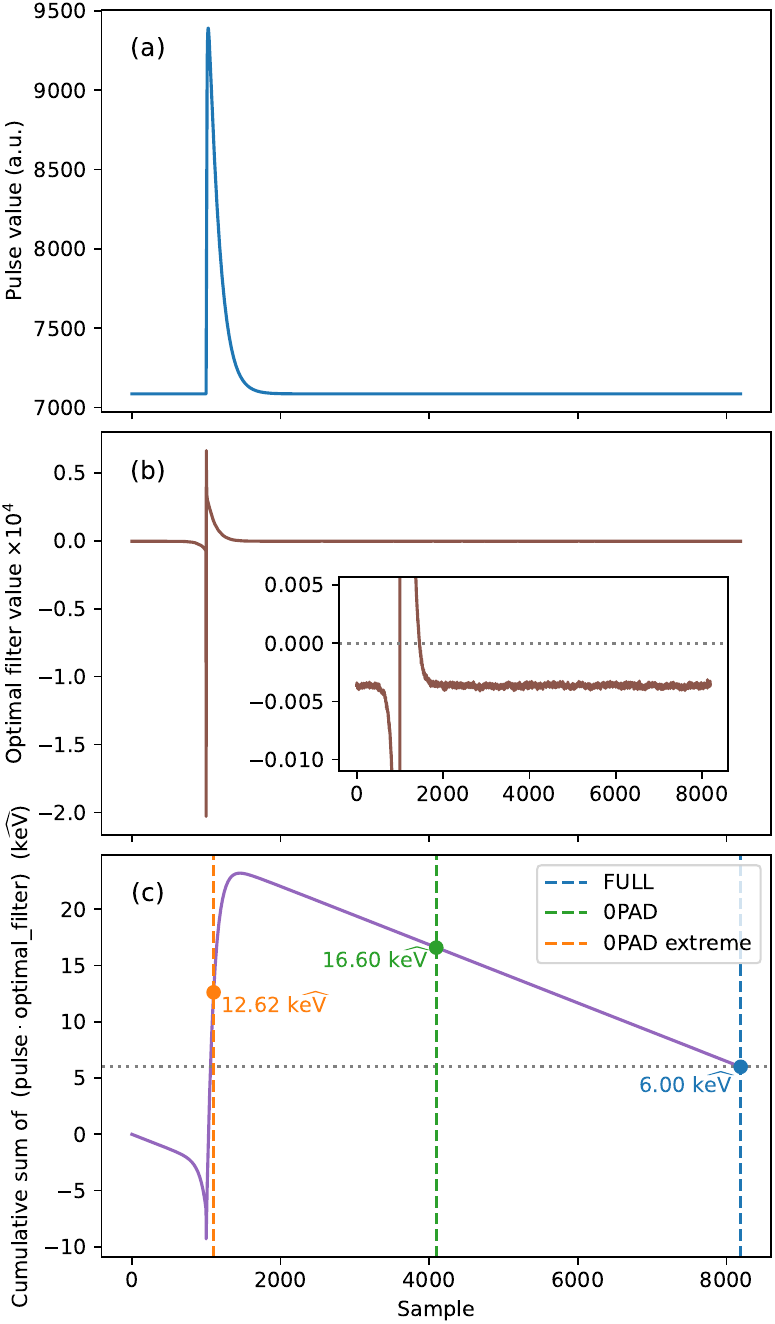}
    \caption{Panel~(a): Example of a noise-free pulse generated with \mbox{\texttt{xifusim}}, corresponding to a photon of 6~keV. The entire pulse comprises \mbox{$N_{\rm FULL}=8192$} samples, with a pre-trigger (i.e., the data signal before the rising of the pulse) of 1000 samples and a baseline of \mbox{$\sim\! 7085$} (arbitrary) units. Panel~(b): Optimal filter constructed from the same pulse displayed in the previous plot. The inset plot is a zoom-in on the vertical axis revealing that the flat part of the optimal filter contains negative numbers (below the horizontal dotted line marking the zero level). Panel~(c): Cumulative sum of the dot product of the pulse and the optimal filter represented above. The filled colored circles in this panel indicate the reconstructed energy obtained when the upper limit of the summation in Eq.~(\ref{eq:energy_estimation_dot_product}) is calculated only up to the sample indicated by the vertical dashed lines: $N_{\rm cut}=1100$ (orange), $N_{\rm cut}=4096$ (green), and $N_{\rm final} = N_{\rm FULL}=8192$ (blue) for a sampling rate of 156.25~kHz.}
    \label{fig:pulse_optfilt_cumsum}
\end{figure*}

As an example, Fig.~\ref{fig:pulse_optfilt_cumsum}(a) illustrates a noise-free 6~keV pulse, which was simulated using the \mbox{\texttt{xifusim}} simulator (v.0.8.3) with an LPA2.5a instrument configuration file (1 pixel) and a sampling rate of 156.25~kHz. This configuration served as the baseline for the X-IFU instrument at the time of writing and it was consistently used throughout this paper.

In Fig.~\ref{fig:pulse_optfilt_cumsum}(b), we can observe the optimal filter computed from that pulse, making use of a noise spectrum derived from 100\,000 noise streams. The cumulative sum of the dot product of the pulse and optimal filter is represented by the purple line in Fig.~\ref{fig:pulse_optfilt_cumsum}(c). As expected, when extending the sum in Eq.~(\ref{eq:energy_estimation_dot_product}) up to $N_{\rm final}=N_{\rm FULL}=8192$ (the full length of the simulated pulse) the reconstructed energy is measured as $6.00\;{\rm k}\widehat{\rm eV}$, as indicated next to the blue filled circle. 

It is important to note that the cumulative sum initiates as negative, reaches a peak around sample $1800$, and then decreases monotonically. This pattern is predictable, as the most significant part of the pulse has been included in the scalar product by the time the cumulative sum reaches the peak. Beyond that sample, the optimal filter remains relatively constant and negative, while  the pulse mainly consists of the baseline value due to the completed exponential decay of the pulse.

In an effort to grasp the impact of filter truncation on the estimation of the pulse energy using the \textit{0-padding} filter, we can decompose the dot product of Eq.~(\ref{eq:energy_estimation_dot_product}) into two components,
\begin{equation}
\label{eq:energy_estimation_dot_product_two_terms}
\begin{split}
    \hat{E} & =
    \sum_{i=1}^{N_{\rm cut}} d(t_i)\cdot \widetilde{o\!f}(t_i) +
    \sum_{i=N_{\rm cut}+1}^{N_{\rm final}} d(t_i)\cdot \widetilde{o\!f}(t_i)\\
    & \equiv \hat{E}_{\textit{0-pad}} + \sum_{i=N_{\rm cut}+1}^{N_{\rm final}} d(t_i)\cdot \widetilde{o\!f}(t_i),
\end{split}
\end{equation}
where $N_{\rm cut}$ is an intermediate time sample that represents the point selected to truncate a FULL filter to construct a \textit{0-padding} filter. $\hat{E}_{\textit{0-pad}}$ corresponds to the reconstructed energy obtained by performing  the dot product using the first $N_{\rm cut}$ samples. Specifically, we emulate the \textit{0-padding} optimal filter as examined by \citet{Cobo2020} by selecting \mbox{$N_{\rm cut}=4096$}. This is illustrated in Fig.~\ref{fig:pulse_optfilt_cumsum}(c) with the vertical green dashed line. In this particular case, the calculated energy is \mbox{$\hat{E}_{\textit{0-pad}}=16.60\;{\rm k}\widehat{\rm eV}$} (green filled circle), which is significantly higher than the expected 6.0~keV value. This discrepancy is in line with the fact that the second term in Eq.~(\ref{eq:energy_estimation_dot_product_two_terms}) is negative. An even more extreme \textit{0-padding} scenario can be achieved by computing the summation only up to $N_{\rm cut}=1100$. This is shown by the vertical orange dashed line in the same figure, resulting in \mbox{$\hat{E}_{\textit{0-pad}}=12.62\;{\rm k}\widehat{\rm eV}$}. 

\begin{figure}[htbp]
    \centering
    \includegraphics[width=1.\linewidth]{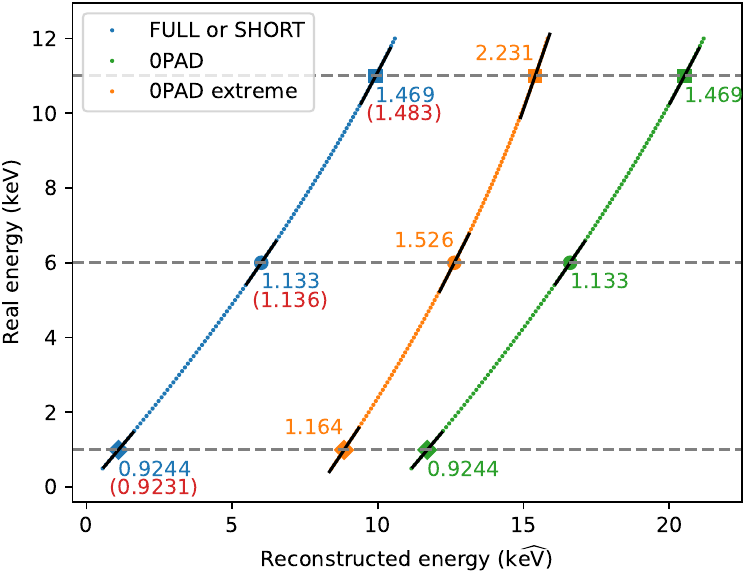}
    \caption{Gain scales computed for different filters: FULL ($N_{\rm FULL}=8192$; blue line), \textit{0-padding} ($N_{\rm cut}=4096$, using the FULL filter data; green line), and extreme \textit{0-padding} ($N_{\rm cut}=1100$, using the FULL filter data; orange line). The gain scale for the SHORT filter ($N_{\rm SHORT}=4096$) is indistinguishable from the FULL filter curve in this representation. The three horizontal grey dashed lines correspond to real energies of~1, 6 and 11~keV, from bottom to top. The tangent lines to the gain scale curves at the intersection with those three horizontal lines are shown with black line segments, and the corresponding slope values are displayed next to the intersection points (the values associated with the SHORT filter are given in red color between parentheses).}
    \label{fig:gainscale_comparison}
\end{figure}

As previously mentioned in the introduction, the optimal filter is computed from a single-energy template, making each reconstructed energy an energy estimation that needs to be converted into a real energy using a gain scale conversion. To achieve this, we simulated noise-free pulses with energy values ranging from 0.5 to 12.0~keV in increments of 0.1~keV using \mbox{\texttt{xifusim}}. By applying the optimal filter computed using the 6.0~keV pulse as the template, we determined the corresponding reconstructed energies. Subsequently, we fitted an eleventh degree polynomial to the relationship between the real (simulated) energies and the reconstructed energies. This relationship is presented in Fig.~\ref{fig:gainscale_comparison} for the three filters depicted in Fig.~\ref{fig:pulse_optfilt_cumsum}(c): FULL ($N_{\rm FULL}=8192$; blue line), \mbox{\textit{0-padding}} ($N_{\rm cut}=4096$; green line), and extreme \mbox{\textit{0-padding}} ($N_{\rm cut}=1100$; orange line). This figure also represents the corresponding gain scale for the SHORT filter ($N_{\rm SHORT}=4096$), although it is visually indistinguishable from the FULL filter curve.

These gain scales, particularly those corresponding to \mbox{\textit{0-padding}} and extreme \mbox{\textit{0-padding}}, are responsible for transforming the reconstructed energies $\hat{E}_{\textit{0-pad}}$ of $16.60\;{\rm k}\widehat{\rm eV}$ and $12.62\;{\rm k}\widehat{\rm eV}$, respectively, into calibrated energies of 6.00~keV in both cases.

The crucial aspect to comprehend here is how the uncertainties in the measurement of the energy are affected by the \mbox{\textit{0-padding}} truncation and how these uncertainties change when the corresponding gain scale correction is applied.

\subsection{Propagation of random uncertainties}
\label{sec_propagation_of_random_undertainties}

To get an initial understanding of how the truncation introduced in the \mbox{\textit{0-padding}} approach affects uncertainties (and consequently impacts the measured energy resolution), we begin by examining the random uncertainty associated to the computation of the energy estimator given by Eq.~(\ref{eq:energy_estimation_dot_product}). It is worth noting that systematic effects arising from our incomplete knowledge of the system will be addressed during the application of the gain scale correction.

In a simplified scenario, we disregard the random uncertainties in the optimal filter in comparison with those in the pulse. This approximation is justified given that the optimal filter, in real-world situations, is derived from an average pulse (achieved by averaging pulses corresponding to photons of a certain energy). The random uncertainty in each sample of this averaged pulse is anticipated to be significantly less than the random uncertainty associated with each sample of individual pulses, the energy of which we aim to determine.

Even though there are specific frequencies with more noise in the system, we will operate under the additional approximation that the noise within different time samples of a specific pulse is uncorrelated, which facilitates the following computation. Applying the law of propagation of uncertainties to Eq.~(\ref{eq:energy_estimation_dot_product}), we obtain
\begin{equation}
\label{eq:uncertainty_energy_estimation_dot_product}
\begin{split}
    (\Delta\hat{E})^2 = & \sum_{i=1}^{N_{\rm final}} \bigl(\widetilde{o\!f}(t_i)\bigr)^2 \cdot \bigl(\Delta d(t_i) \bigr)^2 \simeq (\Delta d)^2 \sum_{i=1}^{N_{\rm final}} \bigl(\widetilde{o\!f}(t_i) \bigr)^2 \\
    = & 
    (\Delta d)^2 \sum_{i=1}^{N_{\rm cut}} \bigl(\widetilde{o\!f}(t_i) \bigr)^2 + 
    (\Delta d)^2 \sum_{i=N_{\rm cut}+1}^{N_{\rm final}} \bigl(\widetilde{o\!f}(t_i) \bigr)^2 \\
    \equiv & 
    (\Delta\hat{E})^2_{\textit{0-pad}} +
    (\Delta d)^2 \sum_{i=N_{\rm cut}+1}^{N_{\rm final}} \bigl(\widetilde{o\!f}(t_i) \bigr)^2.
\end{split}
\end{equation}
In this equation, we have approximated the uncertainty at each sample of the pulse $\Delta d(t_i)$ by its average value $\Delta d$. We have also split the final expression into two terms, similarly to what we have done in Eq.~(\ref{eq:energy_estimation_dot_product_two_terms}), to clarify the computation in the \textit{0-padding} case.
The resulting expression suggests that the expected uncertainty in the reconstructed energy scales with the sum of the squares of the optimal filter values. 

\begin{figure*}[tbp]
    \centering
    \includegraphics[width=0.9\linewidth]{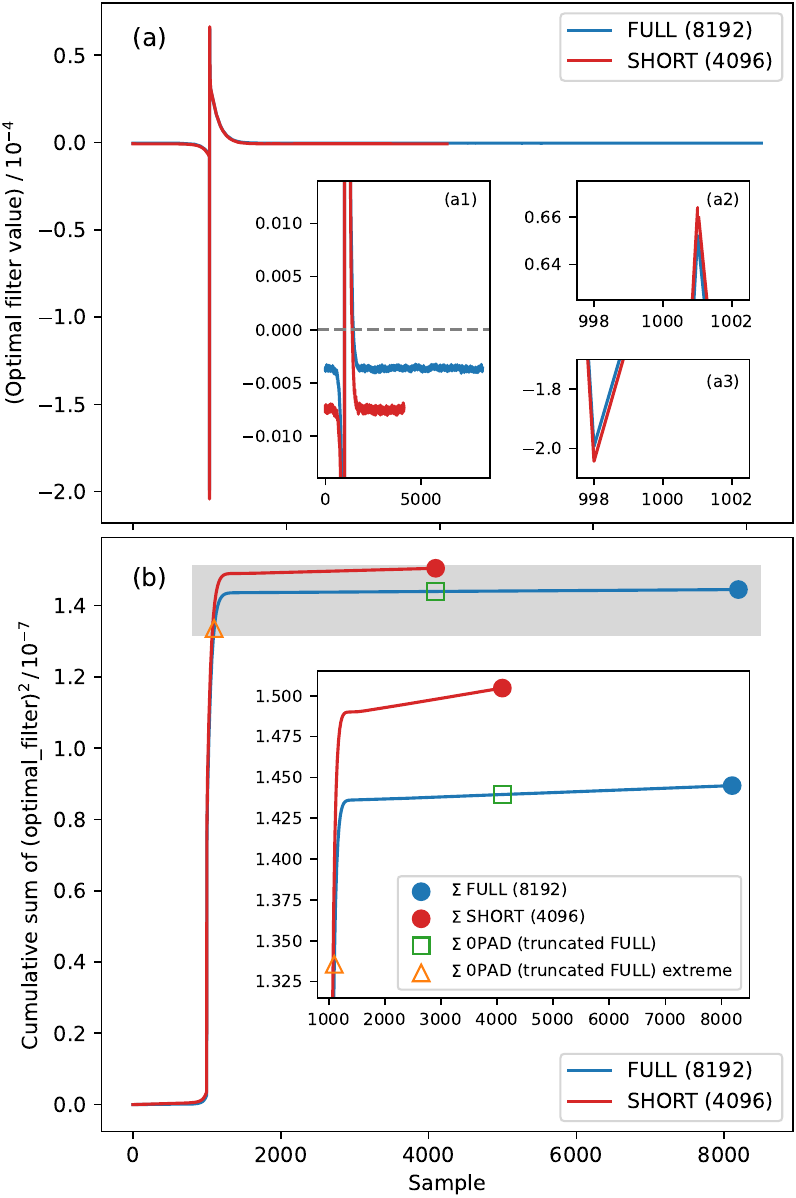}
    \caption{Panel~(a): Comparison of the FULL (\mbox{$N_{\rm FULL}\!=\!8192$}; blue line) and SHORT (\mbox{$N_{\rm SHORT}\!=\!4096$}; red line) optimal filters constructed from a noise-free pulse corresponding to a photon of 6~keV. The inset plots are zooms that highlight the differences between both filters for values close to zero, subplot~(a1), and for values around the maximum and minimum filter peaks, subplots~(a2) and~(a3), respectively. Panel~(b): Cumulative sum of the squared FULL (blue line) and SHORT (red line) optimal filters displayed above. The inset is a zoom of the grey shaded region of the diagram, highlighting the smooth increase of the curves beyond the time samples where most of the information of the pulse is concentrated. The filled symbols indicate the summation factor in $(\Delta\hat{E})^2$ from Eq.~(\ref{eq:uncertainty_energy_estimation_dot_product}) for the filters SHORT (\mbox{$N_{\rm final}\!=\!4096$}; red circle), and FULL (\mbox{$N_{\rm final}\!=\!8192$}; blue circle). The open symbols depict the corresponding summation factor in \mbox{$(\Delta\hat{E})^2_{\textit{0-pad}}$} resulting from the truncation of the FULL filter for \mbox{\textit{0-padding}} (\mbox{$N_{\rm cut}\!=\!4096$}; green open square), and \mbox{\textit{0-padding}} extreme (\mbox{$N_{\rm cut}\!=\!1100$}; orange open triangle).}
    \label{fig:cumsum_opt_filt_squared}
\end{figure*}

Considering the typical shape of the optimal filters, it is clear that the most significant contributions to the uncertainty described by the sum in Eq.~(\ref{eq:uncertainty_energy_estimation_dot_product}) occur at the time samples where the pulse exhibits its abrupt increase and subsequent decline. This behavior is evident in Fig.~\ref{fig:cumsum_opt_filt_squared}.
In particular, panel~\ref{fig:cumsum_opt_filt_squared}(a) compares the FULL (\mbox{$N_{\rm FULL}\!=\!8192$}; blue line) and SHORT (\mbox{$N_{\rm SHORT}\!=\!4096$}; red line) optimal filters, computed from the same 6~keV pulse depicted in Fig.~\ref{fig:pulse_optfilt_cumsum}(a). The insets highlight the differences between them. In addition, panel~\ref{fig:cumsum_opt_filt_squared}(b) represents the cumulative sum of the squared values for both filter data.
These curves demonstrate a drastic change starting around sample 1000 (where the pulse is triggered). They stabilize beyond the sample where most of the pulse's exponential decline has occurred, after which they flatten.

Interestingly, the inset figure in panel~\ref{fig:cumsum_opt_filt_squared}(b) (a zoomed-in view of the graph region demarcated by the shaded grey rectangle) reveals that the displayed cumulative sums indeed exhibit a modest, but not negligible, positive increase after sample \mbox{$\sim\! 1800$}. The filled colored circles indicate the summation factor in $(\Delta\hat{E})^2$ from Eq.~(\ref{eq:uncertainty_energy_estimation_dot_product}) for the FULL and SHORT filters. Analogously, the open symbols represent the summation factor in $(\Delta\hat{E})^2_{\textit{0-pad}}$, resulting from the truncation of the FULL filter for \mbox{\textit{0-padding}} (\mbox{$N_{\rm cut}\!=\!4096$}; green open square), and \mbox{\textit{0-padding}} extreme (\mbox{$N_{\rm cut}\!=\!1100$}; orange open triangle).
The value for the \mbox{\textit{0-padding}} case is slightly below the one corresponding to the FULL filter, as expected considering the second term in the last expression of Eq.~(\ref{eq:uncertainty_energy_estimation_dot_product}), which is always positive but only incorporates optimal filter values very close to zero. In addition, the \mbox{\textit{0-padding}} value is substantially below the one associated with the SHORT filter. This result is also easy to understand looking at the insets of Fig.~\ref{fig:cumsum_opt_filt_squared}(a), where the SHORT filter exhibits larger absolute values than the FULL filter in most the samples.

Given the earlier approximation, where the uncertainty $\Delta d(t_i)$ at each pulse sample can be represented by its mean value $\Delta d$, the previous results imply that the uncertainty in the reconstructed energy obtained with \mbox{\textit{0-padding}} should be marginally smaller than the uncertainty associated with the FULL filter estimate. Moreover, it should be noticeably smaller than the uncertainty linked to the SHORT filter.

It is crucial to note that while the aforementioned results highlight the relative significance of uncertainties in reconstructed energies, these energies still require conversion to a real scale using the appropriate gain scale transformations.

When focusing on pulses produced by photons within a narrow energy interval (like our 6~keV simulated pulses), the application of the gain scale can be approximated by a linear transformation. Under these circumstances, the propagation of uncertainties depends solely on the slope of this transformation. Interestingly, the derivatives illustrated in Fig.~\ref{fig:gainscale_comparison} for the FULL and \textit{0-padding} gain scales at a fixed real energy are identical within four significant figures. This suggests that the gain scale of the \textit{0-padding} filter is the same as the one corresponding to the FULL filter except for a horizontal shift in this diagram. 

As a result, the uncertainties associated with the reconstructed energies are modified by the same factor when converted into real energies upon applying the gain scale correction. This accounts for why the uncertainties in the final energies obtained with the \textit{0-padding} filter remain slightly smaller than those associated with the FULL filter. The same comparison holds true when evaluating the \textit{0-padding} and SHORT filters.

We have quantified this effect using 1\,000\,000 monochromatic noisy 6.0~keV pulses simulated with \mbox{\texttt{xifusim}}, whose reconstructed energies were computed using the four filters FULL, SHORT, \textit{0-padding} and \textit{0-padding} extreme, and later transformed into a real energy scale using their corresponding gain scale corrections. The mean energies obtained in each case, together with the associated dispersion expressed as FWHM, are summarised in Table~\ref{Tab:simulated_6_keV_pulses}. 

\begin{table*}[tbp]
\caption{Statistical summary of monochromatic noisy 6.0~keV pulse simulations.}\label{Tab:simulated_6_keV_pulses}

\begin{tabular*}{\textwidth}{@{\extracolsep\fill}lcccccc}
\toprule
\noalign{\smallskip}

       & Filter length & \multicolumn{2}{@{}c}{Reconstructed energies [$\widehat{\rm eV}$]} &  &\multicolumn{2}{@{}r}{Energies after gain scale correction [$\rm eV$]}\\

\cmidrule{3-4}
\cmidrule{6-7}
Filter & [samples] & $\langle\hat{E}\rangle$ & FHWM & &  $\langle E \rangle$ & FHWM \\
 (1)   &  (2)      &          (3)            & (4)  & &       (5)            & (6) \\
\noalign{\smallskip}
\midrule
\noalign{\smallskip}
FULL & 8192  &\phantom{1}6\,000.001 & 1.706 $\pm 0.001$ & & 6\,000.001 & 1.932 $\pm 0.002$  \\
SHORT & 4096 &\phantom{1}6\,000.001 & 1.734 $\pm 0.001$ & & 6\,000.001 & 1.970 $\pm 0.002$  \\ 
\textit{0-padding} & 4096 &16\,595.455 & 1.691 $\pm 0.001$ & & 6\,000.001 & 1.916 $\pm 0.002$ \\
\textit{0-padding} extreme & 1100 &12\,623.039 & 1.419 $\pm 0.001$ & & 6\,000.001 & 2.166 $\pm 0.002$ \\ 
\noalign{\smallskip}
\botrule
\end{tabular*}
\footnotetext{Note: Mean energies and associated FWHM corresponding to reconstructed (columns~(3) and~(4)) and gain-scale corrected (columns~(5) and~(6)) values computed in a sample of 1\,000\,000 simulated monochromatic noisy 6.0~keV pulses, using the optimal filters listed in the first column. In all cases, these filters have a pre-buffer of 1000~samples. The uncertainty in FWHM was determined by splitting the simulated sample in 100 sub-samples of 10\,000 pulses, computing the standard deviation of the 100 FWHM estimates, and dividing by the square root of 100 to finally provide the expected uncertainty in the mean FWHM}.
\end{table*}

For a given filter, the FWHM corresponding to the reconstructed energies is stretched by the slope values indicated at the locations of the filled circles in Fig.~\ref{fig:gainscale_comparison}: 1.133 (FULL), 1.136 (SHORT), 1.133 (\textit{0-padding}) and 1.526 (\textit{0-padding} extreme). When we move from reconstructed energies to gain-scale corrected energies we observe an increase of $\sim 13$\% in FWHM for the FULL, \textit{0-padding} and SHORT filters. Interestingly, the FWHM of the mean energy reconstructed with the \textit{0-padding} extreme filter was the smallest (1.419~eV); however, when applying its gain scale transformation, this value is stretched by $\sim 53$\% (2.166~eV), making it the worst option.

\begin{figure*}[htbp]
    \centering
    \resizebox{\hsize}{!}
    {\includegraphics{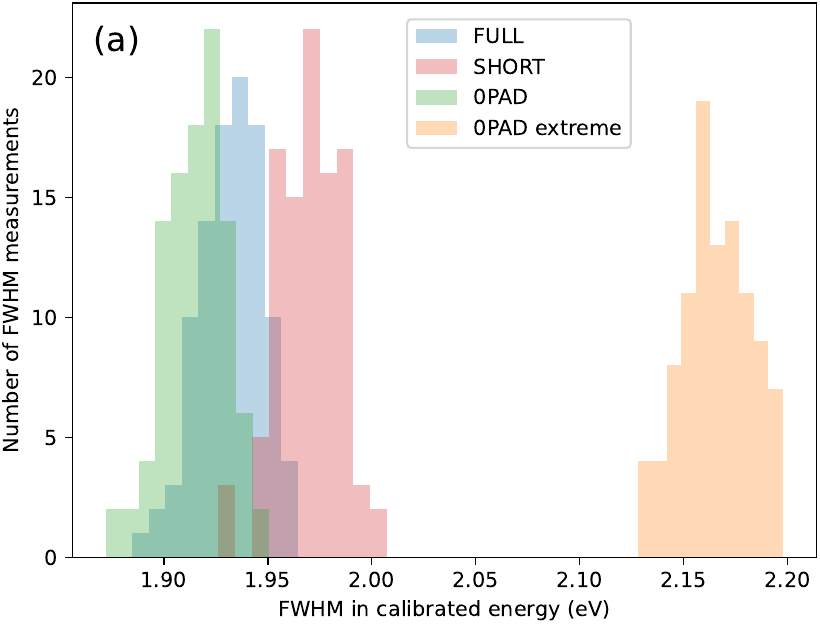}
    \includegraphics{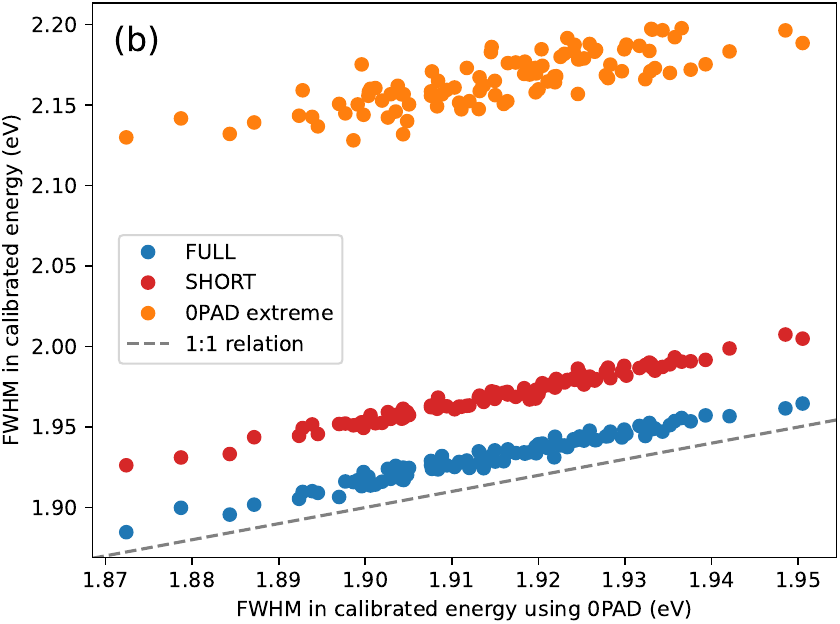}}
    \caption{Panel~(a): histogram of FWHM values corresponding to the energy resolutions obtained with different optimal filters (as indicated in the legend) for 100 simulated TES pixels observing 10\,000 monochromatic 6~keV pulses each. The mean values and standard deviations correspond to the values listed in column (6) of Table~\ref{Tab:simulated_6_keV_pulses}. Panel~(b): comparison of the individual 100~FWHM values whose histograms are displayed in panel~(a), using for the horizontal axis the ones retrieved using the \emph{0-padding} filter. The dashed line indicates the 1:1 relation. Note that all the points appear above this line.}
    \label{fig:FWHM_Ecal_simulated_6keV}
\end{figure*}

It is essential to emphasize that the uncertainty quoted in each FWHM value, as presented in Table~\ref{Tab:simulated_6_keV_pulses} columns (4) and~(6), has been calculated by dividing the simulated dataset into 100 sub-samples, each containing 10\,000 pulses. The standard deviation of the resulting 100 FWHM estimates was then computed and divided by the square root of 100 to obtain the uncertainty in the mean. This process is analogous to having 100 identical TES, each collecting 10\,000 monochromatic 6 keV pulses. Although there may appear to be small differences in FWHM among different filters, a statistical analysis must be conducted considering these expected uncertainties. This analysis must account for the fact that the FWHM values obtained with different optimal filters are paired for a specific simulated TES, meaning each subset of 10\,000 pulses in the simulated dataset. In this regard, although Fig.\ref{fig:FWHM_Ecal_simulated_6keV}(a) shows some overlap in the histogram distributions of the 100 FWHM estimates corresponding to the different optimal filters listed in Table\ref{Tab:simulated_6_keV_pulses}, the mean FWHM values are statistically different. Fig.~\ref{fig:FWHM_Ecal_simulated_6keV}(b) visually represents this difference, indicating that the \emph{0-padding} estimate is consistently lower. A Wilcoxon signed-rank test for paired data \citep{Wilcoxon1945} (non-parametric) rejects the null hypothesis that the FWHM obtained using FULL, SHORT, and \emph{0-padding} extreme are lower than the FWHM obtained with \emph{0-padding}, with a p-value of zero in all three comparisons.

The outcome is not unexpected when we employ \mbox{$N_{\rm cut}\!=\!1100$}. At this cut-off, we are disregarding vital information present in the pulse data, as the exponential decay is still evident at that time sample. Consequently, the signal-to-noise ratio of the reconstructed energy would be significantly lower than when considering all the informative pulse samples. Furthermore, the larger slope in the corresponding gain scale transformation would further degrade the energy resolution.

\begin{figure}[htbp]
    \centering
    \includegraphics[width=1.0\linewidth]{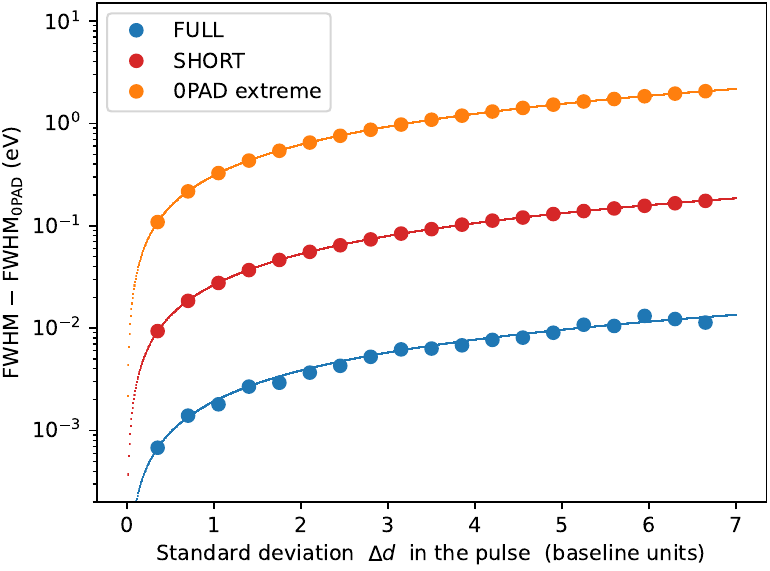}
    \caption{Comparison of energy dispersion values (FWHM) derived from various filters. The differences between the FWHM values derived from FULL, SHORT, and \mbox{\textit{0-padding}} extreme with respect to the FWHM of the \mbox{\textit{0-padding}} filter are plotted against the standard deviation $\Delta d$ in the pulse. Each filled circle represents 100\,000 simulated noisy pulses. Thin lines indicate the theoretical prediction \mbox{$\Delta \hat{E} \propto \Delta d$}, where $\Delta d$ is the assumed standard deviation in the pulse.}
    \label{fig:delta_FWHM_vs_pulse_rms}
\end{figure}

As a final validation of all the approximations leading to Eq.~(\ref{eq:uncertainty_energy_estimation_dot_product}), we have verified the proportionality between the energy uncertainty ($\Delta \hat{E}$) and the noise in the pulse ($\Delta d$) with the help of additional numerical simulations. In particular, we have simulated monochromatic noisy pulses using as starting point the prediction of \mbox{\texttt{xifusim}} for a 6.0~keV noiseless pulse, and adding Gaussian noise with varying standard deviation. After computing the reconstructed energy using the optimal filters FULL, SHORT, \mbox{\textit{0-padding}}, and \mbox{\textit{0-padding}} extreme, we have applied their gain scale transformations to obtain the corresponding corrected energies and associated FWHM. In Fig.~\ref{fig:delta_FWHM_vs_pulse_rms} we represent the difference between the final FWHM values obtained with FULL, SHORT, and \mbox{\textit{0-padding}} extreme, compared to the FWHM corresponding to \mbox{\textit{0-padding}} method, as a function of the noise (standard deviation) in the pulse. Each filled circle represents 100\,000 simulated noisy pulses, whereas the thin lines correspond to the prediction \mbox{$\Delta \hat{E} \propto \Delta d$}, where $\Delta d$ is the assumed standard deviation in the pulse, as shown on the horizontal axis of this figure. We find that the \mbox{\textit{0-padding}} technique consistently performs slightly better than FULL and is significantly superior to both SHORT and \mbox{\textit{0-padding}} extreme. This advantage is particularly pronounced as the noise level in the pulse increases.

In conclusion, the aforementioned discussion has demonstrated that a \mbox{\textit{0-padding}} filter constructed from a truncated FULL filter tends to slightly outperform the latter, as long as the truncation occurs at a time sample where the essential pulse information has already been captured by the first term in Eq.~(\ref{eq:energy_estimation_dot_product_two_terms}). In such cases, the gain scale slope at the reconstructed energy in the FULL and the \textit{0-padding} filters is notably similar. The small increase in noise experienced by the FULL filter due to the inclusion of unnecessary samples in the dot product computation is consequently translated into the gain-scale corrected energies.

\section{X-ray line complex simulation}
\label{sec:mnka_sims}
Before delving into the analysis of real data, the next step after the analysis of the performance of the optimal filters on 6~keV monochromatic simulated pulses, involves running more realistic simulations, generating pulses following a theoretical profile of standard X-ray line complexes used in laboratories: specifically the Mn~K$\alpha$ complex. To achieve this, we also utilized the \mbox{\texttt{xifusim}} simulator.

A laboratory-measured Mn~K$\alpha$ complex is the convolution of the natural Lorentzian profile of the X-ray lines and the Gaussian broadening caused by the instrumental setup. The resulting profile of this Lorentzian-Gaussian convolution is referred to as a Voigt profile \citep{1947ApJ...106..121V}.

The process of generating lists of photon energies within the Mn~K$\alpha$ complex for the simulations, involved following their Lorentzian line profiles with the line parameters described in Table~\ref{Tab:MnKa}. To broaden the lines in a manner similar to the instrument's behaviour, we included an intrinsic controlled Gaussian profile with varying widths. We randomly selected 300~uniform values of FWHM between 0.7 and 2.3~eV for this purpose. This interval was chosen to get final broadened FWHM values in the range from 2.2 to 3.0~eV, similar to the one measured with the laboratory pixels (see Sect.~\ref{sec:fitting}). For each intrinsic width value, we constructed a \mbox{Mn K$\alpha$} complex randomly drawing 10\,000 photon energies with the appropriate distribution (again, this number was selected to reproduce the typical number of photons/pixel in the laboratory data of Sect.~\ref{sec:fitting}).

In practice, to calculate the energy of each photon we inverted the cumulative distribution function (CDF) of the line complex. This was achieved by using a uniformly distributed random number between 0 and 1 as input to the CDF. The CDF of the line complex was computed by adding the expected CDF of each line\footnote{The CDF of each Voigt profile was computed using numerical integration of the \texttt{voigt\_profile} function available in the SciPy Python package \citep[\url{https://scipy.org/}]{2020SciPy-NMeth} within a predefined energy range $[E_{\rm min}, E_{\rm max}]$. In addition, the required integral in the interval $(-\infty, E_{\rm min}]$ was computed using Eq.~(1.19) in \citet{KUMAR2020101986}, which was evaluated employing the hypergeometric function \texttt{hyp2f2}, available in the {\sc mpmath} Python library \citep[\url{https://mpmath.org/}]{mpmath}. This last approach was not used to evaluate the CDF at any arbitrary energy because the use of \texttt{hyp2f2} is slow and the numerical integration of the Voigt profile provided enough accuracy.}.

\begin{table*}[htbp]
\caption{Lorentzian coefficients for \textbf{Mn K$\alpha$} complex.}
\label{Tab:MnKa}
\centering

\begin{tabular*}{0.6\textwidth}{@{\extracolsep\fill}cccc} 
\toprule
\noalign{\smallskip}
 $\rm K{\alpha}$ & $\rm E_0[eV]$ & FWHM [eV] & Amplitude \\ 
\noalign{\smallskip}
\hline
\noalign{\smallskip}
 11  & 5898.882 & 1.7145 & 0.784 \\
 12  & 5897.898 & 2.0442 & 0.263 \\
 13  & 5894.864 & 4.4985 & 0.067 \\
 14  & 5896.566 & 2.6616 & 0.095 \\
 15  & 5899.444 & 0.97669 & 0.071 \\
 16  & 5902.712 & 1.5528 & 0.011 \\
 21  & 5887.772 & 2.3604 & 0.369 \\
 22  & 5886.528 & 4.2168 & 0.100 \\
\noalign{\smallskip}
\hline
\end{tabular*}
\footnotetext{Note: Given by \citet{lab2016}. This complex is constituted by 8~single lines distributed in two sub-complexes. The first column provides a two-digit identification: the sub-complex number and the line number within each sub-complex. The following columns indicate the centroid, width (given as a FWHM) and relative amplitude of the lines.}
\end{table*}

\begin{figure*}[htbp]
    \centering
    \includegraphics[width=0.85\linewidth]{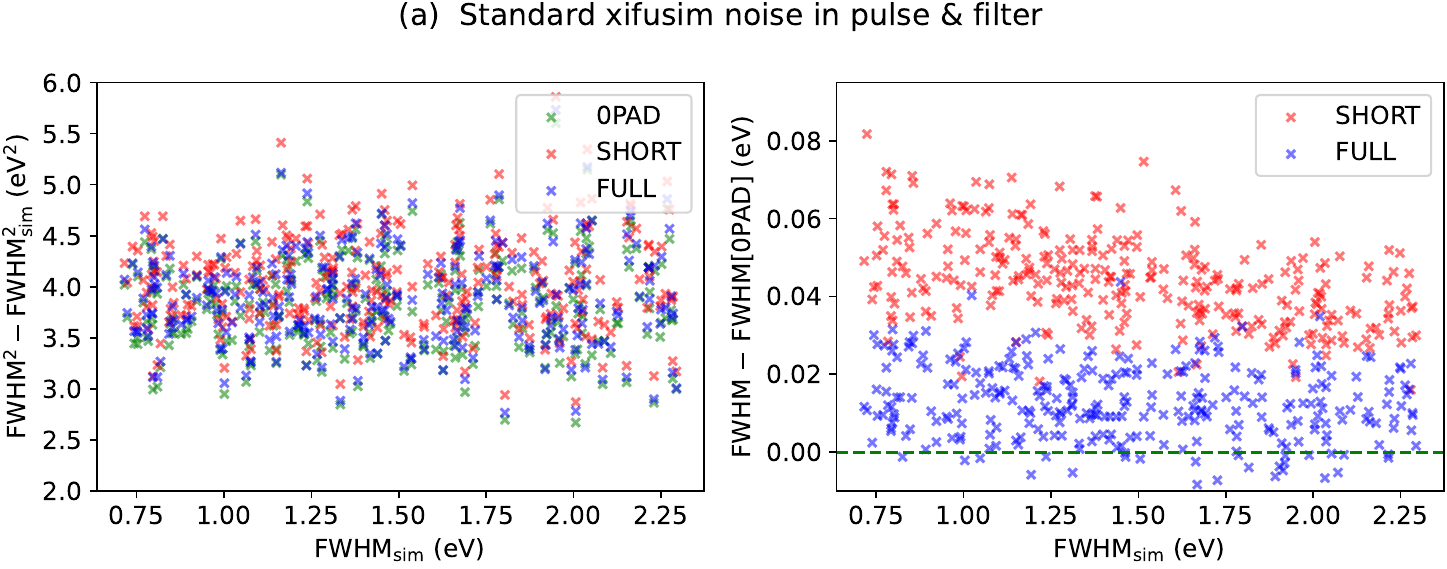}
    \includegraphics[width=0.85\linewidth]{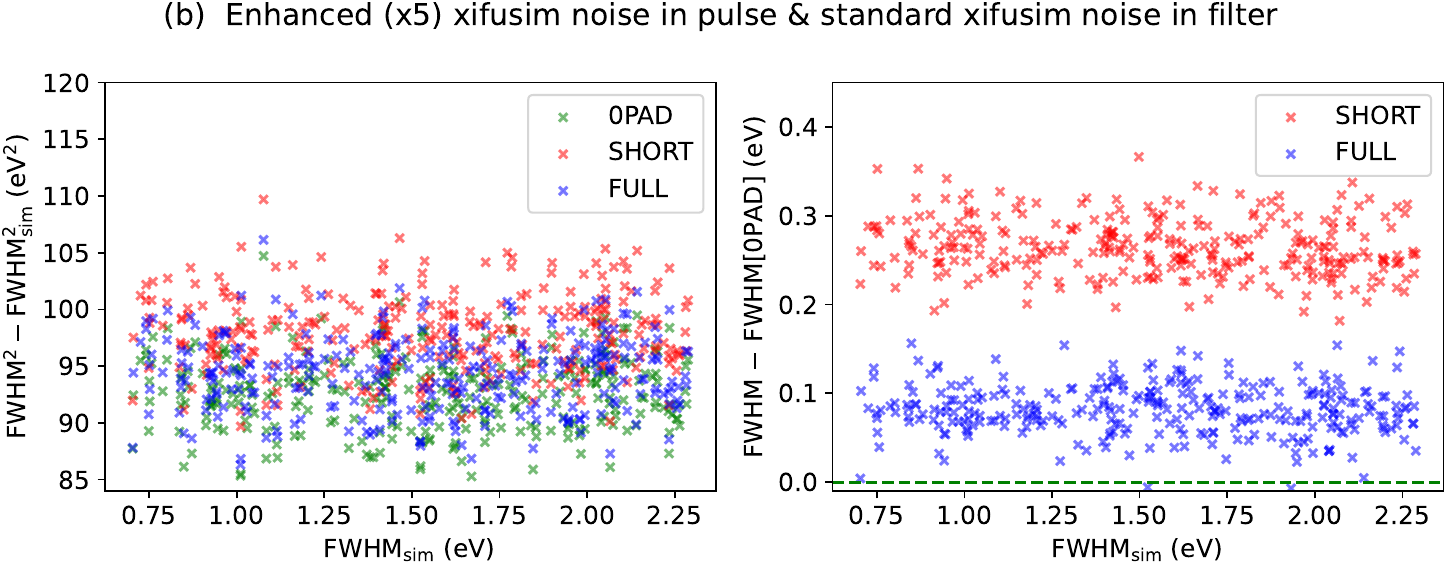}
    \includegraphics[width=0.85\linewidth]{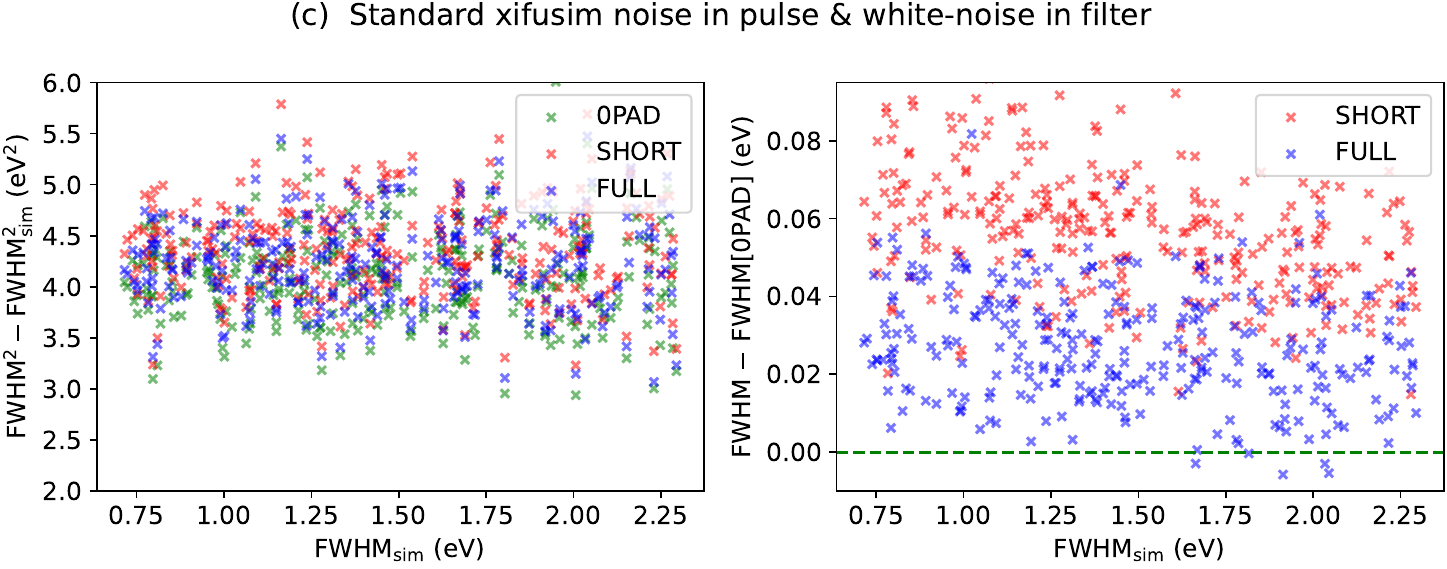}
    \includegraphics[width=0.85\linewidth]{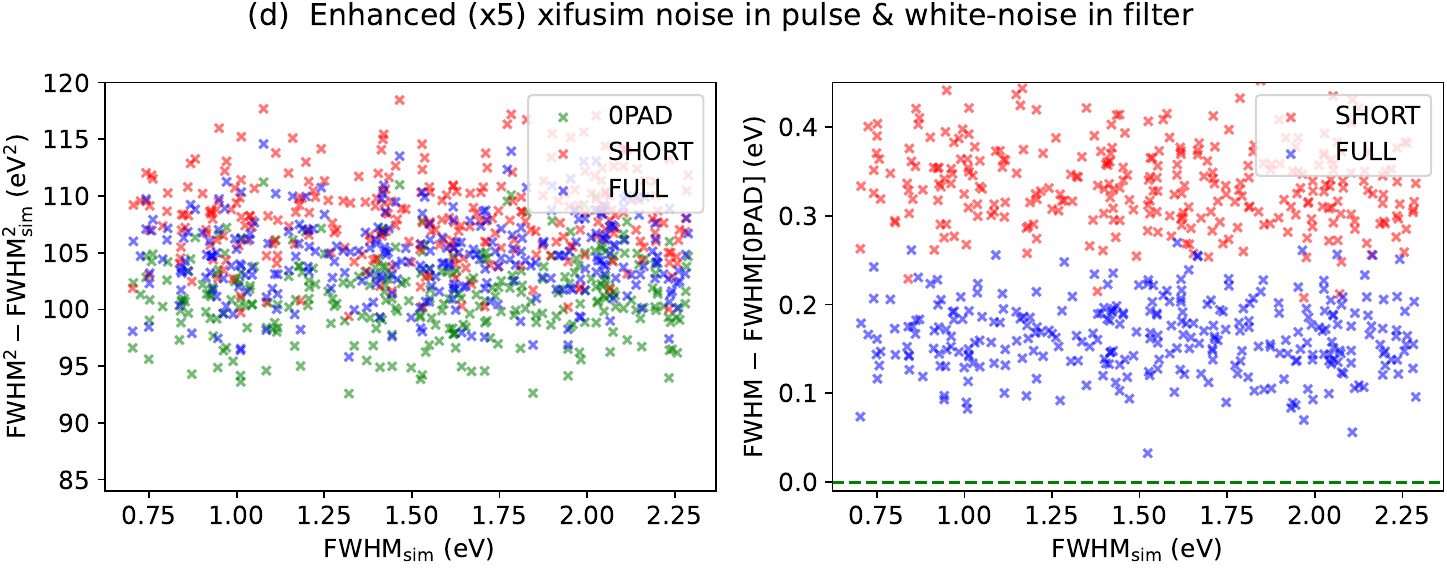}
    \caption{Comparison of Gaussian FWHM values obtained after optimal filter reconstruction of \mbox{\texttt{xifusim}} simulated pulses under different noise conditions. Panels (a) and (c) depict pulses simulated with the expected instrumental noise while panels (b) and (d) show pulses with noise enhanced by a factor of 5. The optimal filter is constructed from a noiseless template in all cases. The noise spectrum part of the filter was generated using  expected-noise streams (for panels (a) and (b)) and white-noise streams (for panels (c) and (d)). In each panel, the left figure displays the quadratic difference between the FWHM value obtained with each filter (blue, red and green symbols for FULL, SHORT and \textit{0-padding} filters respectively) and the simulated FWHM. In panel~(a) this accounts for the squared instrumental resolution in \texttt{xifusim}. The figures on the right of each panel display the difference between the FULL and SHORT filter FWHM values (blue and red symbols respectively) and the \textit{0-padding} FWHM value. In all plots, differences are plotted against the Gaussian FWHM values used in the simulations.}
    \label{fig:resol_sims_table}
\end{figure*}

These lists of photon energies served as inputs for the \mbox{\texttt{xifusim}} simulator which generated a current pulse for each photon. 

To gain further insights into the factors influencing the performance of the filters, we devised an additional analysis to differentiate how noise in the filter and noise in the pulse affected the reconstruction process and the determination of the energy resolution. In this case, the way to determine the energy resolution is by measuring the Gaussian FWHM broadening that affects the Lorentzian profiles of the lines in the complex. The instrument magnifies the simulated Gaussian width, and measuring the final FWHM for each set of simulations allowed us to analyze and compare the impact of the instrument on the performance of the different filters under conditions similar to those of the laboratory data in Sect.~\ref{sec:fitting}.

During the simulations, we generated two sets of pulses: one  with the nominal \texttt{xifusim} noise and the other with the nominal noise enhanced by a factor of~5. For constructing the optimal filters we used a noiseless pulse template at 6~keV,  which is close to the mean energy of the Mn~K$\alpha$ complex. Additionally, we derived a noise spectrum from the average of 100\,000 instrument-expected noise streams or white-noise streams. We included the case of white noise, even though it is not realistic for a real instrument, as it simplifies calculations, and in this scenario, the optimal filter reduces to just the pulse template \citep{Szymkowiak1993}.

The reconstruction of the pulse energies was performed using the three filters introduced in Sect.~\ref{sec:intro} which were also used in the analysis of monochromatic pulses as discussed in Sect.~\ref{sec:zero-padding-explanation}. The \textit{0-padding} extreme filter is no longer relevant in the following discussion as it does not provide reasonable values for the energy resolution. The lengths of these filters were as specified in Table~\ref{Tab:simulated_6_keV_pulses}. For the energy calibration of the Mn~K$\alpha$ photons, the gain scale derived from the monochromatic simulations was utilized.

The results of applying the different filters are presented in Fig.~\ref{fig:resol_sims_table}. This figure displays the recovered Gaussian FWHM values of the four different noise combinations in both the pulses and the filters. As anticipated, the FWHM values obtained are greater than the intrinsic simulated values, indicating the effect of the detector broadening the complex lines. 

From these plots, several conclusions can be drawn. 
When simulating pulses with the nominal noise (panels (a) and (c)), the analysis of the simulations revealed that the \textit{0-padding} filter performed slightly better than the FULL filter and clearly outperformed the SHORT filter. This is true at least under ideal instrumental operational conditions, i.e. in the absence of baseline drifts or jitter effects (see Sect.~\ref{sec:datarecon}).  The right figures of the panels clearly show the relative difference of the FWHM values they produce, with respect to the \textit{0-padding} value.

The slight difference in the FWHM value range between panels (a) and (c) may be attributed to the fact that the noise conditions for the pulses and the optimal filter in panel (a)  are the same, representing realistic instrumental noise. In contrast, in panel (c), the filter was constructed with white noise, which did not fully replicate the conditions of the pulse simulations. As a result, this led to slightly larger resolution values.

Interestingly, when we artificially increased the noise of the pulses (panels (b) and (d)) we observed a similar behavior in the filters, albeit with a more pronounced  difference in the resolution values. This reaffirms our previous observation from the analysis of monochromatic pulses in Sect.~\ref{sec:zero-padding-explanation} that the better performance of \textit{0-padding} scales with the level of  instrumental noise.

Additionally, the similarity of the resolution values regardless of the filter noise conditions (nominal or white-noise) indicates that the dominant factor influencing the filter performance is the noise present in the pulses as already discussed in Sect.~\ref{sec:zero-padding-explanation}.

Given that the ideal non-varying conditions defined during the simulations may  not reflect the realistic conditions encountered during the detector operations (on-board or in the laboratories), it becomes crucial to verify whether the corrections implemented during the reconstruction process on real data are adequate to ensure the differential performance of the filters observed in the simulations.

For this purpose, the next step involves the analysis of realistic TES laboratory data.

\section{The laboratory data}
\label{sec:data}

The measurements used in this analysis were taken on a 1-k pixel prototype X-IFU array developed by NASA/GSFC. Up to 250-pixels in the array can be readout using 8-column $\times$~32-row TDM developed by NIST/Boulder. X-rays are generated by fluorescing different metallic and crystal materials, which enables measurements over the energy range 3.3~keV (K~K$\alpha$) to 12~keV (Br~K$\alpha$). Full details on the design and performance of the detector and readout can be found in \citet{Smith2021} and \citet{Durkin2019}.

Specifically, the data analyzed in this work belong to three datasets with different X-ray line complexes, count rates and bath temperature drifts:

\begin{itemize}
    \item \textbf{10Jan2020} (GSFC): initial dataset with several line complexes and $8\times 32$ channels in the detector.
    \item \textbf{30Sep2020} (GSFC): lower count rate dataset of line complex Mn~K$\alpha$ ($8\times 32$ channels), to avoid an additional effect on the energy resolution caused by a possible imperfect removal of cross-talk events which could contaminate the Mn complex.
    \item \textbf{LargeTdrift} (NIST):  two column measurement ($2\times 32$ channels) taken with the NASA Large Pixel Array (LPA) array at NIST in a cryostat that exhibits much larger drift. This dataset was used to test \textit{0-padding} reconstruction under conditions of worse temperature stability.
    
\end{itemize}

\begin{figure*}[htbp]
    \centering 
    \resizebox{\hsize}{!}
    {\includegraphics[width=\textwidth]{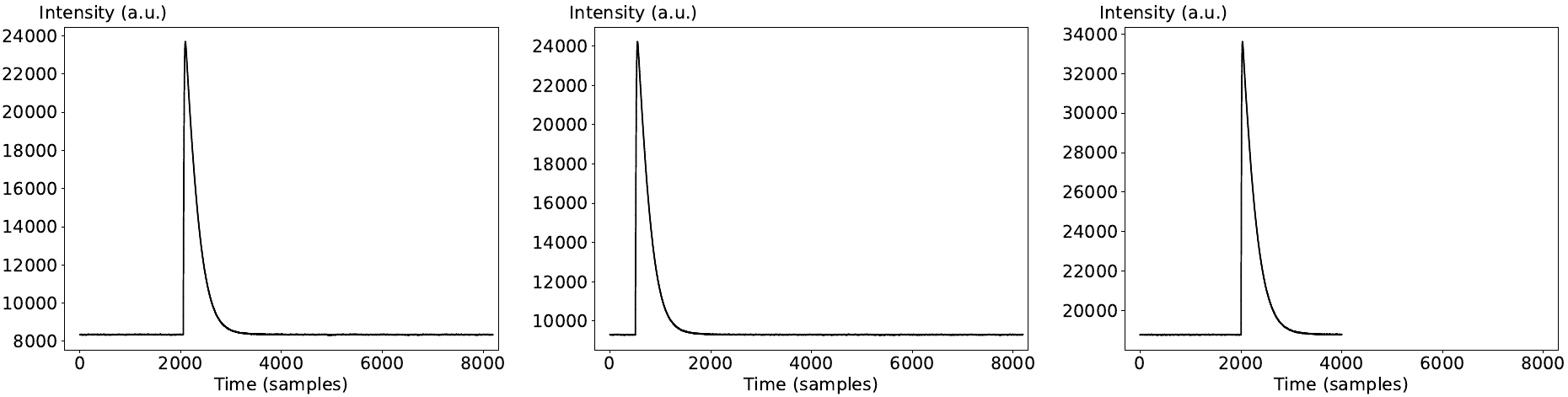}}
    \caption{Data records showing typical pulses for each dataset: 10Jan2020 (left),  30Sep2020 (middle),  LargeTdrift (right). In the X axis, the time in samples for a sampling rate of 195.3125~kHz and in the Y axis the intensity of the pulse in arbitrary units.}
    \label{fig:pulses}
\end{figure*}

Laboratory data are stored in triggered records of data streams containing the current pulses induced by the X-ray photons. A typical record with a single pulse is displayed in Fig.~\ref{fig:pulses} for the three different datasets. As shown, the pulses differ in both the total length of the data record and the pre-trigger length.

To construct the optimal filter used in the data reconstruction, the pulse template used for creating the signal part was obtained by averaging a large number of isolated Mn~K$\alpha$ pulses at 5.9~keV,  as monochromatic as possible. To achieve this, records containing multiple pulses and pulses with heights outside the range of the Mn~K$\alpha$ complex were excluded from consideration. Furthermore, records contaminated by cross-talk events (events produced by a close-in-time impact of another X-ray photon in a different pixel of the same TDM readout column) were also removed from the analysis.

For the noise spectrum, we selected the cleanest set of noise records, ensuring they were free from instrumental artifacts or undesirable effects. Records that produced the largest residuals from the mean noise spectrum were excluded from the selection.

Using the above mentioned pulse template and noise spectrum, we constructed the three types of optimal filters introduced  in Sect.~\ref{sec:intro} and utilized in Sect.~\ref{sec:zero-padding-explanation} and Sect.~\ref{sec:mnka_sims}.

The specific lengths of the filters in the analysis are detailed in Table~\ref{tab:lengths}. It is worth noting that not all the samples in the records were used because the final samples were discarded to avoid alignment problems during template creation. Additionally, for the case of LargeTdrift, the pre-trigger length was reduced due to the shorter pulse length.
 
\begin{table}[htbp]
\caption{Optimal filters pre-trigger lengths and total lengths selected for the analysis.}
\label{tab:lengths}
\centering                          
\begin{tabular}{c c c c c}
\toprule
\noalign{\smallskip}
DATA        & Pre-trigger & FULL & SHORT & \textit{0-padding}  \\
\noalign{\smallskip}
\midrule
\noalign{\smallskip}
10Jan2020   & 2000  & 8000 & 4096 & 4096 \\
30Sep2020   & 450   & 8000 & 4096 & 4096 \\
LargeTdrift & 1000  & 2900 & 1450 & 1450 \\    
\noalign{\smallskip}
\botrule
\end{tabular}
\footnotetext{Note: in samples, for a sampling rate of 195.3125 kHz.}
\end{table}

\section{Data reconstruction}
\label{sec:datarecon}

The energy of the pulses generated by laboratory X-ray photons is estimated using SIRENA through optimal filtering, as described in Sect.~\ref{sec:intro}. 

Initially only photons from the Mn~K$\alpha$ complex, which were used to construct the filter template, were utilized to study the detector's energy resolution. This approach was chosen to minimize any imperfection in the TES non-linearity correction performed by the gain scale transformation. 

\subsection{Energy calibration}
\label{sec:enercal}

Similar to the simulations presented in Sect.~\ref{sec:zero-padding-explanation} and Sect.~\ref{sec:mnka_sims},  a gain scale correction is applied to obtain the real energies from the initial energy estimations. 

\begin{figure*}[tbp]
    \centering
    \includegraphics[width=\linewidth]{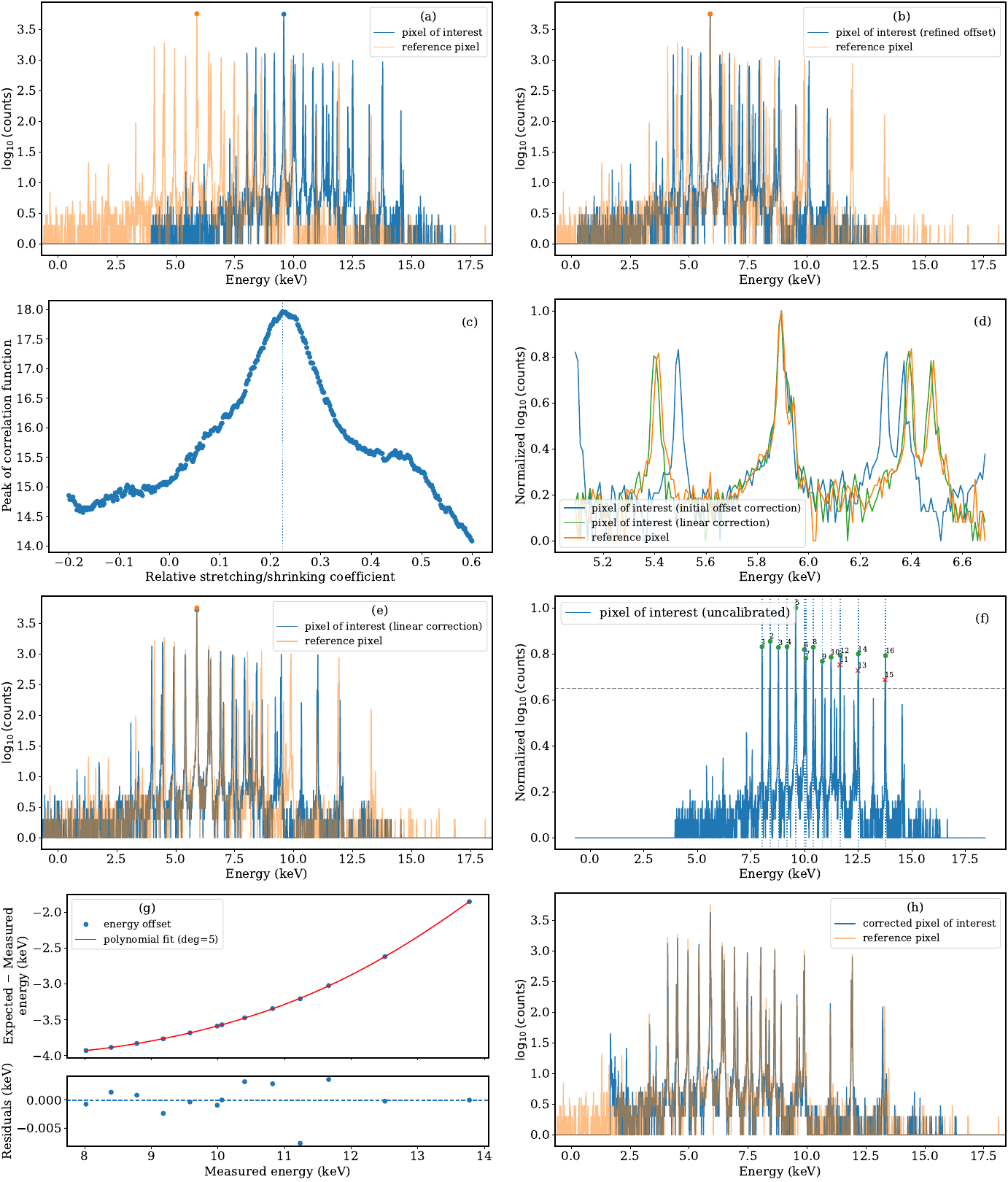}
    \caption{Illustration of the energy calibration procedure. Panel~(a): Determination of a global offset between the pixel of interest (blue, reconstructed with \textit{0-padding} filter) and an already calibrated reference pixel (orange, reconstructed with FULL filter) using the \mbox{Mn K$\alpha$} complex peaks. Panel~(b): Application of the global offset, highlighting the energy scale discrepancy between the reference and pixels of interest. Panel~(c): Initial linear fit to the energy distortion achieved by cross-correlating the reference and pixels of interest in the \mbox{Cr K$\alpha$}--\mbox{Mn K$\beta$} region. Panel~(d) Application of the linear correction to align the problem pixel with the reference data. Panel~(e): Inadequacy of the linear correction for a precise energy calibration across the full energy range. Panel~(f): Identification of line complexes above a predefined relative threshold for a refined calibration. Panel~(g): Computation of the gain scale correction using the initial linear correction for a progressive identification of neighbouring line complexes. Panel~(h): The resulting corrected energy scale after applying the gain scale correction.}
    \label{fig:steps_energy_calibration}
\end{figure*}

The adopted procedure for energy calibration has been developed to ensure its automatic application and it is illustrated in several steps as depicted in Fig.~\ref{fig:steps_energy_calibration}. To begin, we determine a global offset between an already calibrated pixel (orange curve) reconstructed with the FULL filter, and a pixel of interest (blue curve) reconstructed with \textit{0-padding}. The energies of this reference pixel were refined with a gain scale correction obtained from a manual identification of the line complexes whose energies are listed in \citep{lab2016} and tables in Sect.~\ref{sec:other_complexes}. This first step is illustrated in Fig.~\ref{fig:steps_energy_calibration}(a). The offset represents the energy difference between the corresponding \mbox{Mn K$\alpha$} complex peaks. By applying this global offset we can clearly observe the discrepancy in the energy scale between the reference and the pixel of interest, as shown in Fig.~\ref{fig:steps_energy_calibration}(b).

To address the distortion in the energy scale, we initially perform a linear fit by cross-correlating the reference and the pixel of interest within the energy interval containing the \mbox{Cr K$\alpha$} and \mbox{Mn K$\beta$} complexes, using a varying stretching/shrinking coefficient, as shown in Fig.~\ref{fig:steps_energy_calibration}(c). The maximum value in this figure indicates the scale deformation, where a negative coefficient corresponds to energy scale shrinking, and a positive coefficient corresponds to stretching. The linear correction is then applied, as shown in Fig.~\ref{fig:steps_energy_calibration}(d), with the green line representing the energy of the pixel of interest after aligning it with the reference data using the required stretching coefficient.

However, it becomes evident that a linear correction alone is inadequate to achieve precise energy calibration across the entire available energy range, as illustrated in Fig.~\ref{fig:steps_energy_calibration}(e). To obtain a more refined energy calibration, we identify the line complexes above a predefined relative threshold, as shown in Fig.~\ref{fig:steps_energy_calibration}(f). In this process, we use the initial linear correction derived from the \mbox{Cr K$\alpha$}--\mbox{Mn K$\beta$} region to predict the expected location of subsequent line complexes at both lower energies (complexes \mbox{V K$\alpha$}, \mbox{Ti K$\alpha$}, and \mbox{Sc K$\alpha$}) and higher energies (\mbox{Co K$\alpha$}, \mbox{Ni K$\alpha$}, \mbox{Cu K$\alpha$}, \mbox{Zn K$\alpha$}, \mbox{Ge K$\alpha$}, and \mbox{Br K$\alpha$}). This allows us to compute a gain scale correction, as depicted in Fig.~\ref{fig:steps_energy_calibration}(g), which is fitted using a fifth-degree polynomial. The choice of a fifth-degree  polynomial is due to the smaller number of reference energy points compared to the simulations. The application of this gain scale correction results in the corrected energy scale, as seen in Fig.~\ref{fig:steps_energy_calibration}(h).

\subsection{Baseline drift and jitter corrections}
Once the energies of the photons are brought to the correct energy scale, they need to be corrected for instrumental variations that occur during data acquisition. The most notable effects are attributed to baseline time drift caused by instabilities in the TES setup's bath temperature, and the offset between the physical/real arrival time of the photon and its measured/digitized arrival time (jitter) \citep{Fowler2016}.

\begin{figure}[htbp]
    \centering
    \includegraphics[width=1.\linewidth]{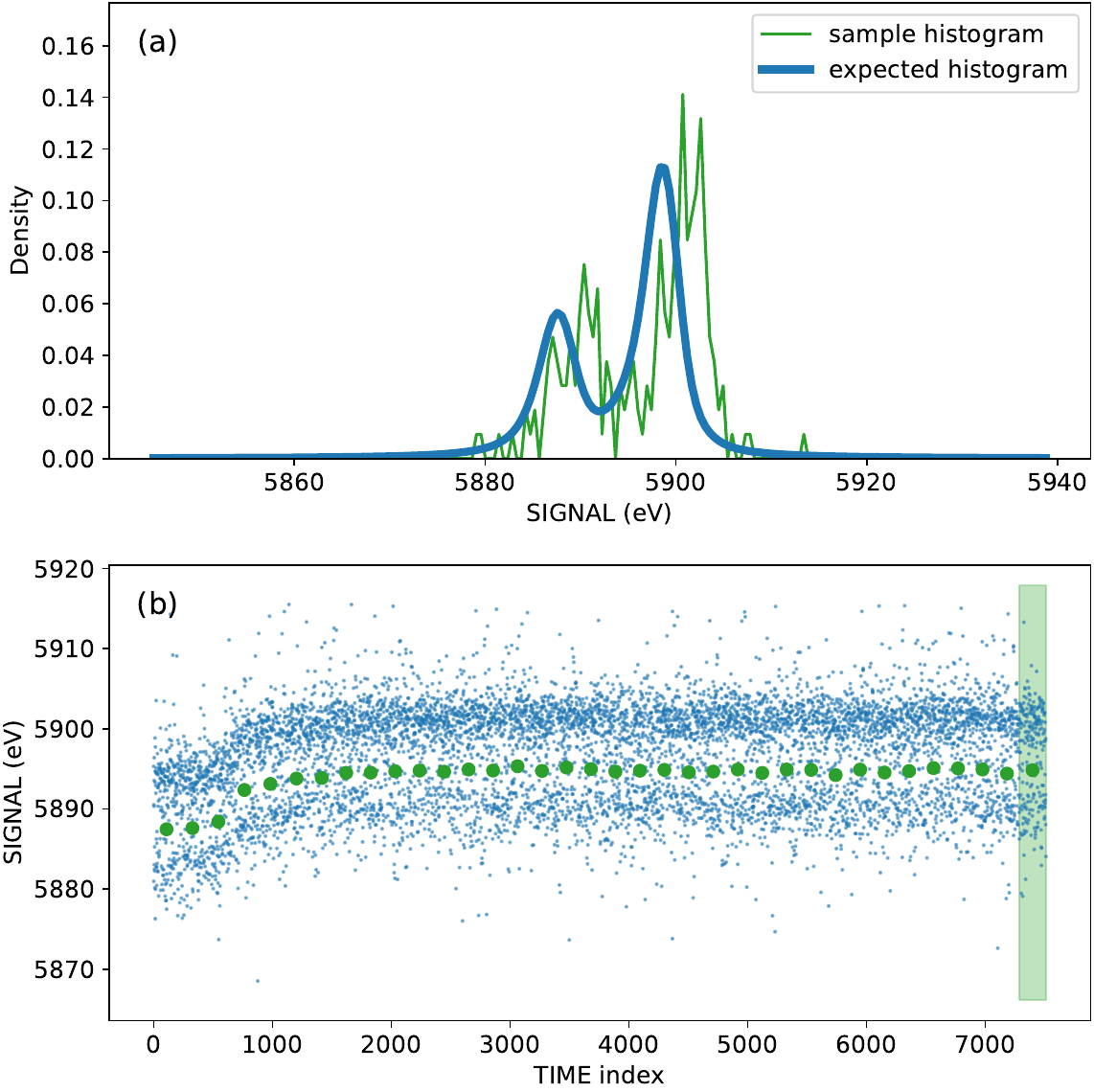}
    \caption{Illustration of cross-correlation method for baseline drift and jitter corrections. Panel~(a): A comparison between the expected energy histogram of the \mbox{Mn K$\alpha$} complex (thick blue line) and a sample histogram (thin green line) obtained by combining photons within a relatively narrow time window. Panel~(b): Gain scale calibrated data (small blue points) displayed as a function of the time index indicating the arrival time of the pulses. The shaded green rectangle exemplifies the width of the moving window used to compute the sample histogram at different times. The green filled circles indicate the relative offsets measured by cross-correlating the two histograms displayed in the top panel.}
    \label{fig:crosscorr_moving_window}
\end{figure}

These two corrections are applied sequentially using a cross-correlation technique, as illustrated in Fig.~\ref{fig:crosscorr_moving_window}. In this example, gain scale calibrated data in the \mbox{Mn K$\alpha$} energy range (shown as the small blue points in panel~\ref{fig:crosscorr_moving_window}(b)) are displayed as a function of the time index indicating the arrival time order of the pulses. 

The first step involves computing an expected energy histogram, represented by the thick blue line in panel~\ref{fig:crosscorr_moving_window}(a), based on an assumed initial Gaussian FWHM for the theoretical line complex Voigt profile. Next, a sample histogram (thin green line) is computed by considering the data values enclosed within a moving window of a fixed width (hereafter referred to as the \texttt{xwidth} parameter), represented by the green shaded rectangle in panel~\ref{fig:crosscorr_moving_window}(b). 

The cross-correlation of both the expected and the sample histograms provides the average energy offset for the data points within the moving window, depicted as green filled circles in panel~\ref{fig:crosscorr_moving_window}(b). As each time window contains \texttt{xwidth} photons and yields only one offset estimate, a final correction for each individual photon is determined by fitting a low-order polynomial to the derived offsets. A Savitzky-Golay interpolation with second degree polynomials and a predefined number of points (referred to as the \texttt{smooth} parameter) can be used for this purpose. In cases where an abrupt energy offset is detected, as observed after the first three computed offsets presented in panel~\ref{fig:crosscorr_moving_window}(b),  the data to be interpolated are split into subsets separated by these energy jumps. This approach is adopted to prevent interpolation attempts across the energy jumps (when present). If the number of data points within a subset falls below the specified \texttt{smooth} parameter, a straightforward linear interpolation is utilized.  

After correcting the energy of each photon, the procedure is repeated for a few iterations. Before each new iteration, the average energy-corrected data histogram is recomputed and its Gaussian FWHM is fitted (for a detailed description of the fitting procedure, refer to Sect.~\ref{sec:fitting}). This fitted FWHM value is then used to create a new expected histogram. Typically, the use of 3~iterations is sufficient to achieve convergence.

Additionally, we tested the cross-correlation method by using the global averaged data histogram as the expected histogram, instead of a theoretical profile. This alternative approach yields the same correction, although its convergence is slower.

The choice of the cross-correlation window size (\texttt{xwidth} parameter) is crucial as using too small a value leads to noisy offset estimates, while too large a value only provides a coarse-grained determination of the energy variation we aim to correct. If we were to unrealistically make \texttt{xwidth} too small, it would result in the offset correction overcompensating for the actual energy displacement. To investigate the potential bias introduced by a too small \texttt{xwidth} parameter, we conducted numerical simulations using 30~series of \mbox{$1.6\times 10^6$} photons following the \mbox{Mn K$\alpha$} complex distribution, assuming a Gaussian FWHM of 2.2~eV and without any distortion in the energy scale. The cross-correlation method was applied with \texttt{xwidth} values ranging from 51 to 601 and \texttt{smooth} values of 11, 31 and~51 points. 

\begin{figure}[htbp]
    \centering
    \includegraphics[width=1.\linewidth]{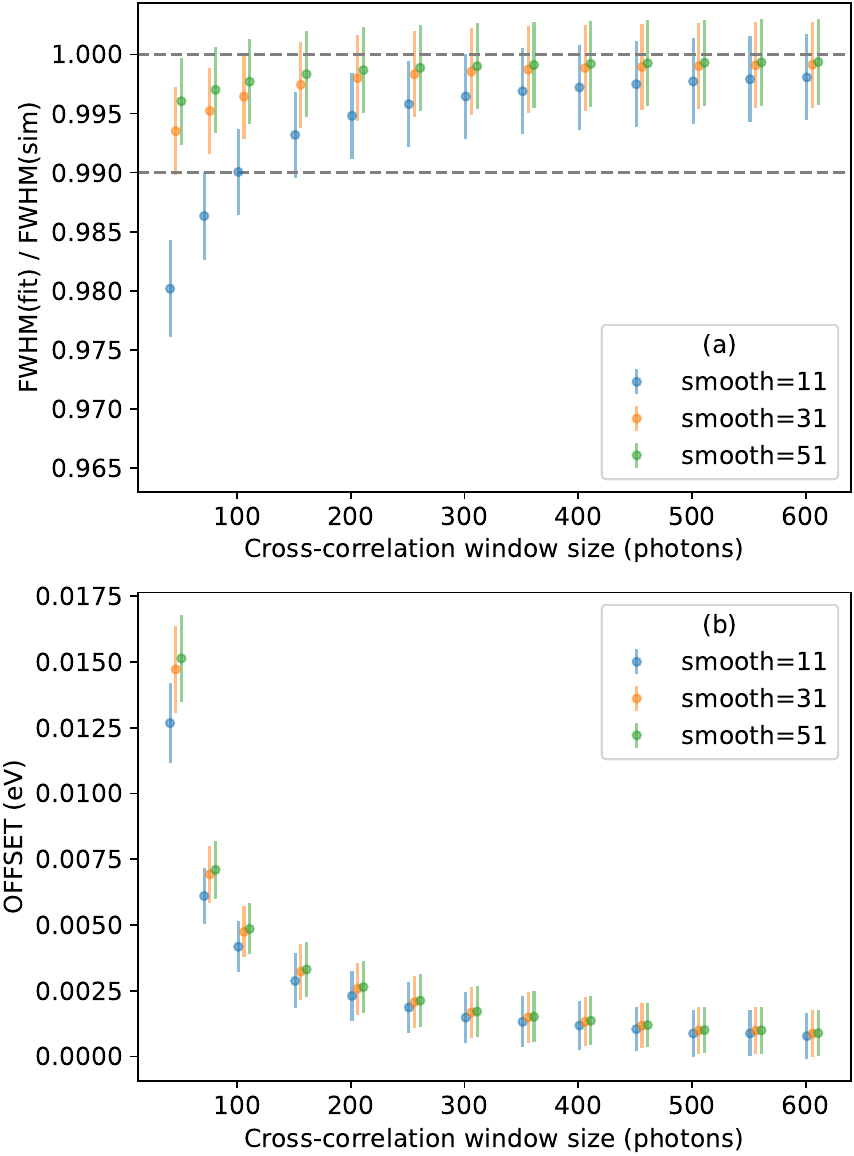}
    \caption{Impact of cross-correlation window size (\texttt{xwidth} parameter) and polynomial smoothing factor (\texttt{smooth} parameter) on fitted Gaussian FWHM. Panel~(a): Ratio of fitted Gaussian FWHM to simulated Gaussian FWHM as a function of cross-correlation window size. Panel~(b): Energy offset in the reconstructed data with varying window sizes for cross-correlation corrections. Both plots consider three smoothing factors: 11 (blue dots), 31 (orange dots), and 51 (green dots) points.}
    \label{fig:crosscorr_xw}
\end{figure}

The results for the Gaussian FWHM and global energy offset of the corrected line complexes are displayed in Fig.~\ref{fig:crosscorr_xw}, revealing that the cross-correlation method tends to slightly over-correct the FWHM, especially for small \texttt{xwidth} and \texttt{smooth} parameters (see panel~\ref{fig:crosscorr_xw}(a)). At the same time, it introduces a minor energy offset in the mean energy of the \mbox{Mn K$\alpha$} complex (see panel~\ref{fig:crosscorr_xw}(b)). A horizontal dashed line at the ratio ${\rm FWHM(fit)}/{\rm FWHM(sim)}=0.99$ is drawn in panel~\ref{fig:crosscorr_xw}(a) to indicate a 1\% threshold for the resolution over-correction introduced by the cross-correlation method. This value is well below the 6\% resolution calibration requirement of the X-IFU. Based on the analysis of these simulations, we have decided to adopt \mbox{${\rm\texttt{xwidth}}=201$} and \mbox{${\rm\texttt{smooth}}=11$} as the default choices for the baseline drift and jitter corrections.

\begin{figure*}[htbp]
    \includegraphics[width=\textwidth]{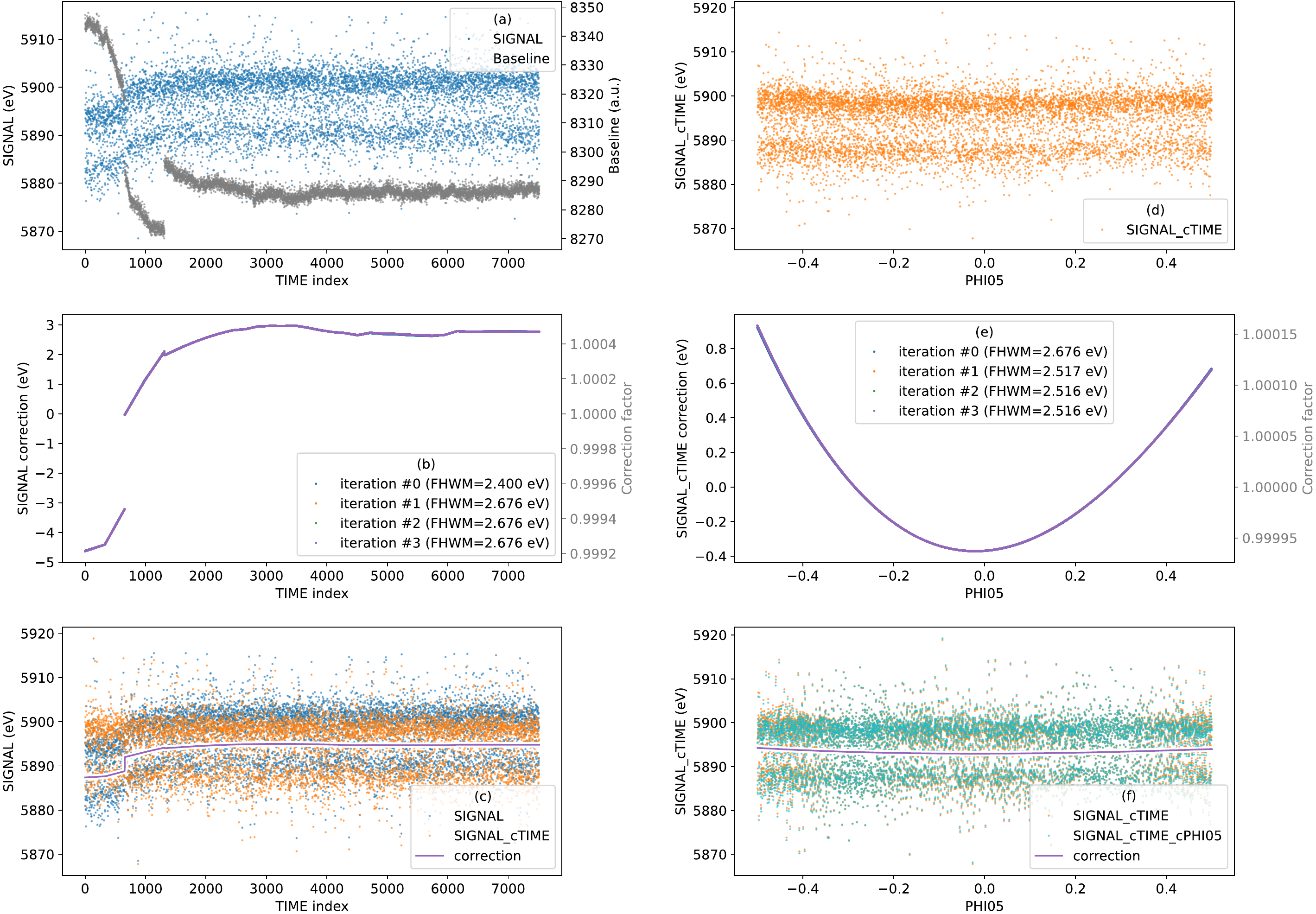}
    
    \caption{Corrections of Mn~K$\alpha$ complex pulses for TIME (linked to baseline) and TIME+JITTER effects, using pixel~1 in the 10Jan2020 dataset. The left column represents the baseline drift correction, while the right column shows the jitter correction. Panel~(a): Gain scale corrected data (small blue dots) and baseline of the corresponding pulses (gray points), plotted as a function of a time index indicating the arrival time of the pulses. Panel~(b): Energy offsets derived by the cross-correlation method as illustrated in Fig.~\ref{fig:crosscorr_moving_window}. Panel~(c): TIME-corrected data (orange) plotted on top of the original data (blue). Panel~(d): TIME-corrected data (SIGNAL\_cTIME) from panel~(c), plotted as a function of the phase in the range of $\pm 0.5$~samples (PHI05). Panel~(e): Energy offsets computed by applying the cross-correlation method again. Panel~(f): TIME+JITTER-corrected data (green) shown on top of the only TIME-corrected data (orange). The magenta lines in panels~(c) and~(f) represent the correction curves already displayed in the middle panels~(b) and~(e), but at the same reconstructed energy scale. This illustration is used to demonstrate the actual magnitude of the applied corrections.}
    \label{fig:timejitter}
\end{figure*}

The graphical illustration of the correction procedure, applied to pixel~1 in the 10Jan2020 dataset, is shown in Fig.~\ref{fig:timejitter}, clearly depicting the baseline variation that occurred during data acquisition and the slight curvature of the energy dependence with the phase (jitter).

\section{Energy resolution determination}
\label{sec:fitting}

To determine the energy resolution (FWHM of the Gaussian component), for both the simulated Mn~K$\alpha$ complex and the laboratory data, we employed two independent approaches: standard fitting of the energy distribution histogram and a new procedure based on the cumulative distribution function \citep[CDF;][]{CardielCDF}.

\subsection{Histogram fitting}
\label{sec:histofit}

The histograms of the corrected-calibrated energies obtained by applying the different filters were fitted using the Fitting module of \textit{AstroPy} \citep{astropy:2013,astropy:2018} employing the Levenberg-Marquardt algorithm and least squares statistic (\texttt{LevMarLSQFitter}). 

For the Mn~K$\alpha$ complex, we utilized a model that simultaneously fits the eight Lorentzian profiles described in Table~\ref{Tab:MnKa} along with an additional Gaussian broadening. The relative intensities of the Lorentzian lines are tied, and the distance between the line centers is also tied relative to the location of a single line. The FWHMs of the Lorentzian profiles are fixed. The Gaussian broadening is a free parameter and is common for all the lines. The FWHM of this Gaussian broadening is used as the figure of merit to quantify the energy resolution of the detector.

During the \textit{AstroPy} fitting procedure, several weight options for the \texttt{LevmarLSQFitter} call have been tested. These weights are defined as the inverse variance ($\sigma^{2}$) of each data bin:

\begin{itemize}
    \item[*] \textbf{\textit{iSig}:} histogram bins are weighted by the number of counts $N$ within each bin ($\sigma^2=N$)
    \item[*] \textbf{\textit{SN}:}  histogram bins are weighted by the Signal-to-Noise ratio in the bin ($\sigma^2 = \frac{1}{\sqrt{N}}$)
    \item[*] \textbf{\textit{None}:} no weight is applied ($\sigma^2=1$)
\end{itemize}

The \textit{iSig} option adopts the iterative approach proposed by \citet{Fowler2014}. It represents one of the alternatives for conducting a Poisson Maximum Likelihood fit (Cash C-statistics) identified in that study as the least biased method.

\begin{figure}[htbp]
    \centering
    \includegraphics[width=1.\linewidth]{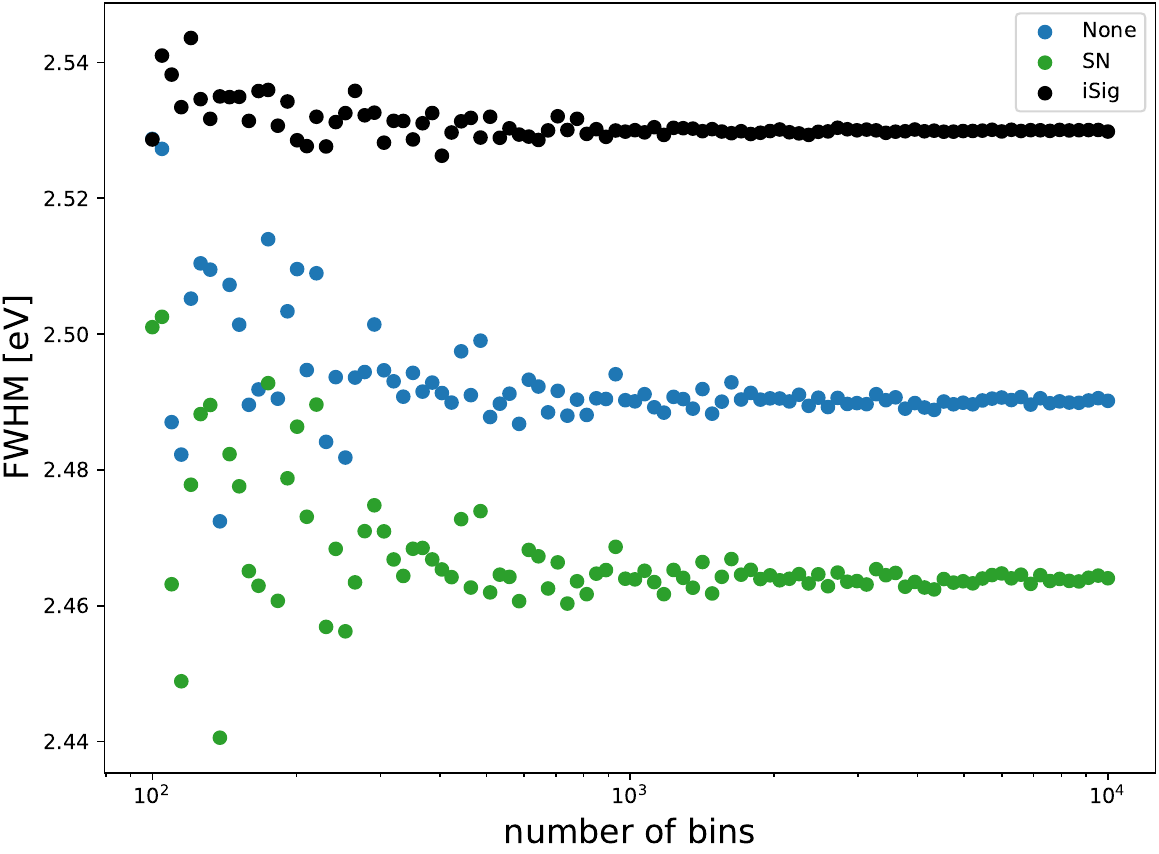}
    \caption{Dependency of energy resolution (Gaussian FWHM) on the number of histogram bins. Different weighting options were considered: \textit{iSig} in black, \textit{None} in blue, and \textit{SN} in green (as described in the text). The results are shown for pixel 1 of the 10Jan2020 dataset.}
    \label{fig:nbins}
\end{figure}

Another crucial parameter in histogram fitting is the number of bins. A study was conducted on the Gaussian FWHM values obtained for different numbers of bins and it was found that using 3000 bins (for a total number of $\sim 8000$ data points spread in the fitted energy range) yields stable results. This is illustrated in Fig.~\ref{fig:nbins} for the case of 10Jan2020 pixel 11. The dispersion shown in the FWHM values provided by the different histogram fittings is consistent with the expected dispersion observed in the simulations (as seen in Sect.~\ref{sec:fittingperfo}).

\subsection{Cumulative Distribution Function Fitting}
\label{sec:CDF}

\begin{figure*}[htbp]
    \centering
    {\includegraphics[width=\textwidth]{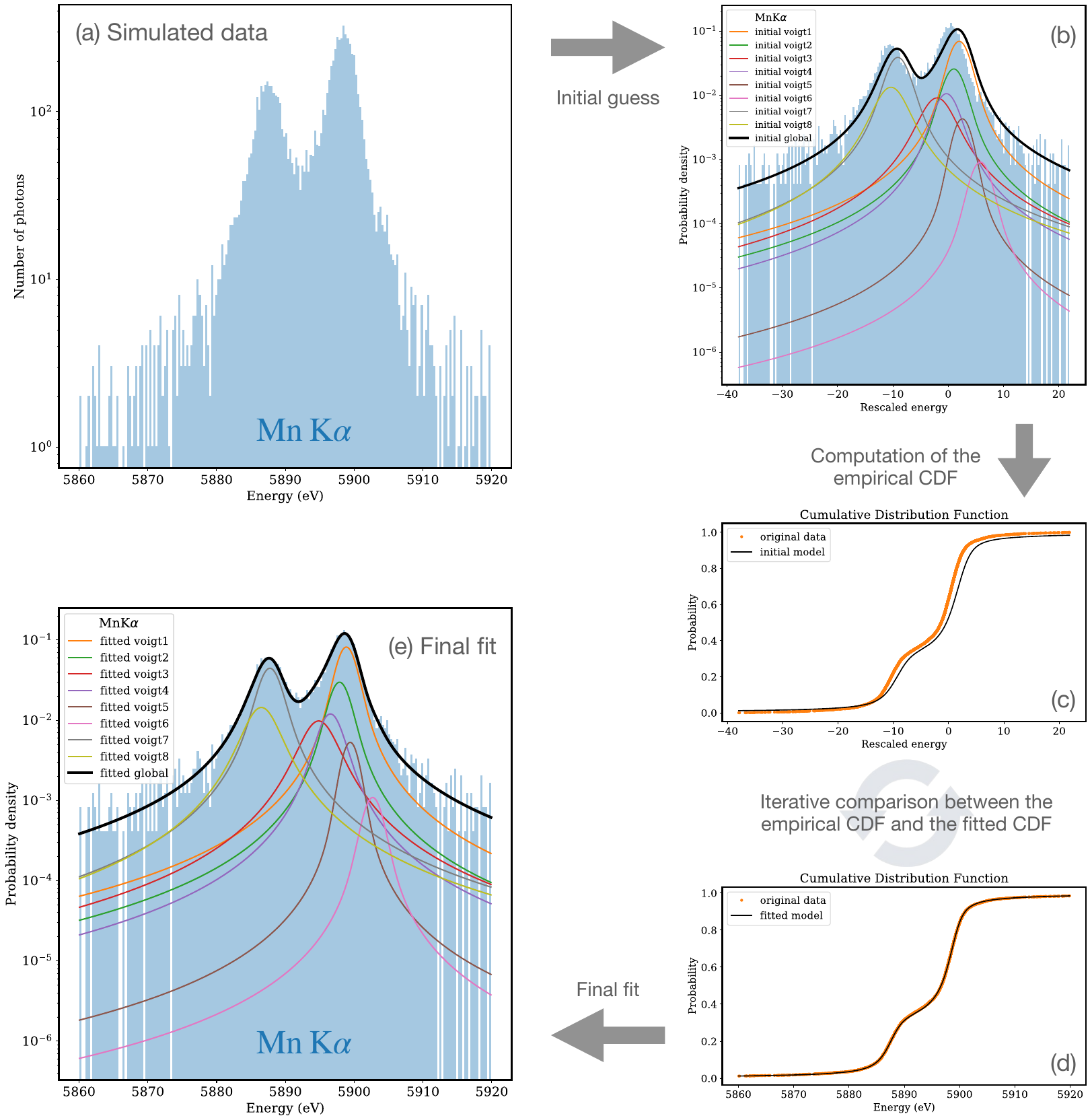}}
    \caption{Schematic of simultaneous fitting of 8~Voigt profiles to the Mn~K$\alpha$ line complex using CDF. Panel~(a): Initial histogram for the data set, consisting of 8300 simulated photons. Panel~(b): Initial guess for the 8~Voigt profiles (thin colored lines) and the expected total probability distribution (thick black line). Panel~(c): Comparison of the empirical CDF of the data to be fitted (orange line) and the temporary fit (black line). Panel~(d): Same as panel~(c) after the iterative numerical fit has converged. Panel~(e): Final fit for the individual 8~Voigt profiles (colored lines) and the global fit (thick black line).}
    \label{fig:CDF_procedure}
\end{figure*}

In order to avoid the need for a priori binning of the data, we explored an alternative approach based on fitting the Cumulative Distribution Function of the photon energies. To test the consistency of this method, we conducted simulations using 8300 photons of the Mn~K$\alpha$ complex energy distribution following the procedure described in Sect.~\ref{sec:mnka_sims}, aiming to have a similar number of pulses as in the laboratory pixels typically analyzed. 

The resulting energy histogram from the simulations displayed the expected double-peak distribution for the Mn~K$\alpha$ complex at our spectral resolution, as shown in panel (a) of Fig.~\ref{fig:CDF_procedure}. Interestingly, a small fraction of the simulated photons fell outside of the displayed energy range, 96~photons below the lower 5860~eV limit and 121~photons above the upper 5920~eV limit.

We demonstrated that the CDF fitting procedure could successfully recover the original parameters used to generate the simulated data set. However, we faced the challenge of having 4 free parameters (i.e., the energy of one reference line, the Gaussian FWHM, and the number of photons below 5860~eV and above 5920~eV) due to the constraints imposed by fixing the Lorentzian FWHM and relative intensity of the eight individual lines, as well as their center-to-center distances. 

To address this challenge we used the following approach: first, we provided an initial guess for the solution, as shown in panel~\ref{fig:CDF_procedure}(b). Next, we compared the empirical CDF of the simulated data (orange line) with the CDF of the temporary solution (black line) in panel~\ref{fig:CDF_procedure}(c). Finally, we used the Levenberg-Marquardt minimization procedure, with the help of the Python package {\sc lmfit}\footnote{\url{https://lmfit.github.io/lmfit-py/}} \citep{newville2021}, to determine a better fit to the empirical CDF, as illustrated in panel~\ref{fig:CDF_procedure}(d). The objective function to be minimized was defined as the weighted difference between the empirical and simulated CDF, using the product $F(x)\,\left(1-F(x)\right)$ as the weight, where $F(x)$ is the empirical CDF. Note that this weighting scheme favours data points nearer the center of the line complex, diminishing the influence of the \mbox{Mn K$\alpha$} complex tails, which are prone to systematically missing photons when correcting energies from pulses severely affected by baseline drift effects. The result of this fitting process was the simultaneous fitting of the eight sought Voigt profiles, as shown in panel~\ref{fig:CDF_procedure}(e).

\subsection{Fitting procedures performance on simulated data}
\label{sec:fittingperfo}

\begin{figure*}[htbp]
    \centering
    \resizebox{\hsize}{!}
    {\includegraphics[width=\textwidth]{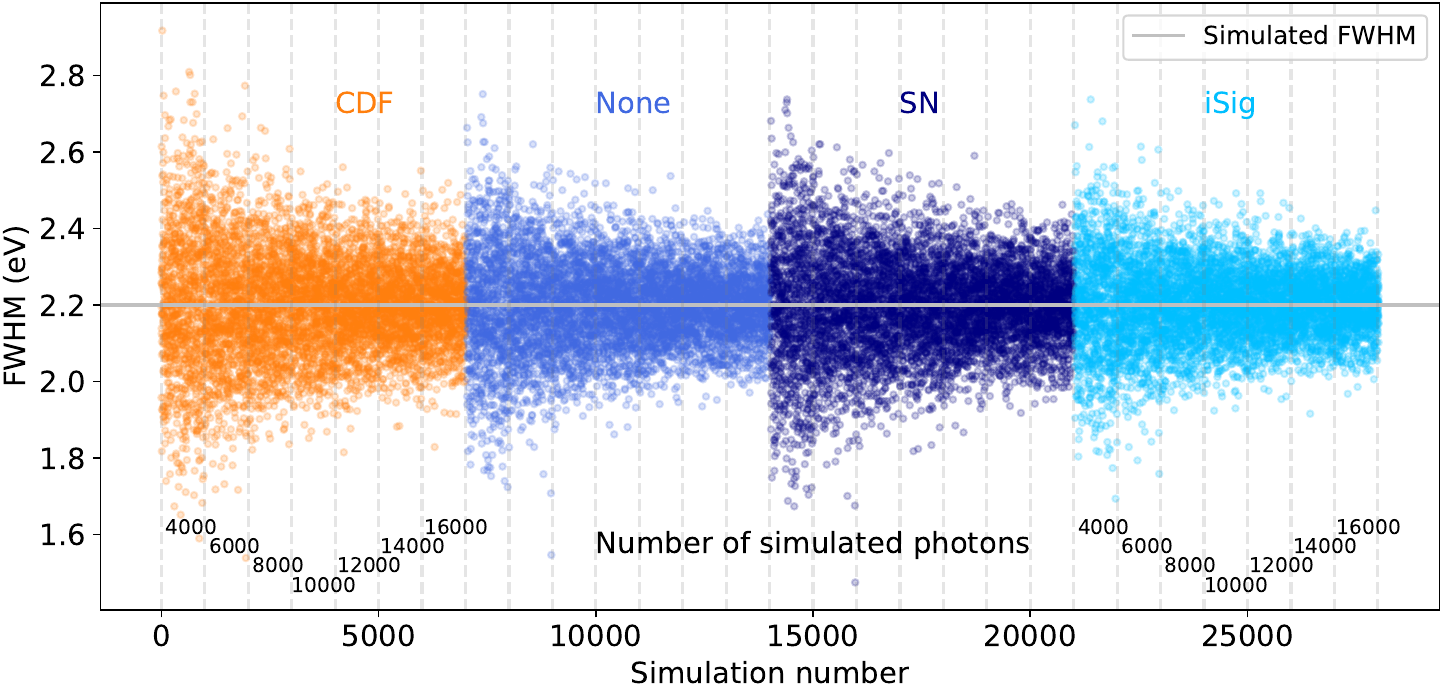}}
    \caption{Distribution of Gaussian FWHM values obtained from simulated data using histogram fitting with three different weights: \textit{None} (light blue), \textit{SN} (dark blue), and \textit{iSig} (cyan), as well as using CDF fitting (orange). The simulations were performed with varying numbers of photons (ranging from 4000 to 16\,000, as specified in the numeric labels). The dashed vertical lines separate sets of~1000 simulations performed with a fixed number of photons. A histogram representation of these data is shown in Fig.~\ref{fig:sims_hist}.}
    \label{fig:sims_cones}
\end{figure*}

\begin{figure*}[htbp]
    \centering
     \resizebox{\hsize}{!}
    {\includegraphics[width=\textwidth]{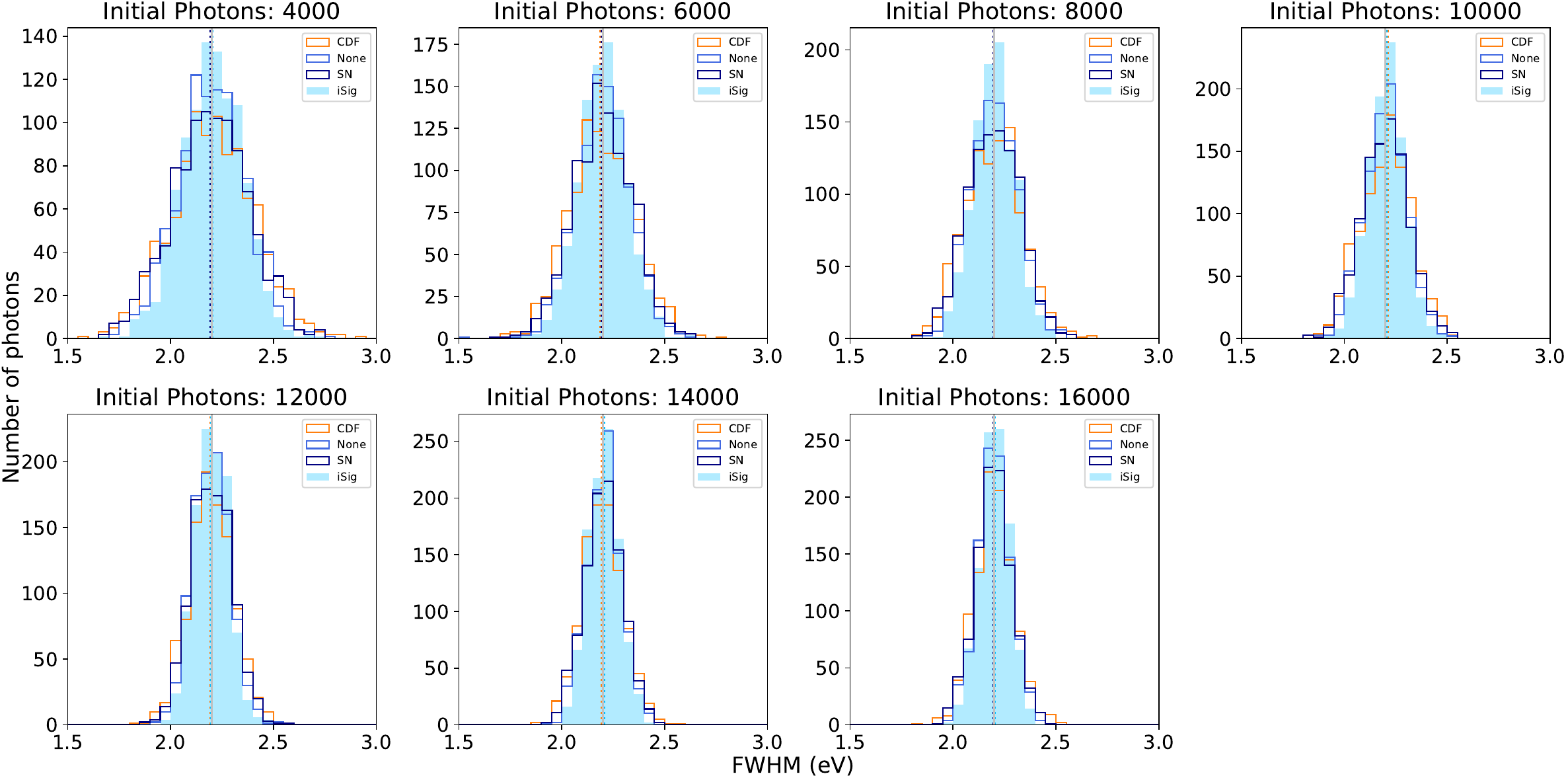}}
    \caption{Histogram representation of data shown in Fig.~\ref{fig:sims_cones}, corresponding to the Gaussian FWHM values obtained from simulated data using histogram fitting with different weights (\textit{None} in light blue, \textit{SN} in dark blue, and \textit{iSig} in shaded cyan) and using the CDF fitting method (in orange). Each panel shows the FWHM distribution for a different number of simulated photons (as indicated at the top of each histogram). The centroid offsets and standard deviations of these histograms are listed in Table~\ref{tab:sigma_sims_hist}.}
    \label{fig:sims_hist}
\end{figure*}

The simulations of the Mn~K$\alpha$ complex were extended to include various numbers of photons, ranging from 4000 to 16\,000, to ensure the accuracy and reliability of the fitting methods when analyzing real data. As previously mentioned, these simulations utilized the eight Voigt profiles defined by the Lorentzian profiles with laboratory parameters from Table~\ref{Tab:MnKa}, and a FWHM Gaussian broadening of 2.2~eV. 

We analyzed these synthetic data using histogram fitting with the three different weights described in Sect.~\ref{sec:histofit}, as well as the CDF fitting method explained in Sect.~\ref{sec:CDF}. The obtained Gaussian FWHM values from the two fitting procedures were compared with the simulated value to evaluate their performance in terms of energy resolution. The distributions of fitted Gaussian FWHMs obtained from the simulated data with varying numbers of photons are presented in Figs.~\ref{fig:sims_cones} and \ref{fig:sims_hist}. 

The results indicate that the distributions are well centered around the simulated resolution value  \mbox{${\rm FWHM}=2.2~{\rm eV}$}, with no systematic deviations at the peak values. Therefore, we can conclude that the fitting methods do not introduce any intrinsic systematic errors to the determination of the Gaussian FWHM of simulated data.

As expected, the dispersion of the measured FWHM decreases as the number of fitted photons in the Mn~K$\alpha$ complex increases, regardless of the adopted fitting procedure. The histogram fitting method with weight \textit{iSig} provides a slightly lower dispersion, as visually evident in Fig.~\ref{fig:sims_hist}. In this figure, the histogram fitting method with weight \textit{iSig} (shaded cyan) appears narrower than histogram fitting with the other weights, and the CDF method yields the widest distribution. This behavior is quantitatively presented in Table~\ref{tab:sigma_sims_hist}, which lists the centroid offsets and standard deviations of the histograms displayed in Fig.~\ref{fig:sims_hist}.

\begin{table*}[htbp]
\caption{Centroid deviations (relative to the simulated FWHM=2.2~eV) and standard deviations of the histograms displayed in Fig.~\ref{fig:sims_hist}.}
\label{tab:sigma_sims_hist}
\centering  
\begin{tabular*}{\textwidth}{@{\extracolsep\fill}crrrrcccc}
\toprule
\noalign{\smallskip}
& \multicolumn{4}{c}{Centroid deviation (eV)} &
\multicolumn{4}{c}{Standard deviation (eV)} \\
\cmidrule{2-5}
\cmidrule{6-9}
$N_{\rm photons}$ & CDF & \textit{iSig}\; & \textit{None} & \textit{SN}\;\; & CDF & \textit{iSig} & \textit{None} & \textit{SN} \\
\noalign{\smallskip}
\hline
\noalign{\smallskip}
\phantom{1}4\,000 &    0.010 &    0.003 & $-$0.003 & $-$0.006 & 0.197 & 0.147 & 0.165 & 0.185 \\
\phantom{1}6\,000 &    0.007 & $-$0.002 & $-$0.003 & $-$0.005 & 0.161 & 0.117 & 0.133 & 0.148 \\
\phantom{1}8\,000 &    0.006 &    0.001 &    0.000 & $-$0.001 & 0.144 & 0.098 & 0.114 & 0.129 \\
          10\,000 & $-$0.003 & $-$0.004 & $-$0.003 & $-$0.003 & 0.128 & 0.094 & 0.105 & 0.117 \\
          12\,000 &    0.005 &    0.004 &    0.003 &    0.002 & 0.116 & 0.085 & 0.097 & 0.107 \\
          14\,000 &    0.002 & $-$0.002 & $-$0.001 & $-$0.001 & 0.107 & 0.075 & 0.085 & 0.096 \\
          16\,000 &    0.004 & $-$0.002 & $-$0.003 & $-$0.003 & 0.103 & 0.070 & 0.080 & 0.089 \\
\noalign{\smallskip}
\hline
\end{tabular*}
\footnotetext{Note: In each case, the different columns indicate the result corresponding to the four fitting methods: CDF, and histogram with three different weights  \textit{iSig}, \textit{None} and \textit{SN}. The first column indicates the number of photons in the simulated Mn~K$\alpha$ line complex.}
\end{table*}

\subsection{Fitting procedures performance on real data}

\begin{figure*}[htbp]
    \centering
    \includegraphics[width=0.63\textwidth]{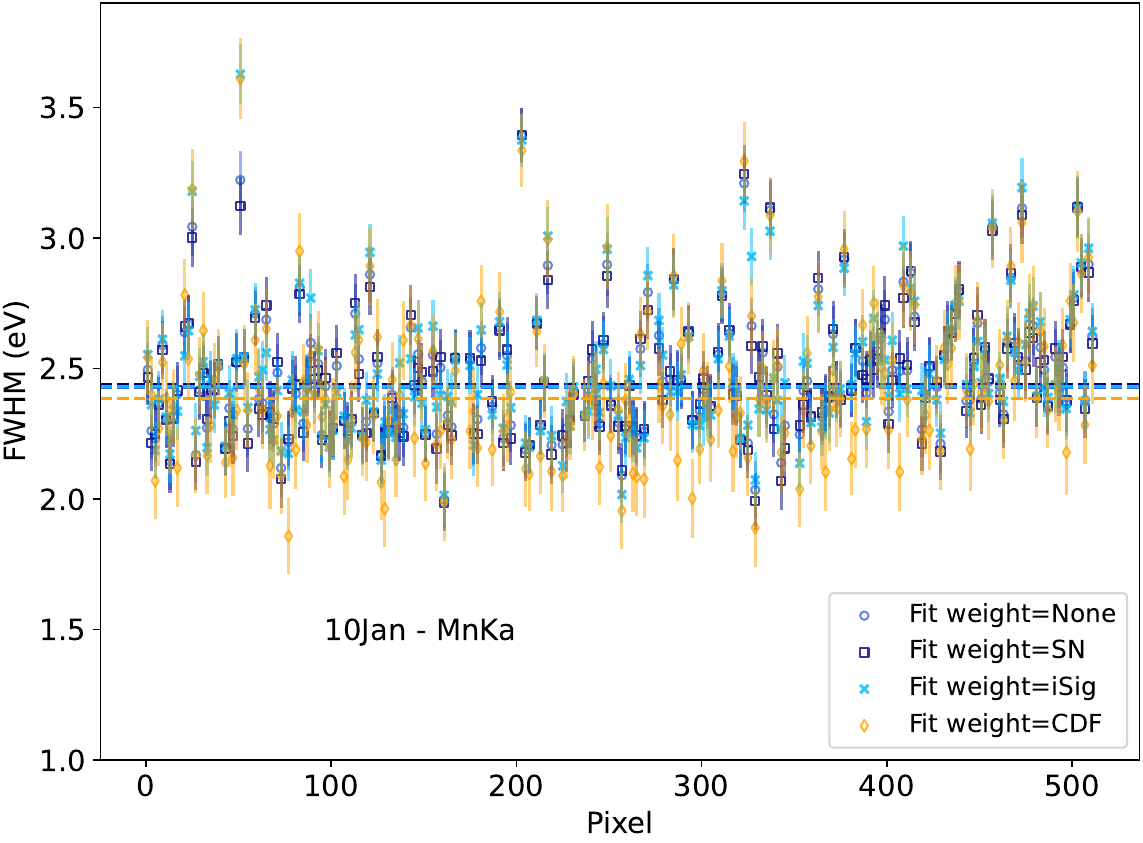}
    \includegraphics[width=0.63\textwidth]{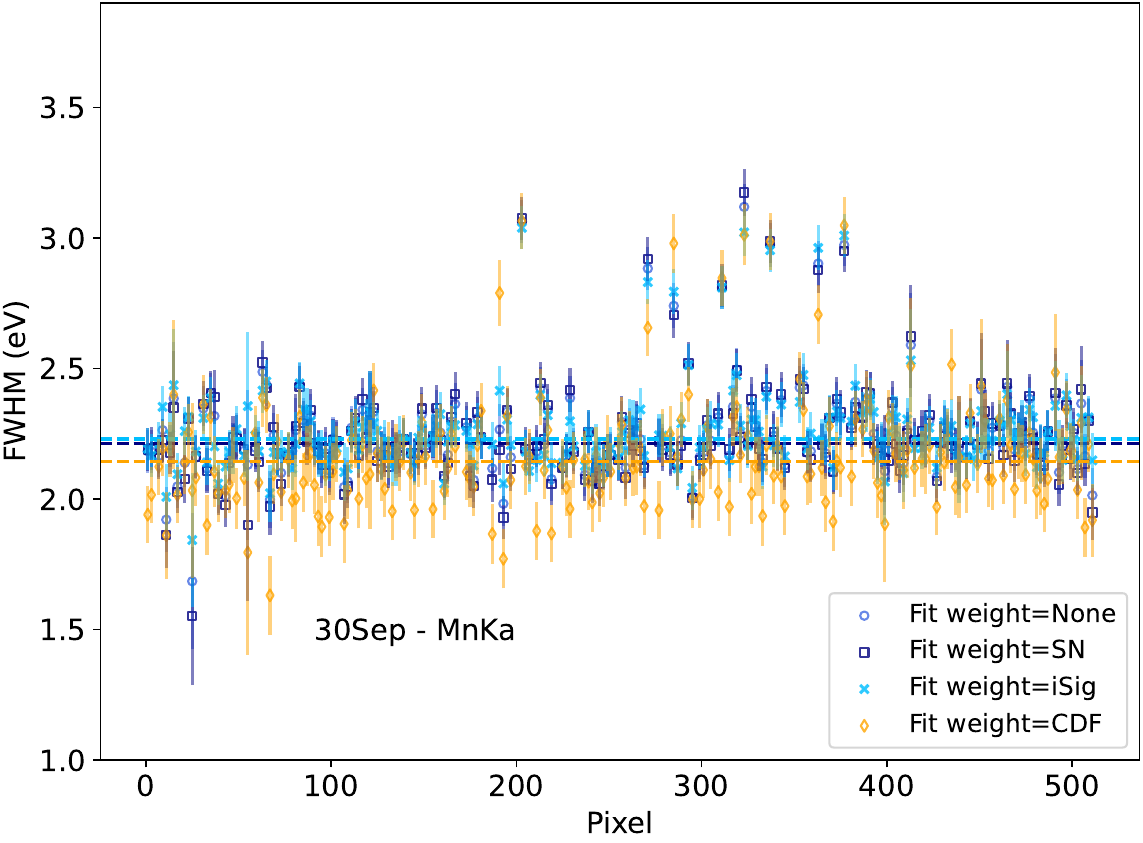}
    \includegraphics[width=0.63\textwidth]{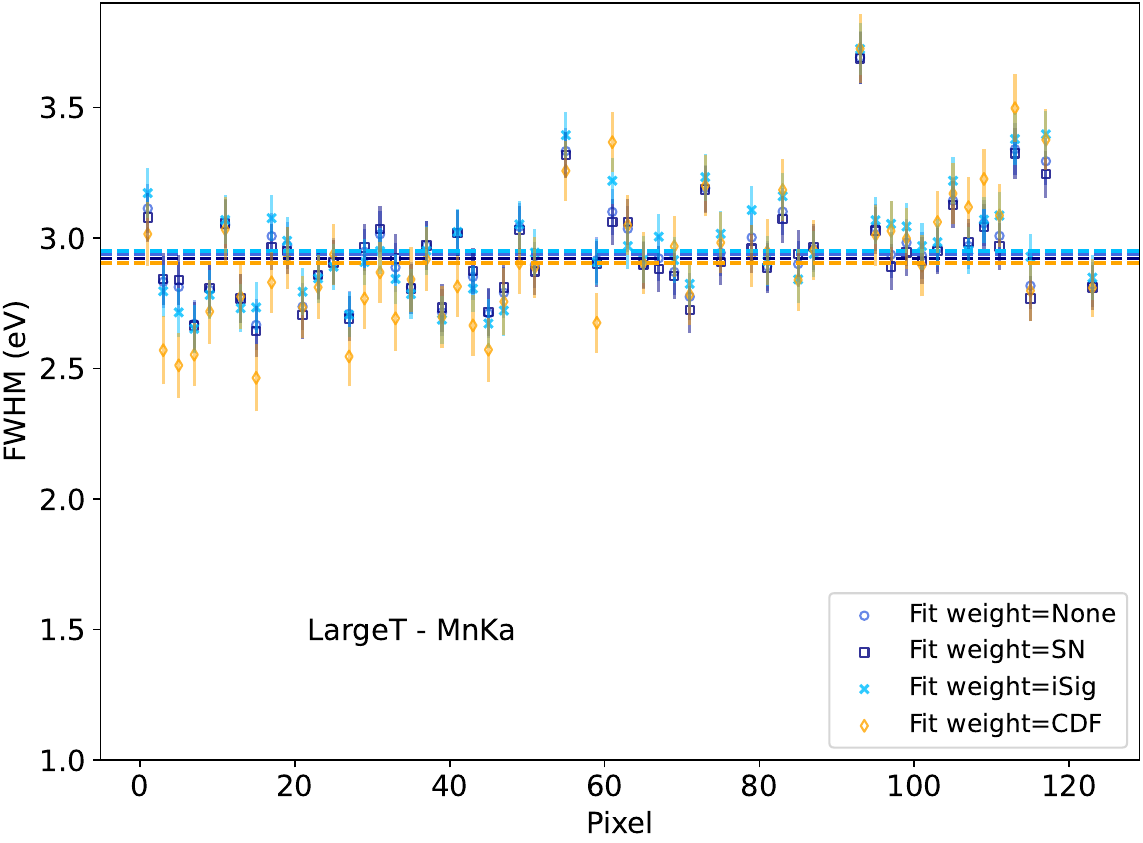}
    \caption{FWHM values for pixels in datasets 10Jan2020 (top), 30Sep2020 (middle), and LargeTdrift  (bottom), using histogram fitting with weights \textit{None} (light blue circles), \textit{SN} (dark blue squares), and \textit{iSig} (cyan $\times$) as well as CDF fitting (orange symbols), all reconstructed using the FULL filter. The dashed lines represent the FWHM obtained by each fitting procedure using the combined information from all pixels.}
    \label{fig:fitweis}
\end{figure*} 

Once we ensured the comparability of the fitting methods on simulated data, we proceeded to analyze real data starting with the individual analysis of each pixel. To do this, we reconstructed the photons from each pixel using the FULL optimal filter, performed gain calibration, and corrected for baseline drifts and jitter. Afterward, we applied both the histogram and the CDF methods to fit the data. 

Note that in this comparative analysis, we are using the FULL filter for the fitting methods, as it serves as the reference method in the literature, allowing us to isolate and evaluate  the performance of the fitting methods independently of the filters' performances.

In contrast to the observations made with the earlier described simulations, we noticed a consistent pattern across all datasets. Specifically, the CDF method consistently yielded slightly lower Gaussian FWHM values (median=2.38~eV) compared to those obtained from the histogram fittings (median=2.43~eV). A Wilcoxon signed-rank test for paired data (non-parametric) was conducted under the null hypothesis that the CDF method yields larger FWHM values than those provided by the histogram fittings. The resulting p-values were $2.2\times 10^{-10}, 9.5\times 10^{-5}$ and $2.8\times 10^{-7}$ for the \textit{iSig}, \textit{SN} and \textit{None} cases, effectively rejecting the null hypothesis and thus confirming the statistical significance of the comparison. This trend is illustrated in Fig.~\ref{fig:fitweis}.

\begin{figure*}[htbp]
    \centering
    \resizebox{\hsize}{!}
    {\includegraphics[width=\textwidth]{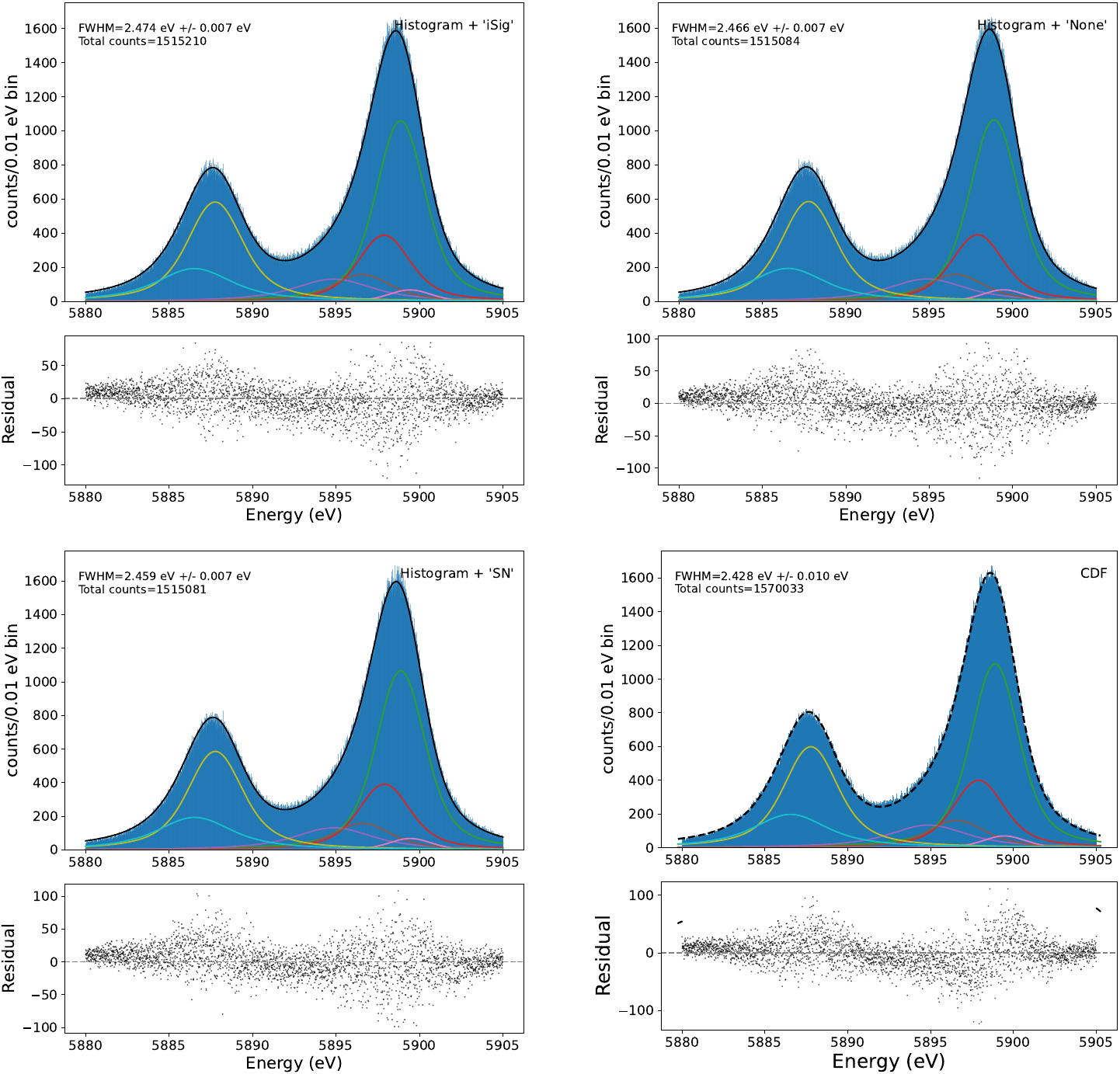}}
    \caption{Global fits to the combined data from all pixels in the 10Jan2020 dataset using different fitting methods. The histogram fitting methods with \textit{iSig}, \textit{None}, and \textit{SN} weights, as well as the fit with the CDF fitting method, are shown. The associated residuals (model - data) are displayed below each particular fit. The plots are arranged from left to right and top to bottom in the order of histogram with \textit{iSig} weights, histogram with \textit{None} weights, histogram with \textit{SN} weights, and CDF fitting. The line complex is fitted using 8 individual Voigt profiles (colored lines), with the black curve representing the co-added result. The fitted Gaussian FWHM value is displayed in the inset text.}
    \label{fig:FWHM_all}
\end{figure*}

To further investigate this discrepancy in the fitting procedures on real data and to assess their performance with improved statistical significance, we conducted a global fit of all the pixels in dataset 10Jan2020. Initially, we processed each pixel individually and subsequently, we combined all the pixels for the global fit using the selected fitting procedure. Figure~\ref{fig:FWHM_all} displays the resulting fits and residuals \mbox{(${\rm model} - {\rm data}$)} obtained for each method. As expected, the FWHM value obtained from the CDF fitting is lower than the values corresponding to the other histogram fitting methods. However, this outcome does not corroborate the results obtained from the analysis of the simulations.

Upon conducting a detailed examination of the residuals in Fig.~\ref{fig:FWHM_all}, it becomes evident that the data on the left wing of the line complex consistently falls below the global fit. As mentioned earlier, the baseline drift correction introduces energy shifts that may cause a slight under-representation of photons at the edges of the energy interval initially chosen for selecting the Mn~K$\alpha$ photons. Consequently, an imperfect photon distribution in the wings of the complex is anticipated to result in variations in the fitted resolution across different fitting methods. Indeed, we have confirmed that eliminating the \mbox{$F(x)\,\left(1 - F(x)\right)$} weight in the CDF method (refer to Sect.~\ref{sec:CDF}), thus giving more prominence to the complex wings in the CDF fit, increases the discrepancy in resolution when comparing histogram fitting with CDF fitting.

After analyzing the results obtained from histogram fitting using three different weights, all of which yielded similar resolution values, we observed that the CDF method consistently produced slightly lower, yet still close results. Appendix~\ref{app:systematics} explores the impact of two known systematic effects, the extended line spread function and the instrumental background, on the different performance of the methods. However none of these factors account for the differences found when analysing real data.

Additionally, we considered the dispersion of the FWHM estimates provided by the \textit{iSig} weight, and found it to be the lowest among the options.

Based on these considerations, we have chosen the histogram fitting method with the \textit{iSig} weight as our baseline approach for analyzing the various datasets. 

\section{Energy resolution analysis: the filters' role}
\label{sec:methods}

In Appendix~\ref{app:std_fwhm}, we present a mathematical expression that quantifies the expected uncertainty in the measured Gaussian FWHM. This expression depends on both the number of photons in the Mn~K$\alpha$ line complex and the FWHM value itself. We will use this derived uncertainty in the upcoming plots of the Gaussian FWHM values measured for real data.

\begin{figure*}[htbp]
    \centering
    \resizebox{\hsize}{!}
    {\includegraphics[width=\textwidth]{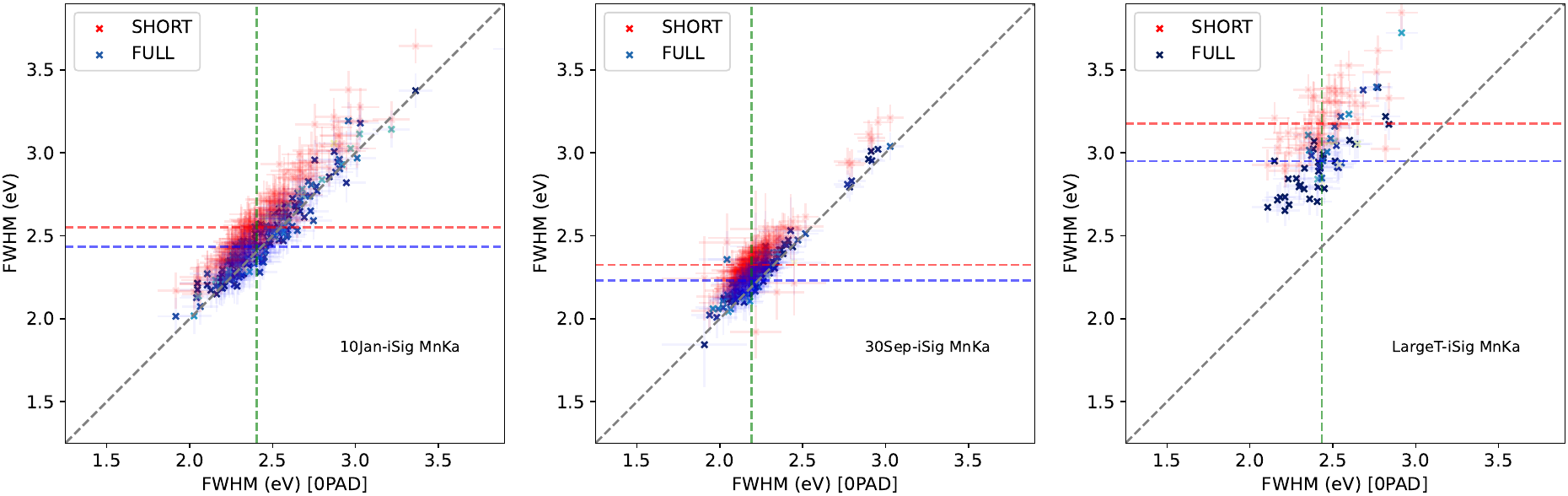}}
    \caption{Comparison of energy resolution values for the FULL filter (blue) and SHORT filter (red) plotted on the Y-axis versus those from the \textit{0-padding} filter reconstruction on the X-axis for datasets 10Jan2020 (left), 30Sep2020 (center), and LargeTdrift (right). The fitting technique applied in the minimization process is the histogram fitting with  \textit{iSig} weight. The dashed lines represent the mean energy resolution values, with red for SHORT, blue for FULL, and green for \textit{0-padding} filters.}

    \label{fig:resol}
\end{figure*}

The comparison of the energy resolution values obtained for all the pixels in the 10Jan2020, LargeTdrift and 30Sep2020 datasets is presented in Fig.~\ref{fig:resol}.

Based on the obtained results, the \textit{0-padding} filter demonstrates comparable performance to the FULL length filter and outperforms the SHORT filter in terms of energy resolution values for all datasets, even under varying instrumental stability conditions and cross-talk levels. Notably, the most significant advantage is observed in  the LargeTdrift dataset, which can be attributed to the shorter length of the LargeTdrift records and filters (see Table~\ref{tab:lengths}). As a result, the $f=0$~Hz bin which is discarded in the construction of the optimal filters contains more information, making its impact more relevant.

Furthermore, the \textit{0-padding} filter not only excels in terms of energy resolution but also provides the added advantage of reduced computational cost.

However, during simulated data tests \citep{Cobo2020}, it was observed that the \textit{0-padding} filter showed heightened sensitivity to baseline fluctuations during data acquisition. This sensitivity can be attributed to the fact that the \textit{0-padding} filter, as explained in Sect.~\ref{sec:intro},  is essentially a truncation of the FULL filter in the time domain. Consequently, it lacks perfect zero-summing when compared to the FULL filter, which leads to increased sensitivity to baseline fluctuations.

To address this issue, we tested a modified \textit{0-padding} technique (called \textit{00-padding}). In this modified approach, we enforced the filter to have a sum of zero in the time domain using different expressions referred to as \texttt{zsum1}, \texttt{zsum2} and \texttt{zsum3} as described in \mbox{Eqs.~(\ref{eq:S1})--(\ref{eq:zsum3})}. We assume \mbox{$N_{\rm final}=8192$}, \mbox{$N_{\rm cut}=N_{\rm final}/2=4096$}, and define
\begin{equation}
S\!1 \equiv \sum_{N_{{\rm cut}+1}}^{N_{\rm final}} \!{\widetilde{o\!f}}_{\text{\textrm{FULL}}}[t_i].
\label{eq:S1}
\end{equation}

The three \textit{00-padding} optimal filters are built using the following prescriptions:
\\[2pt]

\noindent $\bullet$\;\textit{00-padding} with \texttt{zsum1}:
\begin{equation}
\widetilde{o\!f}_{\text{\textrm{00PAD}}}[t_i] = \widetilde{o\!f}_{\text{\textrm{FULL}}}[t_i]+\frac{S\!1}{N_{\rm cut}}, \quad \forall i=1,\ldots,N_{\rm cut}
\label{eq:zsum1}
\end{equation}

\noindent $\bullet$\;\textit{00-padding} with \texttt{zsum2}:
\begin{equation}
\widetilde{o\!f}_{\text{\textrm{00PAD}}}[t_i] = \widetilde{o\!f}_{\text{\textrm{FULL}}}[t_i]+\widetilde{o\!f}_{\text{\textrm{FULL}}}[t_{i\!+\!N_{\rm cut}}], \quad \forall i=1,\ldots,N_{\rm cut}
\label{eq:zsum2}
\end{equation}

\noindent $\bullet$\;\textit{00-padding} with \texttt{zsum3}:
\begin{equation}
\begin{split}
\widetilde{o\!f}_{\text{\textrm{00PAD}}}[t_i] & = \widetilde{o\!f}_{\text{\textrm{FULL}}}[t_i], \quad \forall i=1,\ldots,N_{\rm cut}/2 \\[4pt]
\widetilde{o\!f}_{\text{\textrm{00PAD}}}[t_{i\!+\!N_{\rm cut}/2}] & = \widetilde{o\!f}_{\text{\textrm{FULL}}}[t_{i\!+\!N_{\rm cut}/2}] + \widetilde{o\!f}_{\text{\textrm{FULL}}}[t_{i\!+\!N_{\rm cut}}] \,+ \\
& + \widetilde{o\!f}_{\text{\textrm{FULL}}}[t_{i\!+\!3 N_{\rm cut}/2}]\\
&\quad\quad\quad\quad\quad\quad\quad\forall i=1,\ldots,N_{\rm cut}/2
\label{eq:zsum3}
\end{split}
\end{equation}

\begin{figure*}[htbp]
    \centering
    \includegraphics[width=0.99\textwidth]{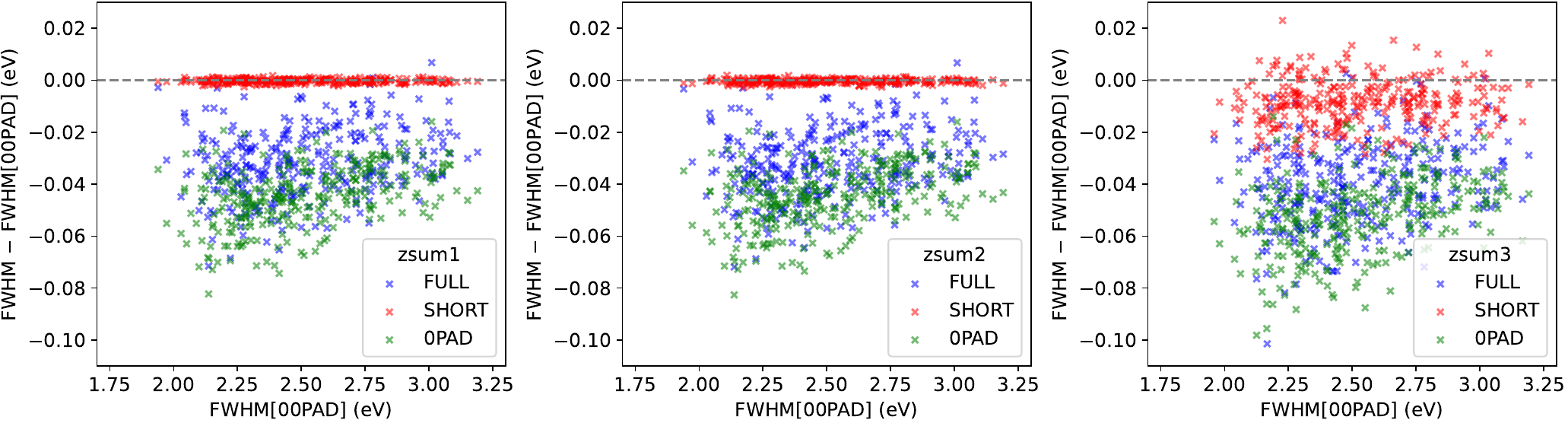}
    \caption{Differential resolution values  obtained with each filter in the analysis  (FULL, blue symbols; SHORT, red symbols and \textit{0-padding}, green symbols), and the zero-summed modified versions of \textit{0-padding}, labeled as 00PAD on the X-axis (\texttt{zsum1}: left, \texttt{zsum2}: center and \texttt{zsum3}: right). Please refer to \mbox{Eqs.~(\ref{eq:S1})--(\ref{eq:zsum3})} for details on the modifications.}
    \label{fig:00-pad}
\end{figure*}

These modified \textit{0-padding} filters were then applied to the \mbox{\texttt{xifusim}} simulated pulses in the Mn~K$\alpha$ complex (as described in  Sect.~\ref{sec:mnka_sims}). Subsequently, the reconstructed energies were gain scale calibrated and fitted using the histogram \textit{iSig} technique. Fig.~\ref{fig:00-pad} illustrates the comparison of the resolution values measured with these zero-summed filters and the FULL, SHORT and \textit{0-padding} filters previously analyzed. 

The results indicate that these modified filters lead to a degraded energy resolution in all the zero-sum designed scenarios making their performance comparable to that of the SHORT filter. As a result, any of these zero-summed \textit{0-padding} filters are unsuitable as a viable option. Consequently, fully harnessing the advantages of the \mbox{\textit{0-padding}} filter will depend on successfully correcting the baseline drift within the limits of the instrument resolution budget. 

In the cases where the baseline drift cannot be properly accounted for, \mbox{\textit{0-padding}} will cause a degradation of the energy resolution. For a simplistic approximation where the variation of the baseline level from pulse to pulse follows a normal distribution with a dispersion $\sigma_{\rm baseline}$, the additional degradation in energy resolution, that should be quadratically added to the expected FWHM of the calibrated energy, can be quantified (following the same reasoning as in Sect.~\ref{sec_propagation_of_random_undertainties}) as
\begin{equation}
{\rm FWHM}_{\rm baseline} = \sigma_{\rm baseline} \, g' \, \sum_{i=1}^{{\rm N}_{\rm cut}} \widetilde{o\!f}(t_i),
\end{equation}
where $g'$ is the first derivative of the gain scale correction evaluated at the energy of the considered photons.

\section{Other line complexes reconstruction}
\label{sec:other_complexes}

\begin{figure*}[htbp]
    \centering
    \includegraphics[width=0.9\textwidth]{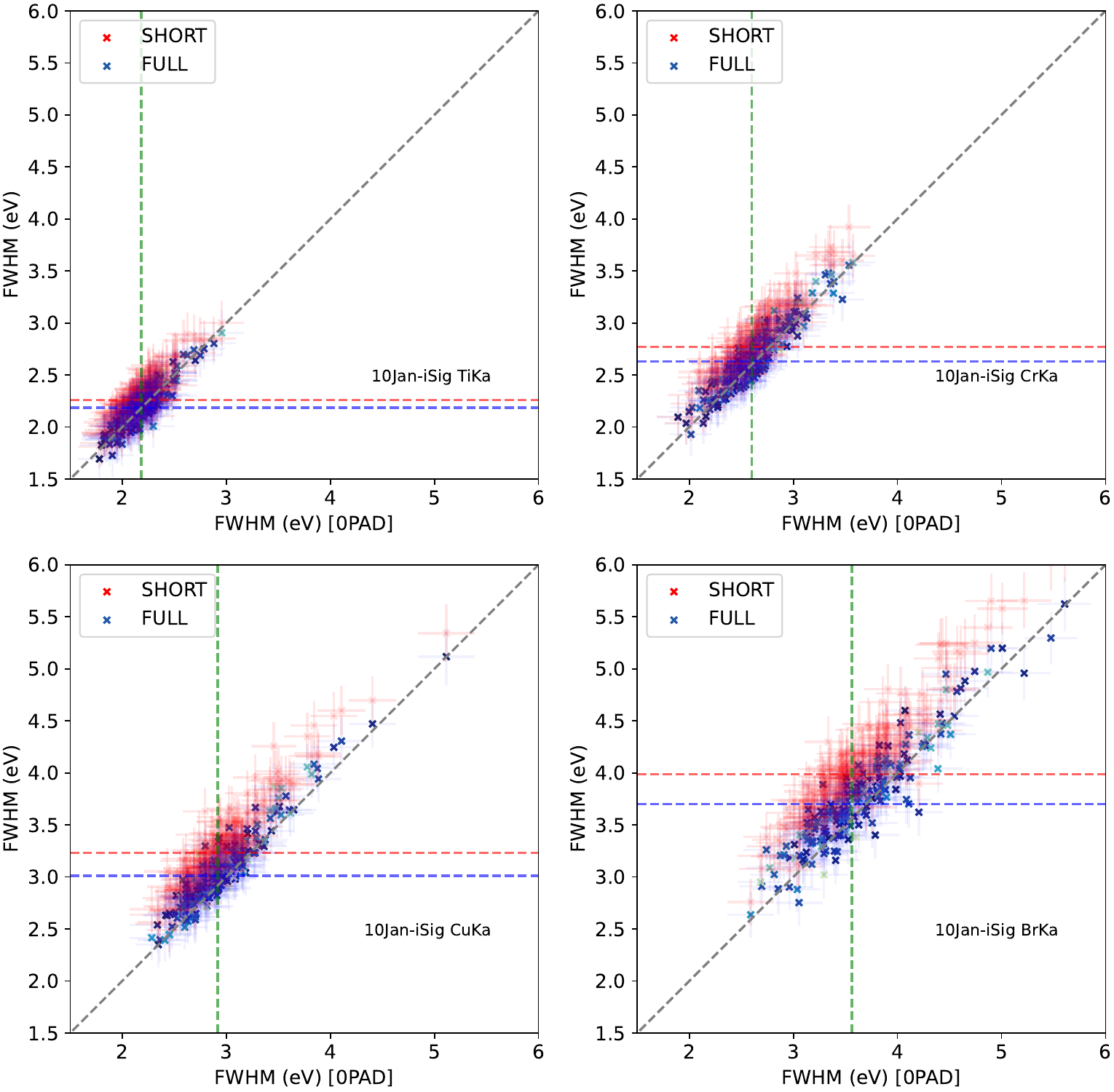}
    
    \caption{Comparison of resolution values for Ti~K$\alpha$ (top-left), Cr-K$\alpha$ (top-right), Cu~K$\alpha$ (bottom-left) and Br~K$\alpha$ (bottom-right) complexes obtained using FULL (blue) and SHORT filters (red) on the Y-axis plotted against those from the \textit{0-padding} reconstruction (X-axis) for dataset 10Jan2020. The resolution values were obtained using histogram fitting with \textit{iSig} weight. Dashed lines represent the mean energy resolution values (red for SHORT, blue for FULL and green for \textit{0-padding} filters).}
    \label{fig:CuCrTiBr_0pad}
\end{figure*}

To address any potential bias in favor of the \textit{0-padding} filter resulting from using the same photons for constructing the optimal filter and resolution analysis, we conducted an additional test. In this test, we reconstructed pulses with energies significantly different from the optimal filter's energy of~$5.9$~keV. By doing so, we aimed to evaluate whether the non-linearity of the detector response influenced the results obtained by the \mbox{\textit{0-padding}} filter.

Specifically, we reconstructed the Ti~K$\alpha$ ($4.9$~keV), Cr~K$\alpha$ ($5.4$~keV), Cu~K$\alpha$ ($8.0$~keV), and Br~K$\alpha$ ($11.9$~keV) complexes found in the 10Jan2020 dataset using the optimal filters constructed from the Mn~K$\alpha$ photons. 

The Lorentzian profiles for each complex are described in the following tables: Table~\ref{Tab:TiKa} (Ti K$\alpha$), \citep{TiKa2006}, Table~\ref{Tab:CrKa} (Cr~K$\alpha$), \citep{lab2016}, Table~\ref{Tab:CuKa} (Cu~K$\alpha$), \citep{lab2016}, and Table~\ref{Tab:BrKa} (Br K$\alpha$).

\begin{table*}
\caption{Lorentzian coefficients for \textbf{Ti K$\alpha$} complex.}
\label{Tab:TiKa}
\centering

\begin{tabular*}{0.6\textwidth}{@{\extracolsep\fill}cccc} 
\toprule
\noalign{\smallskip}
 $\rm K_{\alpha}$ & $\rm E_0[eV]$ & FWHM [eV] & Amplitude \\ 
\noalign{\smallskip}
\hline
\noalign{\smallskip}
 11  & 4510.918 & 1.37 & 1 \\
 12  & 4509.954 & 2.22 & 0.137 \\
 13  & 4507.763 & 3.75 & 0.052 \\
 15  & 4514.002 & 1.7 & 0.031 \\
 21  & 4504.910 & 1.88 & 0.446 \\
 22  & 4503.088 & 4.49 & 0.012 \\
\noalign{\smallskip}
\hline
\end{tabular*}
\footnotetext{Note: Given by \citet{TiKa2006}. Columns as in Table~\ref{Tab:MnKa}.}
\end{table*}

\begin{table*}
\caption{Lorentzian coefficients for \textbf{Cr~K$\alpha$} complex.}
\label{Tab:CrKa}
\centering
\begin{tabular*}{0.6\textwidth}{@{\extracolsep\fill}cccc} 
\toprule
\noalign{\smallskip}
$\rm K_{\alpha}$ & $\rm E_0[eV]$ & FWHM [eV] & Amplitude \\ 
\noalign{\smallskip}
\hline
\noalign{\smallskip}
 11  & 5414.874 & 1.457 & 0.822 \\
 12  & 5414.099 & 1.760 & 0.237 \\
 13  & 5412.745 & 3.138 & 0.085 \\
 14  & 5410.583 & 5.149 & 0.045 \\
 15  & 5418.304 & 1.988 & 0.015 \\
 21  & 5405.551 & 2.224 & 0.386 \\
 22  & 5403.986 & 4.740 & 0.036 \\
 \noalign{\smallskip}
 \hline
\end{tabular*}
\footnotetext{Note: Given by \citet{lab2016}. Columns as in Table~\ref{Tab:MnKa}.}
\end{table*}

\begin{table*}
\caption{Lorentzian coefficients for \textbf{Cu~K$\alpha$} complex.}
\label{Tab:CuKa}
\centering
\begin{tabular*}{0.6\textwidth}{@{\extracolsep\fill}cccc} 
\toprule
\noalign{\smallskip}
$\rm K_{\alpha}$ & $\rm E_0[eV]$ & FWHM [eV] & Amplitude \\ 
\noalign{\smallskip}
\hline
\noalign{\smallskip}
 11  & 8047.837 & 2.285 & 0.957 \\
 12  & 8045.367 & 3.358 & 0.090 \\
 21  & 8027.993 & 2.666 & 0.334 \\
 22  & 8026.504 & 3.571 & 0.111 \\
\noalign{\smallskip} 
 \hline
\end{tabular*}
\footnotetext{Note: Given by \citet{lab2016}. Columns as in Table~\ref{Tab:MnKa}.}
\end{table*}

\begin{table*}
\caption{Lorentzian coefficients for \textbf{Br~K$\alpha$} complex.}
\label{Tab:BrKa}
\centering
\begin{tabular*}{0.6\textwidth}{@{\extracolsep\fill}cccc} 
\toprule
\noalign{\smallskip} 
$\rm K_{\alpha}$ & $\rm E_0[eV]$ & FWHM [eV] & Amplitude \\ 
\noalign{\smallskip} 
\hline
\noalign{\smallskip} 

 1  & 11877.600 & 3.73 & 0.375 \\
 2  & 11924.200 & 3.6 & 1.0  \\
\noalign{\smallskip} 
\hline
\end{tabular*}
\footnotetext{Note: NASA GSFC private communication. Columns as in Table~\ref{Tab:MnKa}.}
\end{table*}

\begin{figure*}[htbp]
    \centering
    \includegraphics[width=0.99\textwidth]{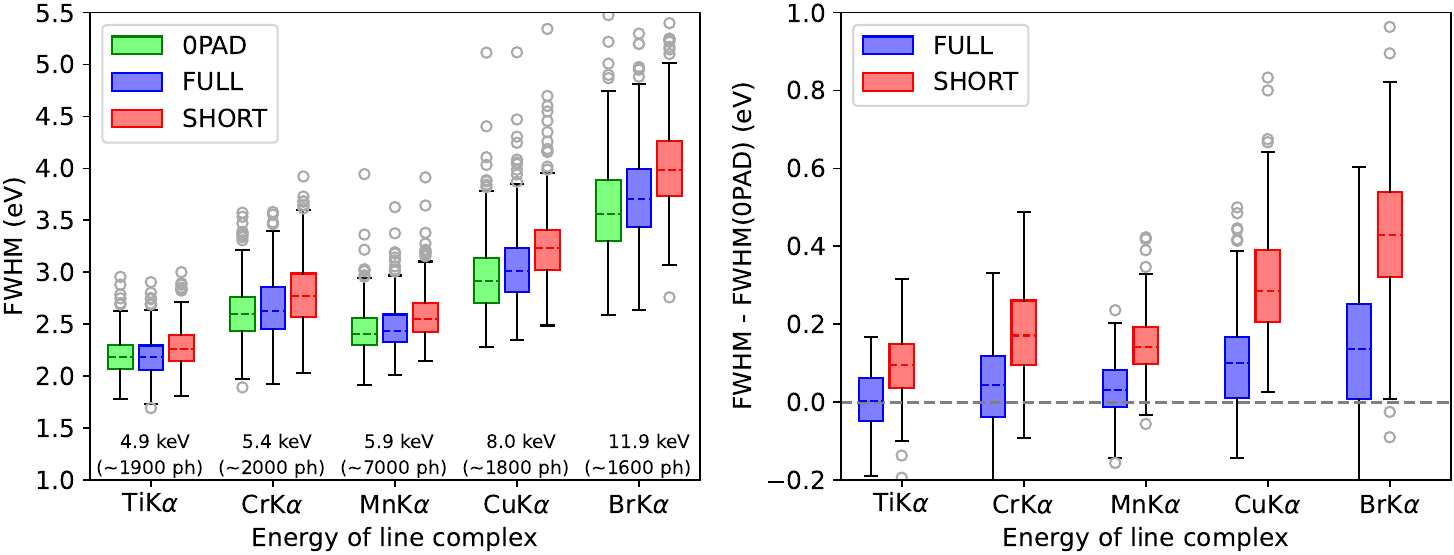}
    \caption{\textit{Boxplot} diagrams showing the improvement in resolution achieved by the \textit{0-padding} filter reconstruction for the analyzed line complexes of the 10Jan2020 data. Dashed central line in the boxes correspond to the median values, the coloured boxes cover the IQR (inter quantile range  ${\rm Q1}-{\rm Q3} \equiv 25\%-75\%$), error bars go from ${\rm Q1}-1.5\times {\rm IQR}$ to ${\rm Q3}+1.5\times {\rm IQR}$ and grey circles are the outliers. Left: Energy resolution values obtained with the \textit{0-padding} (green), FULL (blue) and SHORT (red) filters. Right: Differential energy resolution values of FULL and SHORT filter with respect to \textit{0-padding}}.
    \label{fig:barplot}
\end{figure*}

The reconstruction process was performed using the FULL, SHORT and \mbox{\textit{0-padding}} filters and the energies were gain calibrated as explained in Sect.~\ref{sec:datarecon}. Baseline and jitter corrections were conducted with \mbox{${\rm\texttt{xwidth}}\!=\!101$} (due to the poorer statistics compared to with the Mn~K$\alpha$ case) and \mbox{${\rm\texttt{smooth}}\!=\!11$}, respectively. The histograms were fitted using the \textit{iSig} weight and are displayed in Fig.~\ref{fig:CuCrTiBr_0pad}. Similar to the case of the Mn~K$\alpha$ complex, the \mbox{\textit{0-padding}} reconstruction appears to offer resolution values comparable to the FULL filter and better than the SHORT filter. However, it is important to note that the larger resolution values and dispersion obtained for Cu K$\alpha$ and Br K$\alpha$ lines may be attributed to the non-linearity of the detector, which results in degraded energy resolution at energies far from the optimal filter template.

In the case of these line complexes, where fewer photons are detected, the distribution of pulses along a varying baseline can have a larger effect on the reconstruction. For a few pixels with large variations in baseline during data acquisition, the initial automatic (no baseline-aware) gain scale calibration was not possible for the \mbox{\textit{0-padding}} reconstruction, as line peaks were double-peak shaped due to these different baseline values. Consequently, we removed these pixels from the analysis since they would require a more sophisticated, baseline-accounting gain calibration of the photon energy distribution. For the latest progress on demonstrating the gain scale correction over time, please refer to \citet{Steve2023}.

Figure~\ref{fig:barplot} illustrates the relationship between the gain in resolution and the energy of the complex. The improvement in energy resolution offered by \textit{0-padding} versus FULL and SHORT is statistically significant for all the line complexes, as revealed by the Wilcoxon signed-rank test for paired data, whose p-values are provided in Table~\ref{Tab:Wilcoxon_complexes}. The only  exception is the Ti~K$\alpha$ complex when comparing the \textit{0-padding} and FULL filters. The most substantial improvement occurs for the largest energy complex, Br~K$\alpha$. 

\begin{table*}
\caption{Statistical significance (p-values) of the comparison of the filters for the different line complexes.}
\label{Tab:Wilcoxon_complexes}
\centering
\begin{tabular*}{0.6\textwidth}{@{\extracolsep\fill}ccc} 
\toprule
\noalign{\smallskip} 
Line complex & \textit{0-padding} vs. FULL & \textit{0-padding} vs. SHORT \\ 
\noalign{\smallskip} 
\hline
\noalign{\smallskip} 

Ti~K$\alpha$ & $2.3\times 10^{-1}$ & $0.$ \\
Cr~K$\alpha$ & $1.9\times 10^{-8}$& $0.$  \\
Mn~K$\alpha$ & $2.9\times 10^{-10}$& $0.$ \\
Cu~K$\alpha$ & $3.3\times 10^{-23}$& $0.$ \\
Br~K$\alpha$ & $8.3\times 10^{-17}$& $0.$ \\
\noalign{\smallskip} 
\hline
\end{tabular*}
\footnotetext{Note: Wilcoxon signed-rank test for paired data where the alternative hypothesis is that \textit{0-padding} FWHM is lower than FULL (or SHORT) FWHM.}
\end{table*}

However, it is important to mention that the scatter also increases as we move to higher energies. Therefore, a more extensive investigation with higher statistics would be required to validate this trend. 

\section{Conclusions}
\label{sec:conclusions}

In this study, we take an in-depth look at a variation of the classical optimal filter algorithm to estimate the energy of photons detected by a Transition Edge Sensor device, such as the one to be onboard the \textit{Athena} mission. This approach, initially proposed by \citet{Cobo2020} and called \mbox{\textit{0-padding}}, involves truncating the classical optimal filter in the time domain. 

The results of our analysis, based on both simulated and laboratory data, show that truncating a long optimal filter  (\textit{0-padding}) yields better performance when compared to using a filter constructed from a shorter template but with the same final length as the truncated filter (SHORT). As the information loss resulting from setting the $f\!=\!0$~Hz bin to zero during the construction of the optimal filter diminishes as the filter length increases \citep[as indicated by][]{Doriese2009}, the 0-padding technique experiences less signal degradation as it begins its construction with a filter longer than the final intended size. As a result this approach limits the loss of resolution from shortened filters for high count rate cases.

What is even more relevant is that the resolution values obtained through our \mbox{\textit{0-padding}} approach are comparable, and in some cases slightly better than those achieved with a double-length optimal filter. Additionally our approach offers the advantage of reduced computational cost. As FULL filter and 0-padding only differ by the length the filter (the latter being half length) in terms of on-board computation, we can say that the energy estimation part of the event reconstruction would require half of the operations to be made. It would also require half of the on-board non-volatile memory as only half-length filters would be saved on-board. 
One would expect that computational time should scale at first order with the number of operations required, although this can very much be implementation dependent.

The enhanced performance of the \mbox{\textit{0-padding}} filter
scales with the noise level in the detector's signal, which directly impacts the uncertainty in energy determination. For filters constructed from the same template pulse, where the removal of the \mbox{$f\!=\!0$~Hz} bin has a similar effect (such as a long filter and its corresponding truncated \textit{0-padding} filter obtained by omitting the second half), longer data pulses result in greater uncertainties. 

Making use of the second half of the long filter does introduce a small, albeit non negligible, increase in the uncertainty of energy estimation. This is due to the inclusion of time samples that convey not useful information. This holds true if the cutoff for producing the \textit{0-padding} filter does not disregard relevant time samples where pulses register a signal statistically greater than the baseline. 

To determine the energy resolution of data measured by a TES detector in the laboratory, we compared two different fitting methods: histogram fitting with varying weights and Cumulative Distribution Function fitting. Our analysis confirmed that both methods yield similar results and demonstrated a better performance of the \mbox{\textit{0-padding}} filter.

However, when analyzing the Mn~K$\alpha$ complex, we observed that the CDF method consistently yielded lower resolution values than histogram fitting when applied to laboratory data, but not when applied to simulations. We investigated possible explanations, including the effect of the X-IFU detector's extended line spread function and instrumental background, but none of these factors accounted for the discrepancy. Instead, we found that each fitting procedure responded differently to any discrepancy between the data and the fitted model with the CDF method being more sensitive to photons missed in the tail of the distribution by the line selection process.

One intrinsic characteristic of the \mbox{\textit{0-padding}} filter is that it is no longer zero-summed due to the suppression of the last samples, which makes it more susceptible to baseline and energy scale drifts during data acquisition. To address this issue, one potential solution would be to ensure that the filter is zero-summed by subtracting the value of its sum. However, despite exploring several zero-summed \mbox{\textit{0-padding}} filters, all of them led to a degraded FWHM reaching only the performance of the SHORT filter, which has already the same length and is initially zero-summed. 

The non-linearity of the detector could potentially have a negative impact on the effectiveness of the \mbox{\textit{0-padding}} reconstruction method, especially for photons whose energies significantly differ from the energy of the pulses used to create the optimal filter. To investigate this further, we analyzed distant line complexes (Ti, Cr, Cu, and Br) and compared the energy resolution values obtained using the FULL, SHORT and \mbox{\textit{0-padding}} filters. The results reinforced our earlier findings with the Mn~K$\alpha$ complex, indicating the slightly better performance of the \mbox{\textit{0-padding}} filter. Furthermore, it appears that the degree of improvement in resolution tends to increase with the energy of the complex although it is important to note that a more comprehensive study with increased statistics would be necessary to fully confirm this observation.

It is important to emphasize that the effectiveness of the \mbox{\textit{0-padding}} filter depends on the ability to correct for baseline drift and jitter within the limits set by the X-IFU energy resolution budget. Under such circumstances, the \mbox{\textit{0-padding}} filter emerges as the optimal choice for energy reconstruction of X-ray photons detected by the TES detector. This finding holds great promise especially considering the shorter length of the \mbox{\textit{0-padding}} filter which requires fewer computational resources, a critical advantage for onboard processing. 
Furthermore, this filter would facilitate the analysis of sources with higher count rates at high resolution, limiting the loss of resolution provided by shortened filters.

While the comparative analysis of the energy reconstruction algorithms outlined in this study was initially inspired by the case of the X-IFU instrument, the findings obtained extend far beyond this initial context. They hold relevance for energy reconstruction across a spectrum of TES detectors, encompassing both present configurations and those anticipated in the future.

\bmhead{Acknowledgments}
We express our gratitude to the referee for their meticulous review of the paper and for offering comments and detailed suggestions that significantly contributed to enhancing the manuscript.
 M.T. Ceballos, N. Cardiel and B. Cobo acknowledge grant PID2021-122955OB-C41 funded by MCIN/AEI/ 10.13039/501100011033 and by “ERDF A way of making Europe”. Simulations in this work were made possible by utilizing \mbox{\texttt{xifusim}} \citep{xifusim2022} and {\sc xspec} \citep{XSPEC}. IPython and Jupyter notebooks have been extraordinarily beneficial in our research process \citep{PER-GRA:2007}. This research has also employed {\sc astropy}\footnote{\url{http://www.astropy.org}}, a community-developed core Python package designed specifically for Astronomy \citep{astropy:2013, astropy:2018}, along with the Python packages {\sc numpy} \citep{harris2020array}, {\sc scipy} \citep{2020SciPy-NMeth}, {\sc matplotlib} \citep{4160265}, {\sc mpmath} \citep{mpmath}, and {\sc lmfit} \citep{newville2021}. This research has made use of NASA's Astrophysics Data System Bibliographic Services.

\noindent\textbf{Funding} Open Access funding provided thanks to the CRUE-CSIC agreement with Springer Nature.

\section*{Declarations}

\textbf{Conflict of interest:} The authors declare that they have no conflict of interest.

\noindent\textbf{Availability of data and materials:} Data sets generated during the current study are available from the corresponding author upon request.

\noindent\textbf{Code availability:} Code is available upon request.

\noindent\textbf{Open Access} This article is licensed under a Creative Commons Attribution 4.0 International License, which permits use, sharing, adaptation, distribution and reproduction in any medium or format, as long as you give appropriate credit to the original author(s) and the source, provide a link to the Creative Commons licence, and indicate if changes were made. The images or other third party material in this article are included in the article’s Creative Commons licence, unless indicated otherwise in a credit line to the material. If material is not included in the article’s Creative Commons licence and your intended use is not permitted by statutory regulation or exceeds the permitted use, you will need to obtain permission directly from the copyright holder. To view a copy of this licence, visit http://creativecommons.org/licenses/by/4.0/.

\begin{appendices}

\section{Impact of potential systematic effects}
\label{app:systematics}

In addition to the higher sensitivity of the CDF method to the photons in the distribution's tails, other factors may also contribute to the disparities observed in the performance of fitting procedures between simulations (where no systematics were noticed) and real data (where CDF resolutions consistently turned out to be lower than histogram-fitting resolutions). 

Among these factors, the presence of an instrumental extended Line Spread Function (eLSF) and the background in the detector could be considered. In the following sub-sections, we will analyze how these factors impact the analysis of energy resolution.

\subsection{Extended Line Spread Function}
\label{app:eLSF}

In our previous analysis, we modeled the broadening in the energy distribution of reconstructed events caused by the TES detector and the reconstruction process, as a Gaussian function. The FWHM of this Gaussian function served as the figure of merit for determining the energy resolution. However, it is important to note that the energy dispersion in micro-calorimeters also involves low-level non-Gaussian broadening terms influenced by both the incident photon energy and the physics of the detector \citep{Eckart2019}.

Recent studies conducted by HITOMI/SXS and XRISM teams \citep{Eckart2018Hitomi,Eckart2019} have extensively characterized the X-IFU Line Spread Function (LSF) \citep{Eckart2018XIFU,Eckart2020}. This LSF has been incorporated into the Redistribution Matrix File (RMF) for science simulations. The RMF includes a \textit{core} LSF featuring a Gaussian main peak with a FWHM of $2.5$~eV from $0.05$~keV to $7$~keV, and a linearly increasing dispersion from $7$ to $12$~keV ranging from $2.5$ to $4.8$~eV. Additionally, the RMF incorporates an \textit{extended} LSF (eLSF) which exhibits an exponential shoulder caused by long-lived surface state excitations and the electron loss continuum \citep{RF8}.

To account for the eLSF, we conducted {\sc xspec} \citep{XSPEC} simulations using a unitary Ancillary Response File (ARF), and the RMF file specifically designed for the baseline configuration\footnote{{\tiny XIFU\_CC\_BASELINECONF\_2018\_10\_10\_EXTENDED\_LSF\_x33.rmf}} of X-IFU\footnote{\url{http://x-ifu.irap.omp.eu/resources/for-the-community}} at the time of writing. The adoption of a unitary ARF was aimed at avoiding additional effects of the effective area. As the Mn~K$\alpha$ line complex only spans a short interval of energies, any potential impact is negligible in any case. For comparison, we also utilized the \textit{core} LSF RMF. 

In our {\sc xspec} simulations, we employed the same Lorentzian line profiles as those used in previous simulations. Additionally, the exposure time was adjusted to obtain the required number of photons within the energy interval of the Mn~K$\alpha$ complex.

\begin{figure*}[htbp]
    \centering
    \includegraphics[width=0.8\textwidth]{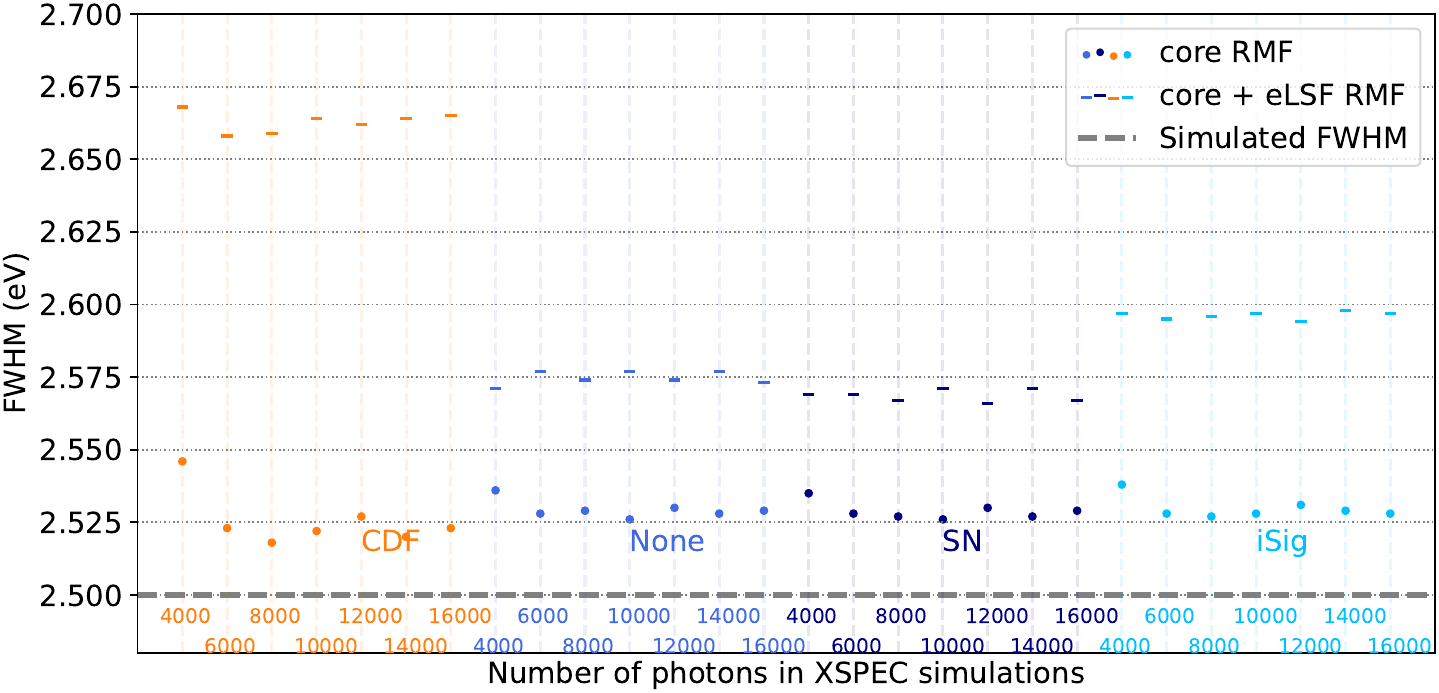}
    \caption{Energy resolution results obtained from {\sc xspec} simulations using both the \textit{core} (dots) and the eLSF RMFs (dashes), with photon counts ranging from~4000 to~16\,000. The simulations were fitted with histogram and CDF methods.}
    \label{fig:eLSF}
\end{figure*} 

As shown in Fig.~\ref{fig:eLSF}, the eLSF has a more noticeable impact on the CDF method compared to the histogram fitting, slightly increasing the resolution values and thus, contradicting the trend observed in real data.

\subsection{Instrumental Background}
\label{app:inst_bgd}

\begin{figure*}[htbp]
    \centering
    \includegraphics[width=0.8\textwidth]{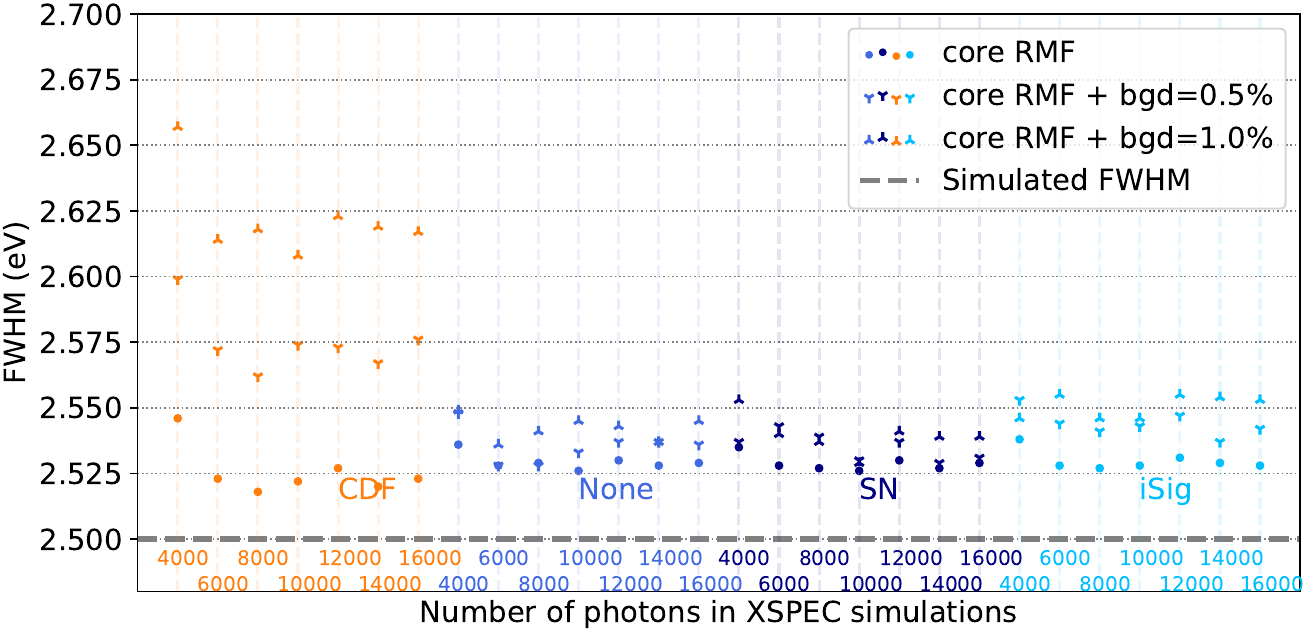}
    \caption{Energy resolution results obtained from {\sc xspec} simulations using the \textit{core} RMF with two levels of added instrumental background: $0.5$\% (down-triangles) and $1$\% (up-triangles). The fitting was performed using histogram and CDF techniques with photon counts ranging from~4000 to~16\,000.}
    \label{fig:eLSF_plus_background}
\end{figure*}

Next, we investigated the influence of instrumental particle background on the measurements. In laboratory data, this background is expected to be present alongside the Mn~K$\alpha$ photons. To simulate this, we randomly introduced a constant background in {\sc xspec}, which was uniformly distributed among the X-ray photons within the narrow energy range where the FWHM is measured. We explored two scenarios, adding $0.5$\% and $1$\% of the X-ray photons representing upper limits for the instrumental background. These levels serve as a generous  overestimation of the background level established for X-IFU \citep{bgd2018}. The results of these tests are depicted  in Fig.~\ref{fig:eLSF_plus_background}.

Similar to the eLSF case, the influence of the instrumental background on the FWHM derived from histogram fitting is negligible, with a slightly more noticeable impact observed in the CDF method. However, the discrepancy in resolution values obtained by the fitting techniques contradicts the behavior observed in the analysis of real data. As a result, we cannot consider these two effects to be relevant contributors to the relative discrepancy of the fitting methods.

\section{Uncertainty in \textit{iSig}-weighted FWHM measurements of Mn~K$\alpha$ complexes}
\label{app:std_fwhm}

This appendix explores in more detail the behavior in the dispersion of the measured FWHM values of the simulated Mn~K$\alpha$ line complexes when using the histogram fitting method with the \textit{iSig} weight. We focus on this specific fitting method because it yields the least dispersion and can be used as a reference to estimate the optimal uncertainty attainable when measuring the FWHM of this line complex. 

For that purpose, we simulated Mn~K$\alpha$ line complexes with varying number of photons $N_{\rm photon}$: 100, 200, 400, 1000, 2000, 4000, 10\,000, 20\,000, 40\,000, 100\,000, 200\,000, 400\,000 and 1\,000\,000, which are approximately equidistant on a logarithmic scale. These line complexes were generated using Gaussian ${\rm FHWM}_{\rm simulated}$ ranging from 1.20 to 5.00~eV, with a step size of 0.20~eV. For each value of $N_{\rm photon}$ and ${\rm FWHM}_{\rm simulated}$, 1000~simulations were conducted, and the line complex was fitted using the histogram method with the \textit{iSig} weight, providing ${\rm FWHM}_{\rm measured}$. 

\begin{figure}[htbp]
    \centering
    \includegraphics[width=0.45\textwidth]{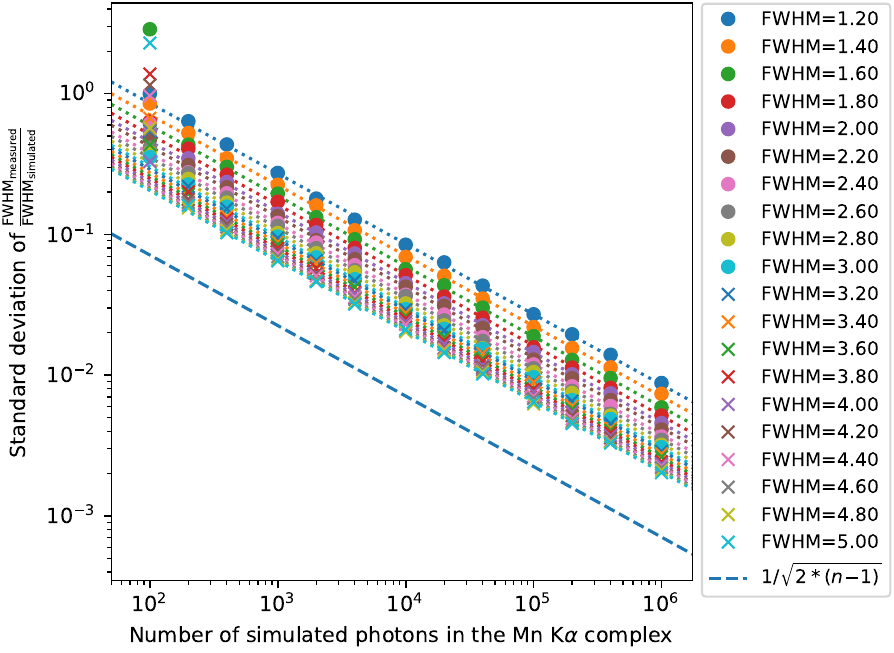}
    \caption{Standard deviation in 1000 estimates of the ratio ${\rm FWHM}_{\rm measured}/{\rm FWHM}_{\rm simulated}$ as a function of the number of photons $N_{\rm photon}$ in simulated Mn~K$\alpha$ line complexes. Different symbols and colors represent distinct Gaussian ${\rm FWHM}_{\rm simulated}$ values (in eV), as indicated in the legend. All measurements were conducted using the histogram fitting method with the \textit{iSig} weight. The blue dashed line corresponds to the simple case of a single Gaussian.}
    \label{fig:std_FWHM_MnKa}
\end{figure}

The results are presented in Fig.~\ref{fig:std_FWHM_MnKa}, which displays the standard deviation in the 1000 estimates of the ratio ${\rm FWHM}_{\rm measured}/{\rm FWHM}_{\rm simulated}$ as a function of $N_{\rm photon}$. Different symbols and colors represent distinct values of ${\rm FWHM}_{\rm simulated}$, as indicated in the legend.

Using a log-log display reveals that the measured standard deviation behaves in a similar manner to a simple single Gaussian (indicated by the dashed blue line), where the fractional uncertainty in the standard deviation of a normally distributed dataset constituted by $N$ data points approximates $1/\sqrt{2(N-1)}$ \citep[see e.g. Appendix~E in][]{093570275X}. For the Mn~K$\alpha$ line complex, the change in standard deviation follows the same trend with the number of photons $N_{\rm photon}$, but with a vertical displacement depending on the value of ${\rm FWHM}_{\rm simulated}$. In particular, for a fixed $N_{\rm photon}$, the represented standard deviation decreases as the simulated ${\rm FWHM}$ increases. This behaviour aligns with expectations since, in the limiting case where ${\rm FWHM}_{\rm simulated}$ continually grows, all the individual lines comprising the Mn~K$\alpha$ complex would merge into a single, very broad Gaussian line.

\begin{figure}[htbp]
    \centering
    \includegraphics[width=0.48\textwidth]{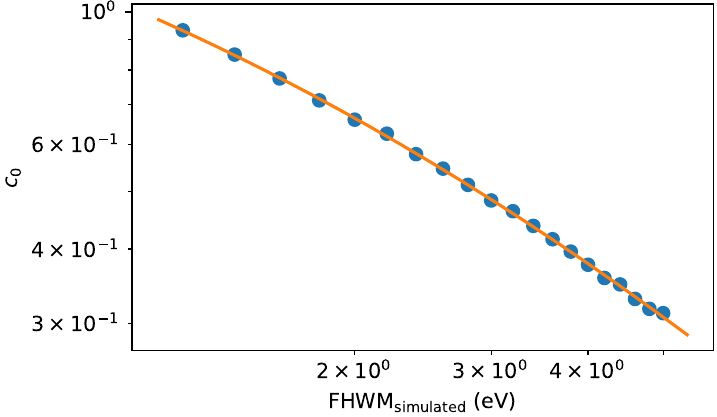}
    \caption{Variation of the $c_0$ coefficient introduced in Eq.~(\ref{eq:log_FWHM_Nphotons}), as a function of the ${\rm FWHM}_{\rm simulated}$ value employed in the simulated Mn~K$\alpha$ line complex. The orange line is the second-order polynomial fit given in Eq.~(\ref{eq:c0_FWHM}).}
    \label{fig:c0_FWHM_MnKa}
\end{figure}

The data points represented in Fig.~\ref{fig:std_FWHM_MnKa} for each fixed ${\rm FWHM}_{\rm simulated}$ have been fitted to straight lines with a slope $-1/2$ (depicted as dotted lines matching the color of the symbols) in this log-log representation. The resulting equation can be expressed as
\begin{equation}
    \label{eq:log_FWHM_Nphotons}
    \log_{10}\left[{\rm std}\left( 
    \frac{{\rm FWHM}_{\rm measured}}{{\rm FWHM}_{\rm simulated}} \right)\right] = c_0 -\frac{1}{2}\,\log_{10} 
    \left( N_{\rm photon} \right),
\end{equation}
where the $c_0$ coefficient depends on ${\rm FWHM}_{\rm simulated}$, as illustrated in Fig.~\ref{fig:c0_FWHM_MnKa}. In this log-log figure, the variation of $c_0$ fits well using the following second-order polynomial
\begin{equation}
\label{eq:c0_FWHM}
\begin{split}
    \log_{10}c_0 = & +0.01488989 \\
     & -0.55470163 \cdot {\rm FWHM}_{\rm simulated} \\
     & -0.28524671 \cdot {\rm FWHM}_{\rm simulated}^2\;.
\end{split}
\end{equation}

\begin{figure}[htbp]
    \centering
    \includegraphics[width=0.48\textwidth]{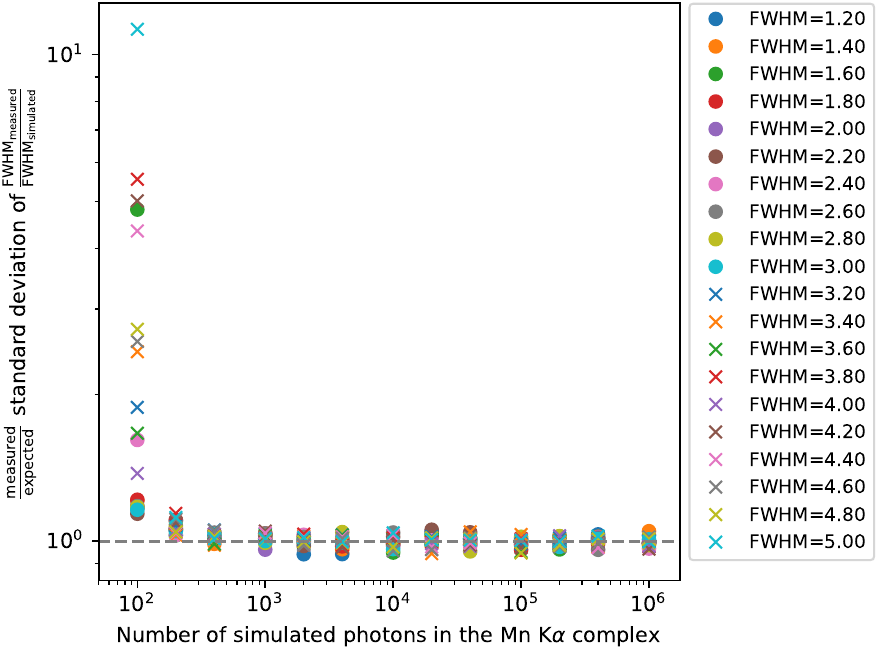}
    \caption{Ratio between the measured standard deviations plotted in Fig.~\ref{fig:std_FWHM_MnKa} and the expected values predicted by Eq.~(\ref{eq:log_FWHM_Nphotons}) and~(\ref{eq:c0_FWHM}). We are using the same symbols and colors as in Fig.~\ref{fig:std_FWHM_MnKa}.}
    \label{fig:residuals_std_FWHM_MnKa}
\end{figure}

The application of Eq.~(\ref{eq:log_FWHM_Nphotons}), using the $c_0$ value predicted by Eq.~(\ref{eq:c0_FWHM}), enables the determination of the expected standard deviation in  ${\rm FWHM}_{\rm measured}/{\rm FWHM}_{\rm simulated}$. Fig.~\ref{fig:residuals_std_FWHM_MnKa} depicts the ratio between the measured and expected standard deviation across the entire simulated dataset. It is evident from this figure that for $N_{\rm photon}>200$, the estimated uncertainties in the measured FWHM are well reproduced by the aforementioned equations.

In practical terms, when working with real Mn~K$\alpha$ data and considering that $\langle {\rm FWHM}_{\rm measured}/{\rm FWHM}_{\rm simulated}\rangle \simeq 1$ within the dispersion given by its standard deviation, the expected uncertainty in ${\rm FWHM}_{\rm measured}$ can be approximated as
\begin{equation}
    \Delta\, {\rm FWHM}_{\rm measured} \simeq 10^{c_0 - \frac{1}{2}\log_{10}N_{\rm photon}},
\end{equation}
where $c_0$ can be obtained from Eq.~(\ref{eq:c0_FWHM}) by substituting ${\rm FWHM}_{\rm simulated}$ with ${\rm FWHM}_{\rm measured}$.

\end{appendices}

\bibliography{zeropadding} 

\begin{thebibliography}{40}
\providecommand{\natexlab}[1]{#1}
\providecommand{\url}[1]{{#1}}
\providecommand{\urlprefix}{URL }
\providecommand{\doi}[1]{\url{https://doi.org/#1}}
\providecommand{\eprint}[2][]{\url{#2}}
 \bibcommenthead

\bibitem[{{Arnaud}(1996)}]{XSPEC}
{Arnaud} KA (1996) {XSPEC: The First Ten Years}. In: {Jacoby} GH, {Barnes} J
  (eds) Astronomical Data Analysis Software and Systems V, p~17

\bibitem[{{Astropy Collaboration} et~al(2013){Astropy Collaboration},
  {Robitaille}, {Tollerud}, {Greenfield}, {Droettboom}, {Bray}, {Aldcroft},
  {Davis}, {Ginsburg}, {Price-Whelan}, {Kerzendorf}, {Conley}, {Crighton},
  {Barbary}, {Muna}, {Ferguson}, {Grollier}, {Parikh}, {Nair}, {Unther},
  {Deil}, {Woillez}, {Conseil}, {Kramer}, {Turner}, {Singer}, {Fox}, {Weaver},
  {Zabalza}, {Edwards}, {Azalee Bostroem}, {Burke}, {Casey}, {Crawford},
  {Dencheva}, {Ely}, {Jenness}, {Labrie}, {Lim}, {Pierfederici}, {Pontzen},
  {Ptak}, {Refsdal}, {Servillat}, and {Streicher}}]{astropy:2013}
{Astropy Collaboration}, {Robitaille} TP, {Tollerud} EJ, et~al (2013) {Astropy:
  A community Python package for astronomy}. \aap 558:A33.
  \doi{10.1051/0004-6361/201322068},
  {\href{https://arxiv.org/abs/1307.6212}{{arXiv:1307.6212}}} {[astro-ph.IM]}

\bibitem[{{Astropy Collaboration} et~al(2018){Astropy Collaboration},
  {Price-Whelan}, {Sip{\H{o}}cz}, {G{\"u}nther}, {Lim}, {Crawford}, {Conseil},
  {Shupe}, {Craig}, {Dencheva}, {Ginsburg}, {Vand erPlas}, {Bradley},
  {P{\'e}rez-Su{\'a}rez}, {de Val-Borro}, {Aldcroft}, {Cruz}, {Robitaille},
  {Tollerud}, {Ardelean}, {Babej}, {Bach}, {Bachetti}, {Bakanov}, {Bamford},
  {Barentsen}, {Barmby}, {Baumbach}, {Berry}, {Biscani}, {Boquien}, {Bostroem},
  {Bouma}, {Brammer}, {Bray}, {Breytenbach}, {Buddelmeijer}, {Burke},
  {Calderone}, {Cano Rodr{\'\i}guez}, {Cara}, {Cardoso}, {Cheedella}, {Copin},
  {Corrales}, {Crichton}, {D'Avella}, {Deil}, {Depagne}, {Dietrich}, {Donath},
  {Droettboom}, {Earl}, {Erben}, {Fabbro}, {Ferreira}, {Finethy}, {Fox},
  {Garrison}, {Gibbons}, {Goldstein}, {Gommers}, {Greco}, {Greenfield},
  {Groener}, {Grollier}, {Hagen}, {Hirst}, {Homeier}, {Horton}, {Hosseinzadeh},
  {Hu}, {Hunkeler}, {Ivezi{\'c}}, {Jain}, {Jenness}, {Kanarek}, {Kendrew},
  {Kern}, {Kerzendorf}, {Khvalko}, {King}, {Kirkby}, {Kulkarni}, {Kumar},
  {Lee}, {Lenz}, {Littlefair}, {Ma}, {Macleod}, {Mastropietro}, {McCully},
  {Montagnac}, {Morris}, {Mueller}, {Mumford}, {Muna}, {Murphy}, {Nelson},
  {Nguyen}, {Ninan}, {N{\"o}the}, {Ogaz}, {Oh}, {Parejko}, {Parley}, {Pascual},
  {Patil}, {Patil}, {Plunkett}, {Prochaska}, {Rastogi}, {Reddy Janga},
  {Sabater}, {Sakurikar}, {Seifert}, {Sherbert}, {Sherwood-Taylor}, {Shih},
  {Sick}, {Silbiger}, {Singanamalla}, {Singer}, {Sladen}, {Sooley},
  {Sornarajah}, {Streicher}, {Teuben}, {Thomas}, {Tremblay}, {Turner},
  {Terr{\'o}n}, {van Kerkwijk}, {de la Vega}, {Watkins}, {Weaver}, {Whitmore},
  {Woillez}, {Zabalza}, and {Astropy Contributors}}]{astropy:2018}
{Astropy Collaboration}, {Price-Whelan} AM, {Sip{\H{o}}cz} BM, et~al (2018)
  {The Astropy Project: Building an Open-science Project and Status of the v2.0
  Core Package}. \aj 156(3):123. \doi{10.3847/1538-3881/aabc4f},
  {\href{https://arxiv.org/abs/1801.02634}{{arXiv:1801.02634}}} {[astro-ph.IM]}

\bibitem[{{Barret} et~al(2023){Barret}, {Albouys}, {Herder}, {Piro}, {Cappi},
  {Huovelin}, {Kelley}, {Mas-Hesse}, {Paltani}, {Rauw}, {Rozanska}, {Svoboda},
  {Wilms}, {Yamasaki}, {Audard}, {Bandler}, {Barbera}, {Barcons}, {Bozzo},
  {Ceballos}, {Charles}, {Costantini}, {Dauser}, {Decourchelle}, {Duband},
  {Duval}, {Fiore}, {Gatti}, {Goldwurm}, {Hartog}, {Jackson}, {Jonker},
  {Kilbourne}, {Korpela}, {Macculi}, {Mendez}, {Mitsuda}, {Molendi}, {Pajot},
  {Pointecouteau}, {Porter}, {Pratt}, {Pr{\^e}le}, {Ravera}, {Sato}, {Schaye},
  {Shinozaki}, {Skup}, {Soucek}, {Thibert}, {Vink}, {Webb}, {Chaoul}, {Raulin},
  {Simionescu}, {Torrejon}, {Acero}, {Branduardi-Raymont}, {Ettori},
  {Finoguenov}, {Grosso}, {Kaastra}, {Mazzotta}, {Miller}, {Miniutti},
  {Nicastro}, {Sciortino}, {Yamaguchi}, {Beaumont}, {Cucchetti}, {D'Andrea},
  {Eckart}, {Ferrando}, {Kammoun}, {Lotti}, {Mesnager}, {Natalucci}, {Peille},
  {de Plaa}, {Ardellier}, {Argan}, {Bellouard}, {Carron}, {Cavazzuti},
  {Fiorini}, {Khosropanah}, {Martin}, {Perry}, {Pinsard}, {Pradines}, {Rigano},
  {Roelfsema}, {Schwander}, {Torrioli}, {Ullom}, {Vera}, {Villegas},
  {Zuchniak}, {Brachet}, {Cicero}, {Doriese}, {Durkin}, {Fioretti}, {Geoffray},
  {Jacques}, {Kirsch}, {Smith}, {Adams}, {Gloaguen}, {Hoogeveen}, {van der
  Hulst}, {Kiviranta}, {van der Kuur}, {Ledot}, {van Leeuwen}, {van Loon},
  {Lyautey}, {Parot}, {Sakai}, {van Weers}, {Abdoelkariem}, {Adam}, {Adami},
  {Aicardi}, {Akamatsu}, {Alonso}, {Amato}, {Andr{\'e}}, {Angelinelli},
  {Anon-Cancela}, {Anvar}, {Atienza}, {Attard}, {Auricchio}, {Balado},
  {Bancel}, {Barusso}, {Bascu{\~n}an}, {Bernard}, {Berrocal}, {Blin}, {Bonino},
  {Bonnet}, {Bonny}, {Boorman}, {Boreux}, {Bounab}, {Boutelier}, {Boyce},
  {Brienza}, {Bruijn}, {Bulgarelli}, {Calarco}, {Callanan}, {Campello},
  {Camus}, {Canourgues}, {Capobianco}, {Cardiel}, {Castellani}, {Cheatom},
  {Chervenak}, {Chiarello}, {Clerc}, {Clerc}, {Cobo}, {Coeur-Joly}, {Coleiro},
  {Colonges}, {Corcione}, {Coriat}, {Coynel}, {Cuttaia}, {D'Ai}, {D'anca},
  {Dadina}, {Daniel}, {Dauner}, {DeNigris}, {Dercksen}, {DiPirro}, {Doumayrou},
  {Dubbeldam}, {Dupieux}, {Dupourqu{\'e}}, {Durand}, {Eckert}, {Eiriz},
  {Ercolani}, {Etcheverry}, {Finkbeiner}, {Fiocchi}, {Fossecave}, {Franssen},
  {Frericks}, {Gabici}, {Gant}, {Gao}, {Gastaldello}, {Genolet}, {Ghizzardi},
  {Gil}, {Giovannini}, {Godet}, {Gomez-Elvira}, {Gonzalez}, {Gonzalez},
  {Gottardi}, {Granat}, {Gros}, {Guignard}, {Hieltjes}, {Hurtado}, {Irwin},
  {Jacquey}, {Janiuk}, {Jaubert}, {Jim{\'e}nez}, {Jolly}, {Jourdan}, {Julien},
  {Kedziora}, {Korb}, {Kreykenbohm}, {K{\"o}nig}, {Langer}, {Laudet},
  {Laurent}, {Laurenza}, {Lesrel}, {Ligori}, {Lorenz}, {Luminari}, {Maffei},
  {Maisonnave}, {Marelli}, {Massonet}, {Maussang}, {Melchor}, {Le Mer},
  {Millan}, {Millerioux}, {Mineo}, {Minervini}, {Molin}, {Monestes},
  {Montinaro}, {Mot}, {Murat}, {Nagayoshi}, {Naz{\'e}}, {Nogu{\`e}s}, {Pailot},
  {Panessa}, {Parodi}, {Petit}, {Piconcelli}, {Pinto}, {Plaza}, {Plaza},
  {Poyatos}, {Prouv{\'e}}, {Ptak}, {Puccetti}, {Puccio}, {Ramon}, {Reina},
  {Rioland}, {Rodriguez}, {Roig}, {Rollet}, {Roncarelli}, {Roudil}, {Rudnicki},
  {Sanisidro}, {Sciortino}, {Silva}, {Sordet}, {Soto-Aguilar}, {Spizzi},
  {Surace}, {Fern{\'a}ndez S{\'a}nchez}, {Taralli}, {Terrasa}, {Terrier},
  {Todaro}, {Ubertini}, {Uslenghi}, {de Vaate}, {Vaccaro}, {Varisco},
  {Varni{\`e}re}, {Vibert}, {Vidriales}, {Villa}, {Vodopivec}, {Volpe}, {de
  Vries}, {Wakeham}, {Walmsley}, {Wise}, {de Wit}, and
  {Wo{\'z}niak}}]{barret2023}
{Barret} D, {Albouys} V, {Herder} JWd, et~al (2023) {The Athena X-ray Integral
  Field Unit: a consolidated design for the system requirement review of the
  preliminary definition phase}. Experimental Astronomy 55(2):373--426.
  \doi{10.1007/s10686-022-09880-7},
  {\href{https://arxiv.org/abs/2208.14562}{{arXiv:2208.14562}}} {[astro-ph.IM]}

\bibitem[{Boyce et~al(1999)Boyce, Audley, Baker, Dumonthier, Fujimoto,
  Gendreau, Ishisaki, Kelley, Stahle, Szymkowiak, and Winkert}]{Boyce1999}
Boyce KR, Audley MD, Baker RG, et~al (1999) {Design and performance of the
  ASTRO-E/XRS signal processing system}. In: Siegmund OHW, Flanagan KA (eds)
  EUV, X-Ray, and Gamma-Ray Instrumentation for Astronomy X, International
  Society for Optics and Photonics, vol 3765. SPIE, pp 741 -- 750,
  \doi{10.1117/12.366557}

\bibitem[{Cardiel et~al(2023)Cardiel, Ceballos, and Cobo}]{CardielCDF}
Cardiel N, Ceballos MT, Cobo B (2023) {Using cumulative distribution functions
  to characterize X-ray line complexes}. In: Manteiga M, Bellot L, Benavidez P,
  et~al (eds) Highlights of Spanish Astrophysics XI, Proceedings of the XV
  Scientific Meeting of the Spanish Astronomical Society, p~NA

\bibitem[{{Ceballos} et~al(2019{\natexlab{a}}){Ceballos}, {Cobo}, and
  {Peille}}]{Ceballos2019}
{Ceballos} MT, {Cobo} B, {Peille} P (2019{\natexlab{a}}) {Jitter and Readout
  Sampling Frequency Impact on the Athena/X-IFU Performance}. In: {Teuben} PJ,
  {Pound} MW, {Thomas} BA, et~al (eds) Astronomical Data Analysis Software and
  Systems XXVII, p 547

\bibitem[{{Ceballos} et~al(2019{\natexlab{b}}){Ceballos}, {Cobo}, {Peille},
  {Wilms}, {Brand}, {Dauser}, {Bandler}, and {Smith}}]{Sirena2019}
{Ceballos} MT, {Cobo} B, {Peille} P, et~al (2019{\natexlab{b}}) {SIRENA:
  Software for Athena X-IFU Event Reconstruction}. In: {Molinaro} M,
  {Shortridge} K, {Pasian} F (eds) Astronomical Data Analysis Software and
  Systems XXVI, p 663

\bibitem[{{Ceballos} et~al(2022){Ceballos}, {Cardiel}, {Cobo}, {Peille},
  {Smith}, {Witthoeft}, and {Durkin}}]{Ceballos2022}
{Ceballos} MT, {Cardiel} N, {Cobo} B, et~al (2022) {When less is more: the
  truncation of the optimal filter to reconstruct events in X-IFU/Athena-like
  TES detectors}. In: {den Herder} JWA, {Nikzad} S, {Nakazawa} K (eds) Society
  of Photo-Optical Instrumentation Engineers (SPIE) Conference Series, p
  1218145, \doi{10.1117/12.2631165}

\bibitem[{{Ceballos} et~al(2023){Ceballos}, {Cobo}, {Peille}, {Wilms}, {Brand},
  {Dauser}, {Bandler}, and {Smith}}]{sirenaASCL}
{Ceballos} MT, {Cobo} B, {Peille} P, et~al (2023) {SIRENA: Energy
  reconstruction of X-ray photons for Athena X-IFU}. Astrophysics Source Code
  Library, record ascl:2307.013, \eprint{2307.013}

\bibitem[{Chantler et~al(2006)Chantler, Kinnane, Su, and Kimpton}]{TiKa2006}
Chantler CT, Kinnane MN, Su CH, et~al (2006) Characterization of
  $k\ensuremath{\alpha}$ spectral profiles for vanadium, component
  redetermination for scandium, titanium, chromium, and manganese, and
  development of satellite structure for $z=21$ to $z=25$. Phys Rev A
  73:012508. \doi{10.1103/PhysRevA.73.012508}

\bibitem[{Cobo et~al(2020)Cobo, Cardiel, Ceballos, and Peille}]{Cobo2020}
Cobo B, Cardiel N, Ceballos MT, et~al (2020) {Pulse processing in TES
  detectors: comparative of different short filter methods based on optimal
  filtering: case study for Athena X-IFU}. In: den Herder JWA, Nikzad S,
  Nakazawa K (eds) Space Telescopes and Instrumentation 2020: Ultraviolet to
  Gamma Ray, International Society for Optics and Photonics, vol 11444. SPIE,
  pp 1333 -- 1344, \doi{10.1117/12.2562733}

\bibitem[{Consortium(2018)}]{RF8}
Consortium XI (2018) {X-IFU response matrices (2018)}

\bibitem[{{Cucchetti} et~al(2018){Cucchetti}, {Pointecouteau}, {Barret},
  {Lotti}, {Macculi}, {Molendi}, {Pajot}, {Peille}, {Piro}, and
  {Pratt}}]{bgd2018}
{Cucchetti} E, {Pointecouteau} E, {Barret} D, et~al (2018) {Reproducibility and
  monitoring of the instrumental particle background for the x-ray integral
  field unit}. In: {den Herder} JWA, {Nikzad} S, {Nakazawa} K (eds) Space
  Telescopes and Instrumentation 2018: Ultraviolet to Gamma Ray, p 106994N,
  \doi{10.1117/12.2312179}, \eprint{1807.01573}

\bibitem[{{Doriese} et~al(2009){Doriese}, {Adams}, {Hilton}, {Irwin},
  {Kilbourne}, {Schima}, and {Ullom}}]{Doriese2009}
{Doriese} WB, {Adams} JS, {Hilton} GC, et~al (2009) {Optimal filtering, record
  length, and count rate in transition-edge-sensor microcalorimeters}. In:
  {Young} B, {Cabrera} B, {Miller} A (eds) The Thirteenth International
  Workshop on Low Temperature Detectors - LTD13, pp 450--453,
  \doi{10.1063/1.3292375}

\bibitem[{Durkin et~al(2019)Durkin, Adams, Bandler, Chervenak, Chaudhuri,
  Dawson, Denison, Doriese, Duff, Finkbeiner, FitzGerald, Fowler, Gard, Hilton,
  Irwin, Joe, Kelley, Kilbourne, Miniussi, Morgan, O'Neil, Pappas, Porter,
  Reintsema, Rudman, Sakai, Smith, Stevens, Swetz, Szypryt, Ullom, Vale,
  Wakeham, Weber, and Young}]{Durkin2019}
Durkin M, Adams JS, Bandler SR, et~al (2019) Demonstration of athena x-ifu
  compatible 40-row time-division-multiplexed readout. IEEE Transactions on
  Applied Superconductivity 29(5):1--5. \doi{10.1109/TASC.2019.2904472}

\bibitem[{Eckart(2018)}]{Eckart2018XIFU}
Eckart M (2018) {X-IFU LSF Equation based on Astro-H LSF Measurements}.
  \doi{XCaT–TN–005, v1.4}

\bibitem[{Eckart(2020)}]{Eckart2020}
Eckart M (2020) {X-IFU LSF parameter estimates for science simulations based on
  Astro-H SXS measurements and preliminary X-IFU measurements}.
  \doi{XCaT–TN–NNN, v0.1}

\bibitem[{Eckart et~al(2016)Eckart, Kilbourne, and Porter}]{lab2016}
Eckart M, Kilbourne C, Porter F (2016) {Instrument Calibration Report –
  Natural Line shapes of SXS onboard calibration sources}.
  \urlprefix\url{https://heasarc.gsfc.nasa.gov/docs/hitomi/calib/caldb_doc/asth_sxs_caldb_linefit_v20161223.pdf}

\bibitem[{Eckart et~al(2018)Eckart, Adams, Boyce, Brown, Chiao, Fujimoto, Haas,
  den Herder, Hoshino, Ishisaki, Kilbourne, Kitamoto, Leutenegger, McCammon,
  Mitsuda, Porter, Sato, Sawada, Seta, Sneiderman, Szymkowiak, Takei, Tashiro,
  Tsujimoto, de~Vries, Watanabe, Yamada, and Yamasaki}]{Eckart2018Hitomi}
Eckart ME, Adams JS, Boyce KR, et~al (2018) {Ground calibration of the Astro-H
  (Hitomi) soft x-ray spectrometer}. Journal of Astronomical Telescopes,
  Instruments, and Systems 4(2):021406. \doi{10.1117/1.JATIS.4.2.021406}

\bibitem[{Eckart et~al(2019)Eckart, Adams, Bandler, Beaumont, Chervenak,
  Datesman, Finkbeiner, Hummatov, Kelley, Kilbourne, Leutenegger, Miniussi,
  Moseley, Porter, Sadleir, Sakai, Smith, Wakeham, and Wassell}]{Eckart2019}
Eckart ME, Adams JS, Bandler SR, et~al (2019) Extended line spread function of
  tes microcalorimeters with au/bi absorbers. IEEE Transactions on Applied
  Superconductivity 29(5):1--5. \doi{10.1109/TASC.2019.2903420}

\bibitem[{Fowler et~al(2015)Fowler, Alpert, Doriese, Young, O'Neil, Ullom, and
  Swetz}]{Fowler2016}
Fowler J, Alpert B, Doriese W, et~al (2015) The practice of pulse processing.
  Journal of Low Temperature Physics
  \urlprefix\url{https://tsapps.nist.gov/publication/get_pdf.cfm?pub_id=919341}

\bibitem[{Fowler(2014)}]{Fowler2014}
Fowler JW (2014) {Maximum-Likelihood Fits to Histograms for Improved Parameter
  Estimation}. {Journal of Low Temperature Physics} 176:414.
  \doi{10.1007/s10909-014-1098-4}

\bibitem[{Harris et~al(2020)Harris, Millman, van~der Walt, Gommers, Virtanen,
  Cournapeau, Wieser, Taylor, Berg, Smith, Kern, Picus, Hoyer, van Kerkwijk,
  Brett, Haldane, del R{'{\i}}o, Wiebe, Peterson, G{'{e}}rard-Marchant,
  Sheppard, Reddy, Weckesser, Abbasi, Gohlke, and Oliphant}]{harris2020array}
Harris CR, Millman KJ, van~der Walt SJ, et~al (2020) Array programming with
  {NumPy}. Nature 585(7825):357--362. \doi{10.1038/s41586-020-2649-2}

\bibitem[{{Hunter}(2007)}]{4160265}
{Hunter} JD (2007) Matplotlib: A 2d graphics environment. Computing in Science
  Engineering 9(3):90--95

\bibitem[{Kirsch et~al(2022)Kirsch, Lorenz, Peille, Dauser, Ceballos, Cobo,
  Merino-Alonso, Cucchetti, Smith, Gottardi, Hartog, Miniussi, Durkin, Prêle,
  and Wilms}]{xifusim2022}
Kirsch C, Lorenz M, Peille P, et~al (2022) The athena x-ifu instrument
  simulator xifusim. Journal of Low Temperature Physics 209:988--997.
  \doi{10.1007/s10909-022-02700-4}

\bibitem[{Kumar(2020)}]{KUMAR2020101986}
Kumar R (2020) The generalized modified bessel function and its connection with
  voigt line profile and humbert functions. Advances in Applied Mathematics
  114:101986. \doi{10.1016/j.aam.2019.101986},
  \urlprefix\url{https://www.sciencedirect.com/science/article/pii/S0196885819301708}

\bibitem[{{Nandra} et~al(2013){Nandra}, {Barret}, {Barcons}, {Fabian}, {den
  Herder}, {Piro}, {Watson}, {Adami}, {Aird}, {Afonso}, {Alexander},
  {Argiroffi}, {Amati}, {Arnaud}, {Atteia}, {Audard}, {Badenes}, {Ballet},
  {Ballo}, {Bamba}, {Bhardwaj}, {Stefano Battistelli}, {Becker}, {De Becker},
  {Behar}, {Bianchi}, {Biffi}, {B{\^\i}rzan}, {Bocchino}, {Bogdanov}, {Boirin},
  {Boller}, {Borgani}, {Borm}, {Bouch{\'e}}, {Bourdin}, {Bower}, {Braito},
  {Branchini}, {Branduardi-Raymont}, {Bregman}, {Brenneman}, {Brightman},
  {Br{\"u}ggen}, {Buchner}, {Bulbul}, {Brusa}, {Bursa}, {Caccianiga},
  {Cackett}, {Campana}, {Cappelluti}, {Cappi}, {Carrera}, {Ceballos},
  {Christensen}, {Chu}, {Churazov}, {Clerc}, {Corbel}, {Corral}, {Comastri},
  {Costantini}, {Croston}, {Dadina}, {D'Ai}, {Decourchelle}, {Della Ceca},
  {Dennerl}, {Dolag}, {Done}, {Dovciak}, {Drake}, {Eckert}, {Edge}, {Ettori},
  {Ezoe}, {Feigelson}, {Fender}, {Feruglio}, {Finoguenov}, {Fiore}, {Galeazzi},
  {Gallagher}, {Gandhi}, {Gaspari}, {Gastaldello}, {Georgakakis},
  {Georgantopoulos}, {Gilfanov}, {Gitti}, {Gladstone}, {Goosmann}, {Gosset},
  {Grosso}, {Guedel}, {Guerrero}, {Haberl}, {Hardcastle}, {Heinz}, {Alonso
  Herrero}, {Herv{\'e}}, {Holmstrom}, {Iwasawa}, {Jonker}, {Kaastra}, {Kara},
  {Karas}, {Kastner}, {King}, {Kosenko}, {Koutroumpa}, {Kraft}, {Kreykenbohm},
  {Lallement}, {Lanzuisi}, {Lee}, {Lemoine-Goumard}, {Lobban}, {Lodato},
  {Lovisari}, {Lotti}, {McCharthy}, {McNamara}, {Maggio}, {Maiolino}, {De
  Marco}, {de Martino}, {Mateos}, {Matt}, {Maughan}, {Mazzotta}, {Mendez},
  {Merloni}, {Micela}, {Miceli}, {Mignani}, {Miller}, {Miniutti}, {Molendi},
  {Montez}, {Moretti}, {Motch}, {Naz{\'e}}, {Nevalainen}, {Nicastro}, {Nulsen},
  {Ohashi}, {O'Brien}, {Osborne}, {Oskinova}, {Pacaud}, {Paerels}, {Page},
  {Papadakis}, {Pareschi}, {Petre}, {Petrucci}, {Piconcelli}, {Pillitteri},
  {Pinto}, {de Plaa}, {Pointecouteau}, {Ponman}, {Ponti}, {Porquet}, {Pounds},
  {Pratt}, {Predehl}, {Proga}, {Psaltis}, {Rafferty}, {Ramos-Ceja}, {Ranalli},
  {Rasia}, {Rau}, {Rauw}, {Rea}, {Read}, {Reeves}, {Reiprich}, {Renaud},
  {Reynolds}, {Risaliti}, {Rodriguez}, {Rodriguez Hidalgo}, {Roncarelli},
  {Rosario}, {Rossetti}, {Rozanska}, {Rovilos}, {Salvaterra}, {Salvato}, {Di
  Salvo}, {Sanders}, {Sanz-Forcada}, {Schawinski}, {Schaye}, {Schwope},
  {Sciortino}, {Severgnini}, {Shankar}, {Sijacki}, {Sim}, {Schmid}, {Smith},
  {Steiner}, {Stelzer}, {Stewart}, {Strohmayer}, {Str{\"u}der}, {Sun}, {Takei},
  {Tatischeff}, {Tiengo}, {Tombesi}, {Trinchieri}, {Tsuru}, {Ud-Doula},
  {Ursino}, {Valencic}, {Vanzella}, {Vaughan}, {Vignali}, {Vink}, {Vito},
  {Volonteri}, {Wang}, {Webb}, {Willingale}, {Wilms}, {Wise}, {Worrall},
  {Young}, {Zampieri}, {In't Zand}, {Zane}, {Zezas}, {Zhang}, and
  {Zhuravleva}}]{Nandra2013}
{Nandra} K, {Barret} D, {Barcons} X, et~al (2013) {The Hot and Energetic
  Universe: A White Paper presenting the science theme motivating the Athena+
  mission}. arXiv e-prints
  {\href{https://arxiv.org/abs/1306.2307}{{arXiv:1306.2307}}} {[astro-ph.HE]}

\bibitem[{Newville et~al(2021)Newville, Otten, Nelson, Ingargiola, Stensitzki,
  Allan, Fox, Carter, Michał, Osborn, Pustakhod, lneuhaus, Weigand, Glenn,
  Deil, Mark, Hansen, Pasquevich, Foks, Zobrist, Frost, Beelen, Stuermer,
  azelcer, Hannum, Polloreno, Nielsen, Caldwell, Almarza, and
  Persaud}]{newville2021}
Newville M, Otten R, Nelson A, et~al (2021) lmfit/lmfit-py: 1.0.3.
  \doi{10.5281/zenodo.5570790}

\bibitem[{P\'erez and Granger(2007)}]{PER-GRA:2007}
P\'erez F, Granger BE (2007) {IP}ython: a system for interactive scientific
  computing. Computing in Science and Engineering 9(3):21--29.
  \doi{10.1109/MCSE.2007.53}, \urlprefix\url{https://ipython.org}

\bibitem[{Ravera et~al(2014)Ravera, Cara, Ceballos, Barcons, Barret,
  Clédassou, Clénet, Cobo, Doumayrou, den Hartog, van Leeuwen, van Loon,
  Mas-Hesse, Pigot, and Pointecouteau}]{Ravera2014}
Ravera L, Cara C, Ceballos MT, et~al (2014) {The DRE: the digital readout
  electronics for ATHENA X-IFU}. In: Takahashi T, den Herder JWA, Bautz M (eds)
  Space Telescopes and Instrumentation 2014: Ultraviolet to Gamma Ray,
  International Society for Optics and Photonics, vol 9144. SPIE, pp 1734 --
  1741, \doi{10.1117/12.2055750}

\bibitem[{{Ravera} et~al(2018){Ravera}, {Gumuchian}, {Cl{\'e}net}, {Bertrand},
  {Murat}, {Oziol}, {Camus}, {Pointecouteau}, and {Barret}}]{Ravera2018}
{Ravera} L, {Gumuchian} P, {Cl{\'e}net} A, et~al (2018) {First results of the
  ATHENA/X-IFU digital readout electronics prototype}. In: {den Herder} JWA,
  {Nikzad} S, {Nakazawa} K (eds) Space Telescopes and Instrumentation 2018:
  Ultraviolet to Gamma Ray, p 106994V, \doi{10.1117/12.2313519}

\bibitem[{Smith et~al(2021)Smith, Adams, Bandler, Beaumont, Chervenak, Denison,
  Doriese, Durkin, Finkbeiner, Fowler, Hilton, Hummatov, Irwin, Kelley,
  Kilbourne, Leutenegger, Miniussi, Porter, Reintsema, Sadleir, Sakai, Swetz,
  Ullom, Vale, Wakeham, Wassell, and Witthoeft}]{Smith2021}
Smith SJ, Adams JS, Bandler SR, et~al (2021) Performance of a broad-band,
  high-resolution, transition-edge sensor spectrometer for x-ray astrophysics.
  IEEE Transactions on Applied Superconductivity 31(5):1--6.
  \doi{10.1109/TASC.2021.3061918}

\bibitem[{{Smith} et~al(2023){Smith}, {Witthoeft}, {Adams}, {Bandler},
  {Beaumont}, {Chervenak}, {Cumbee}, {Eckart}, {Finkbeiner}, {Hull}, {Kelley},
  {Kilbourne}, {Leutenegger}, {Porter}, {Sakai}, {Wakeham}, and
  {Wassell}}]{Steve2023}
{Smith} SJ, {Witthoeft} MC, {Adams} JS, et~al (2023) {Correcting Gain Drift in
  TES Detectors for Future X-Ray Satellite Missions}. IEEE Transactions on
  Applied Superconductivity 33(5):3258908. \doi{10.1109/TASC.2023.3258908}

\bibitem[{{Szymkowiak} et~al(1993){Szymkowiak}, {Kelley}, {Moseley}, and
  {Stahle}}]{Szymkowiak1993}
{Szymkowiak} AE, {Kelley} R, {Moseley} SH, et~al (1993) {Signal processing for
  microcalorimeters}. Journal of Low Temperature Physics, 93(3) 93:281--285

\bibitem[{Taylor(1996)}]{093570275X}
Taylor JR (1996) An Introduction to Error Analysis: The Study of Uncertainties
  in Physical Measurements, 2nd edn. University Science Books

\bibitem[{mpmath~development team(2023)}]{mpmath}
mpmath~development team T (2023) mpmath: a {P}ython library for
  arbitrary-precision floating-point arithmetic (version 1.3.0). {\tt
  http://mpmath.org/}

\bibitem[{{van de Hulst} and {Reesinck}(1947)}]{1947ApJ...106..121V}
{van de Hulst} HC, {Reesinck} JJM (1947) {Line Breadths and Voigt Profiles.}
  The Astrophysical Journal 106:121. \doi{10.1086/144944}

\bibitem[{Virtanen et~al(2020)Virtanen, Gommers, Oliphant, Haberland, Reddy,
  Cournapeau, Burovski, Peterson, Weckesser, Bright, {van der Walt}, Brett,
  Wilson, Millman, Mayorov, Nelson, Jones, Kern, Larson, Carey, Polat, Feng,
  Moore, {VanderPlas}, Laxalde, Perktold, Cimrman, Henriksen, Quintero, Harris,
  Archibald, Ribeiro, Pedregosa, {van Mulbregt}, and {SciPy 1.0
  Contributors}}]{2020SciPy-NMeth}
Virtanen P, Gommers R, Oliphant TE, et~al (2020) {{SciPy} 1.0: Fundamental
  Algorithms for Scientific Computing in Python}. Nature Methods 17:261--272.
  \doi{10.1038/s41592-019-0686-2}

\bibitem[{Wilcoxon(1945)}]{Wilcoxon1945}
Wilcoxon F (1945) Individual comparisons by ranking methods. Biometrics
  Bulletin 1(6):80--83. \urlprefix\url{http://www.jstor.org/stable/3001968}

\end{thebibliography}

\end{document}